\documentclass[fleqn,usenatbib]{mnras}

\usepackage{newtxtext,newtxmath}

\usepackage[T1]{fontenc}

\DeclareRobustCommand{\VAN}[3]{#2}
\let\VANthebibliography\thebibliography
\def\thebibliography{\DeclareRobustCommand{\VAN}[3]{##3}\VANthebibliography}

\usepackage{graphicx}	
\usepackage{amsmath}	

\usepackage{lipsum}
\usepackage[english]{babel}
\usepackage{color}
\usepackage[dvipsnames]{xcolor}

\usepackage{multirow}
\usepackage{cleveref}
\usepackage{subcaption}
\crefformat{section}{\S#2#1#3} 
\crefformat{subsection}{\S#2#1#3}
\crefformat{subsubsection}{\S#2#1#3}
\usepackage{lscape} 
\usepackage{romannum}

\title[The DBL Survey I]{The DBL Survey I: discovery of 34 double-lined double white dwarf binaries}

\author[J.\ Munday et al.]{James Munday,$^{1}$\thanks{Email: james.munday98@gmail.com}
Ingrid Pelisoli,$^{1}$
P.-E. Tremblay,$^{1}$
T. R. Marsh,$^{1}$
Gijs Nelemans,$^{2,3,4}$
Antoine B\'edard,$^{1}$
\newauthor
Silvia Toonen,$^{5}$
Elm\'e Breedt,$^{6}$
Tim Cunningham,$^{7\thanks{NASA Hubble Fellow}}$
Mairi W. O'Brien,$^{1}$
Harry Dawson$^{8}$\\
$^{1}$Department of Physics, Gibbet Hill Road, University of Warwick, Coventry CV4 7AL, United Kingdom\\
$^{2}$ Department of Astrophysics/IMAPP, Radboud University, P.O. Box 9010, 6500 GL Nijmegen, The Netherlands\\
$^{3}$ Institute for Astronomy, KU Leuven, Celestijnenlaan 200D, 3001 Leuven, Belgium\\
$^{4}$ SRON, Netherlands Institute for Space Research, Niels Bohrweg 4, 2333 CA Leiden, The Netherlands\\
$^{5}$ Anton Pannekoek Institute for Astronomy, University of Amsterdam, 1090 GE Amsterdam, The Netherlands\\
$^{6}$Institute of Astronomy, University of Cambridge, Madingley Road, Cambridge CB3 0HA, UK\\
$^{7}$Center for Astrophysics, Harvard \& Smithsonian, 60 Garden Street, Cambridge, MA 02138, USA\\
$^{8}$Institute for Physics and Astronomy, University of Potsdam, Karl-Liebknecht-Str. 24/25, 14476 Potsdam, Germany
}

\date{Accepted 2024 July 02. Received 2024 July 02; in original form 2024 May 18}

\pubyear{2024}

\begin{document}
\label{firstpage}
\pagerange{\pageref{firstpage}--\pageref{lastpage}}
\maketitle

\begin{abstract}
We present the first discoveries of the double-lined double white dwarf (DBL) survey that targets over-luminous sources with respect to the canonical white dwarf cooling sequence according to a set of well-defined criteria. The primary goal of the DBL survey is to identify compact double white dwarf binary star systems from a unique spectral detection of both stars, which then enables a precise quantification of the atmospheric parameters and radial velocity variability of a system. Our search of 117 candidates that were randomly selected from a magnitude limited sample of 399 yielded a 29\% detection efficiency with 34 systems exhibiting a double-lined signature. A further 38 systems show strong evidence of being single-lined or potentially-double-lined double white dwarf binaries and 7 single-lined sources from the full observed sample are radial velocity variable. The 45 remaining candidates appear as a single WD with no companion or a non-DA white dwarf, bringing the efficiency of detecting binaries to 62\%. Atmospheric fitting of all double-lined systems reveals a large fraction that have two similar mass components that combine to a total mass of 1.0--1.3\,$\mathrm{M}_\odot$ -- a class of double white dwarf binaries that may undergo a sub-Chandrasekhar mass type~\Romannum{1}a detonation or merge to form a massive O/Ne WD, although orbital periods are required to infer on which timescales. One double-lined system located 49\,pc away, WDJ181058.67+311940.94, is super-Chandrasekhar mass, making it the second such double white dwarf binary to be discovered.
\end{abstract}




\section{Introduction}
White dwarfs (WDs) are the final evolutionary stage of approximately 95\% of stars in the universe and come in multiple flavours, with their most common spectral types being: DA (Balmer lines), DB (HeI lines), DC (continuum only, featureless), DO (HeII lines), DQ (carbon features) or DZ (metal lines). As a large fraction of stars form in a binary pair, there exists a population of wide double WD (DWD) binaries that formed at large orbital separations where both stars have evolved in near-isolated conditions without ever coming into contact \citep[][]{Heintz2022WDageDWD, Heintz2024}, and DWDs that exist in compact configurations that have survived at least one phase of mass transfer \citep{Webbink1984DWDprogenitorsRCrB, IbenTutukov1984}. DWDs in compact configurations are of particular galactic interest, long suspected to contribute towards a large fraction of type \Romannum{1}a supernovae \citep[see][for a review]{Maoz2014Type1aProgenitors} that are responsible for the enrichment of the interstellar medium, and to be an explanation for the existence of exotic merger remnants \citep[e.g.][]{Webbink1984DWDprogenitorsRCrB, Zhang2012FormationOfHeRichHotSubdwarfs}. Furthermore, compact DWD binaries serve as surviving test subjects of the common envelope phase \citep{Nandez2015RecombinationEnergyDWDformation}, which can be used to probe the conditions that may lead to the survival or demise of a binary through population modelling \citep{Toonen2012type1aCommonEnvelope, Toonen2017, Korol2022}. As an endpoint of binary evolution, the overall percentage of DWDs in the Milky Way compared to the number of WDs is predicted to be $\approx$5--10\% \citep{MaxtedMarsh1999, Maoz2017spy,MaozHallakounBadenes2018,Napiwotzki2020spy,Korol2022gapDWDseparationGaia, OBrien2024} and thousands are detectable with space-based gravitational wave detectors in the mHz regime \citep{Lamberts2019, LISAwhitepaper2023}.

The first definitive discovery of a compact DWD binary was L870$-$2, identified as a DA+DA and double-lined system with an orbital period of 1.6\,d \citep{Saffer1988}, sparking the inspiration for dedicated searches in the decade that followed \citep{1987ApJ...322..296R, 1990ApJ...365L..13B, 1991ApJ...374..281F, 1995MNRAS.275..828M, 1995MNRAS.275L...1M, 1997MNRAS.288..538M}. Since then and predominantly in the last two decades, there has been a boom in the number of compact DWD binaries discovered. Many stem from the pioneering work of two surveys: the ESO supernovae type Ia progenitor survey (SPY) \citep{Napiwotzki2020spy} and the Extremely Low Mass (ELM) survey \citep{Brown2020elmNorthFinal, Kosakowski2023elmSouth}. Of the 643 DA WDs with measured RVs observed within the SPY survey, 39 double degenerate binaries were identified; 20 being double-lined and 19 showing single-lined RV variability \citep{2001A&A...378L..17N, 2002A&A...386..957N, 2003A&A...410..663K, 2005A&A...440.1087N, 2010A&A...515A..37G, 2011A&A...528L..16G, Napiwotzki2020spy}. From the ELM survey, \citet{Brown2020elmNorthFinal} report a final northern sample of 98 DWD binaries and thusfar a further 34 ELM WD binaries have been found in the southern sky \citep{KosakowskiELMsouth1, Kosakowski2023elmSouth}. Considering that the current sample size of compact DWD binaries comprises approximately 270 systems, together these surveys constitute the discovery of half of the sample, highlighting the need to be careful when arriving at conclusions about the population. Both surveys have clear observational biases that, while possible to include, can be difficult to incorporate into model comparisons \citep{Kilic2011a, Li2019formationOfELMs}. For instance, the SPY sample relied on white dwarf catalogues built from surveys targeting blue objects; as a result, it is biased against reddened regions and against cooler white dwarfs. The ELM survey has well defined colour (and hence temperature) selection criteria, however there is a priority in follow-up campaigns for the shorter orbital periods and their selection on orbital periods is less documented.

Observational biases also affect photometric surveys, which have been the leading means of identifying short-period double-degenerate binaries through their photometric variability \citep[e.g.][]{Burdge2020systematic, Keller2022ztfGaia, Roestel2022, Ren2023}. Such systems probe the environment of DWDs on the cusp of merger, detonation or the survival of the binary. Yet naturally, photometric searches are more sensitive to stars that have larger radii (therefore lower WD masses) and are hotter due to a combination of reflection effects, eclipse likelihood and the improved photometric precision of bright systems. In the context of DWDs, this leads to an unavoidable deficit of high-mass binaries in photometric searches and the only means to overcome large observational biases is through complete, volume-limited samples \citep{Toonen2017}.

The drawback is that volume-limited samples depend on a smaller subset of systems since complete samples out to large distances are observationally expensive. Recently, all-sky, multi-object spectrograph surveys have provided a means to improve the statistics for synthetic population comparison due to the sheer number of WDs of all types that can be observed \citep[]{BadenesMaoz2012}. Through a combined comparison with the Sloan Digital Sky Survey (SDSS) and the SPY survey, \citet{MaozHallakounBadenes2018} speculate that 10\% of the single WD population are a result of DWDs merger events, serving as a means to explain a `broad shoulder' of WDs between 0.7--0.9\,\(\textup{M}_\odot\) \citep[see also][for other interpretations on the shoulder in the DA mass distribution]{Tremblay2016,Cunningham2024}. Meanwhile, \citet{Kilic2020sdss100pc} find that mergers of all kinds \citep{Temmink2020} are a contributing factor but are insufficient to reproduce the broad shoulder in volume-limited populations.

A new means of detecting thousands of wide binary companions to a WD has lately become possible through astrometric solutions using \textit{Gaia} \citep{GaiaDR1,GaiaDR3_2023}. \citet{ElBadry2018commonProperMotion, ElBadry2021edr3properMotion} revealed hundreds of wide DWD binaries through the means of a common proper motion and parallax between the two stars ($\gtrapprox$100\,au). As for the more compact systems, signatures of binarity in the local population can be predicted through a low-amplitude astrometric wobble of the centre of light, which has often been used to select strong candidates through the so called Renormalised Unit Weight Error \citep[RUWE,][]{Belokurov2020ruwe}. Of the likes, \citet{Korol2022gapDWDseparationGaia} analyse RUWE-excessive candidates to identify a gap in the separation distribution of DWDs at approximately 1\,au while finding a consistent DWD fraction with the spectroscopic sample analyses and indicating a sensitivity in the detection of DWDs with separations $a$ of 0.01\,au~<~a~<~2\,au.

Looking towards the future, upcoming \textit{Gaia} data releases should begin to expand the time baseline of astrometric measurements to probe DWDs separations <0.01\,au and reveal many more DWD candidates at further distances through time-series photometry \citep{Steen2024}. An ever growing completeness in the spectroscopically observed sample will build better sensitivity for radial velocity (RV) variations over all orbital periods, and the gravitational-wave signature of ultra-compact binaries across the entire Milky Way will be unlocked through the launch of the Laser Interferometer Space Antenna \citep[LISA,][]{AmaroSeoane2023lisaWhitePaper}, providing a few thousands of systems \citep{Korol2017prospects, Korol2018detectabilityDWDs, Korol2022, LISAwhitepaper2023,Kupfer2024}. However, while each of these methods have great potential to interpret the orbital parameters of the binary alone, separating one star from the other is far more challenging and relies on a unique signature from both of the stars in the data, best identifiable through eclipsing systems and/or spectroscopically double-lined DWDs.

Of the full confirmed and compact DWD sample, where we have created and made accessible a database that will be regularly maintained at \url{github.com/JamesMunday98/CloseDWDbinaries}, just 46 are known to be double-lined and 20 have a mass precision for both stars of $<0.1\,\mathrm{M}_\odot$. Only with a large sample of precise constraints for both stars can we unravel the nuances of binary star evolution and effectively construct evolutionary models suitable for all initial conditions of two stars. In this pilot study, we introduce the double-lined double white dwarf (DBL) survey which aims to expand the double-lined and well-constrained sample of compact DWDs significantly, constructed with clearly defined selection cuts based on the position of candidate systems in the Hertzsprung-Russell (HR) diagram. As well, we strive to provide a well-constrained sample of compact DWDs surveying a range of orbital separations/periods to facilitate better mapping of post-common envelope binary stars. The undertaken search indicates an approximately one in three success rate of a candidate being double-lined, outlining a clear means to expand to an observable population of precisely constrained DWD binaries and especially so with the addition of phase-resolved spectroscopy (Munday et al, in prep).

\section{The double-lined double white dwarf survey sample}
\label{sec:theSample}
\subsection{Sample selection}
\label{subsec:SampleSelection}
\begin{figure*}
    \centering
    \includegraphics[keepaspectratio, trim={2.5cm 0cm 2.75cm 1.5cm},clip,width=\textwidth]{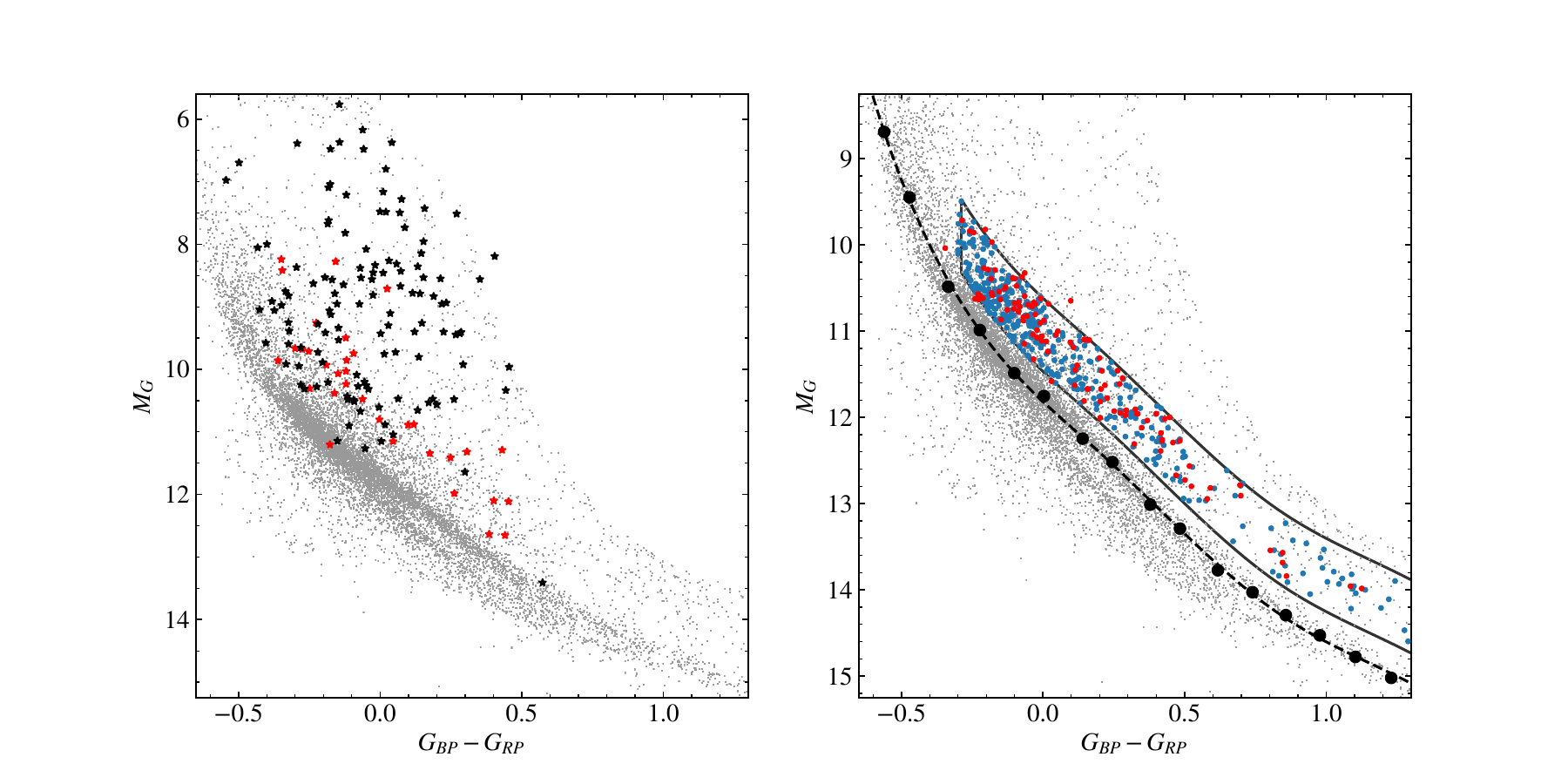}
    \caption{Our sample selection cut on the HR diagram. Underplot on both graphs is the 100\,pc sample of WDs from \citet{NicolaDR2} with $\textsc{Pwd}>0.3$, $\textsc{Gmag}<18$, $\textsc{parallax/parallax\_error}>6$. Left: the full sample of confirmed DWD binaries that were published before this work plotted on the HR diagram. Double-lined systems are denoted by red stars while single-lined systems are shown in black. Right: the 399 DWD candidate sources available for our WHT identification spectra with $G<17$\,mag and $\delta>-20^\circ$ in blue and in red the candidates that we observed. Both use an updated HR diagram location using \textit{Gaia}~DR3 astrometric solutions, which is why some of the observed candidates now fall slightly outside of the search box. In filled circles are evenly spaced colour bins with a dashed spline passing through, which we used to define the boxed space slightly above with step (2) of the selection criteria.}
    \label{fig:sampleSelection}
\end{figure*}
Beginning the survey in 2018, we initiated the target selection of the DBL survey using the latest \textit{Gaia} data release available at the time: \textit{Gaia} DR2 \citep{GaiaDR2}. We started with the full catalogue of 486\,641 candidate WD sources presented in \citet{NicolaDR2}, all of which assigned a probability of being a WD ($\textsc{Pwd}$).

In search of double-lined DWD binaries, we considered a few steps. Firstly, we consider that the majority of the population of WDs fall along a typical cooling track of a 0.6\,\(\textup{M}_\odot\) DA WD and use this as a baseline for the expected luminosity of a single star. As shown in Fig.~\ref{fig:sampleSelection}, this track traverses the A-branch of WDs that is dominated by a population of hydrogen-rich atmosphere DA WDs \citep{Esteban2023}, implying that the majority of systems were expected to have Balmer absorption lines usable for RV extraction.
Considering that we sought to find double-lined DWDs where the dimmer companion is not washed out of the spectrum by the brighter star, we aimed for the temperature of the brighter star to be less than 20\,000\,K. Such a 0.6\,\(\textup{M}_\odot\) WD has $G_{BP}-G_{RP}\approx-0.3$\,mag, therefore, all targets in the selection required $G_{BP}-G_{RP} > -0.3$\,mag.

Next, we required that the flux ratio between the companion and the brighter star is greater than a third to show a clear spectroscopic signature, meaning that a second star would cause the binary to be 0.31\,mag above the single-star cooling track.
Factoring in the uncertainty of a target's \textit{Gaia} HR diagram location from errors in the parallax and photometry, we slightly increase this minimum value and required the brightness of the binary to be 0.35\,mag above the 0.6\,\(\textup{M}_\odot\) WD cooling track to improve the search efficiency. 

Going too far off of the cooling track would also contaminate our selection and its efficiency by introducing an abundance of WD--MS star candidates of the likes of cataclysmic variables \citep[e.g.][]{Abril020CV_HRdiagram}, and with this in mind we limited the selection to systems with an absolute magnitude less than 1.2\,mag brighter than the 0.6\,\(\textup{M}_\odot\) WD cooling track. This upper boundary coincides with the typical cooling track for a helium-core WD of mass 0.3\,\(\textup{M}_\odot\), hence excluded DWD binaries from this upper brightness limit are those with an approximately 0.35\,\(\textup{M}_\odot\) or less massive component. The cuts applied to the \citet{NicolaDR2} catalogue conclude as written in Table~\ref{tab:SelectionCuts}.
\begin{table}
\normalsize
\begin{tabular}{l|r}
\multicolumn{1}{l}{\textbf{Starting DR2 WD catalogue:} 486\,641 sources}\\
\hline
Selection for creating spline:\\
(\textsc{Pwd} > 0.3) \textsc{and} \   (\textsc{Gmag} < 18) \textsc{and} \    (\textsc{Plx} > 6.*\textsc{e\_Plx}) & \\
22\,107 sources\\
\\
Sample 20 evenly spaced bins between BP\_RP = $-$0.65 \& \\
1.65, fit with a spline \textsc{spl\_MG}~=~f(BP\_RP).\\
\hline
Select for a `clean' sample of candidate targets:\\
(\textsc{Gmag} < 18.) \textsc{and} \  (\textsc{Plx} > 10.*\textsc{e\_Plx}) & (1) \\  
22756 sources\\
\\
(\textsc{M\_G} < \textsc{spl\_MG}$-$0.35) \textsc{and} (\textsc{M\_G} > \textsc{spl\_MG}$-$1.2)  & (2)\\  
3494 sources\\
\\
(\textsc{BP\_RP} > $-$0.3)   & (3)\\
2697 sources\\
\hline
Select for an observable sample on the WHT telescope:\\
(\textsc{Gmag} < 17.) & (4)\\
625 sources \\
\\
($\textsc{DEdeg}>-20$) & (5)\\
399 sources\\
\hline
\end{tabular}
\caption{The selection cuts applied to target DWD candidates along with the number of sources after each cut. Top: spline of the typical WD cooling sequence. Middle: selection of DWD candidates. Bottom: observable targets for the WHT.}
\label{tab:SelectionCuts}
\end{table}

After applying a selection cut of \textsc{Gmag} < 17 to select targets that would grant a signal-to-noise (S/N) ratio greater than 15 per spectral element at the centre of H$\alpha$ in less than a 30\,min exposure time with the 4.2\,m William Herschel Telescope (WHT, see Section~\ref{sec:Observations}), we were left with 625 sources. Lastly, we selected targets that were visible on site for prolonged durations (\textsc{$\delta$} > $-$20$^\circ$) to arrive at a final sample that was suitable for identification spectra of 399 sources.

We note that after applying the same cuts to \textit{Gaia} DR3, each step of the sample selection had the following number of sources utilising the same numbering scheme: 1) 26\,616, 2) 3729, 3) 2862, 4) 695, 5) 451. The primary cause of the increase in size of the final sample stems from the improved parallax uncertainties in \textit{Gaia} DR3.

\begin{figure}
    \centering
    \includegraphics[trim={0.5cm 0.5cm 0.5cm 1.5cm},clip,width=\columnwidth,keepaspectratio]{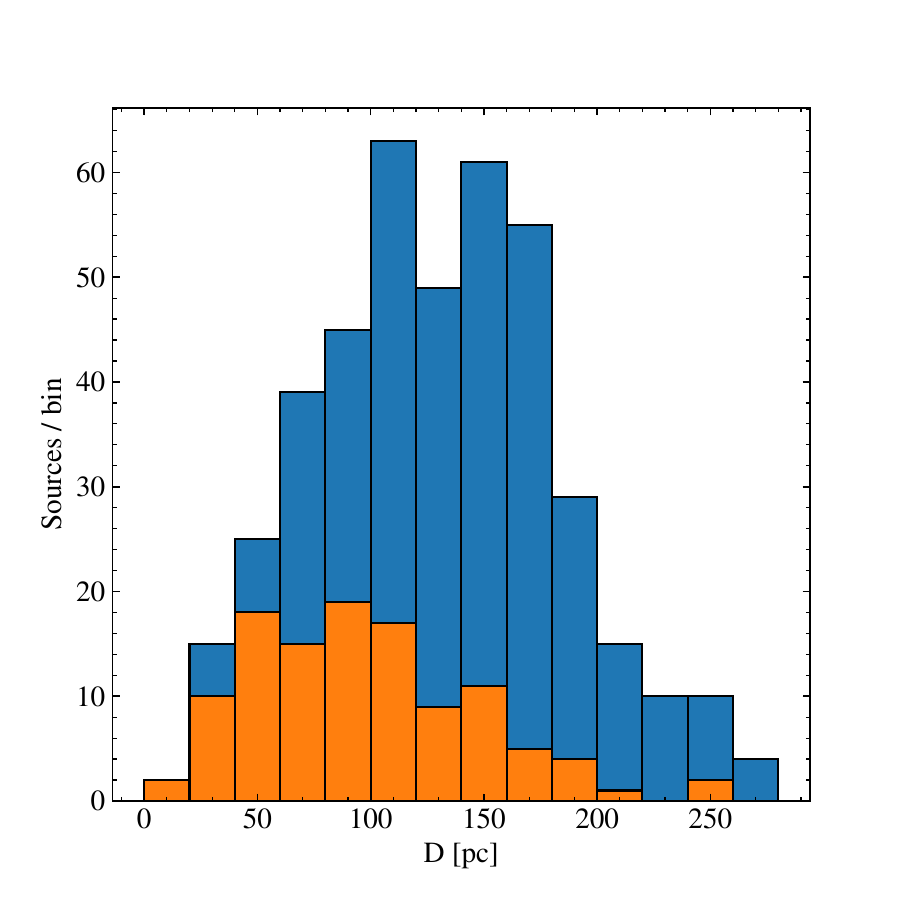}
    \caption{The distance distribution of the observable WD sample for the WHT identification sample, following steps (1) to (5) of the selection cuts mentioned in Section~\ref{subsec:SampleSelection} when applied to the \textit{Gaia}~DR3 catalogue. Distances were obtained using \textit{Gaia} DR3 parallaxes. Displayed in blue is the full DR3 candidate sample and in orange the sample that we have observed.}
    \label{fig:candidateDistances}
\end{figure}

\subsection{Sample biases}
The sample cuts outlined in steps (1) to (3) in Table~\ref{tab:SelectionCuts} cause the sample to be slightly biased towards brighter systems owing to the fact that brighter systems have more precise astrometric solutions, hence smaller parallax uncertainties. This is particularly relevant given the initial \textit{Gaia} DR2 selection of the candidate sample, but far less so for \textit{Gaia} DR3. The fact that the selection relies on a 0.6\,M$_\odot$ WD cooling track and that this is the median WD mass induces a natural bias towards at least one of the components being a WD of mass 0.6\,\,M$_\odot$ ($\log\,g\approx8$). This is primarily the case for wide binaries where the two stars evolve in near-isolated conditions. WDs with smaller masses have larger radii and are hence more luminous for all effective temperatures compared to a WD with higher mass. With this and the fact that the upper M$_G$ selection cut of 1.2\,mag above the 0.6\,M$_\odot$ cooling track coincides with that of a 0.3\,M$_\odot$ helium-core WD, we expect no DWD binaries that host an extremely-low-mass WD.

By applying a magnitude limit to the sample in steps (1) and (4), we induce a strong bias towards the frequency of hotter systems in the sample, which is clearly noticeable by the grouping of candidate systems with $\mathrm{M}_G<12$\,mag in Fig.~\ref{fig:sampleSelection}. The magnitude limit imposed in the spline creation barely induces bias in the sample due to the large number of sources used in the spline fit, and the premise of introducing the magnitude limit in the step of creating the sample was to ensure good photometry across the BP-, RP- and G-bands.

The target selection, especially near the lining of the Milky Way and towards the Galactic bulge, is minorly biased towards brighter systems due to the impact of reddening since we applied no reddening correction to the sample selection. The most distant sources in the sample are found at 300\,pc and the mean distance is $\approx$120\,pc (Fig.~\ref{fig:candidateDistances}), meaning that the impact of extinction is small and that even though the targets in our selection cuts will shift in colour to become bluer where some may lie slightly outside of the selection, the number that will is very small.

Limiting to a northern declination in step (5) is suspected to have negligible effects owing to the similarity between the results of volume-limited WD surveys in each hemisphere \citep{McCleery2020_40pcNorth, OBrien2023_40pcSouth}. A final bias in the results of the targets in this sample, which does not stem from the sample selection itself, are introduced by the detection efficiency in the observing strategy (see Section~\ref{sec:Observations}).

\section{Observations}
\label{sec:Observations}
\begin{figure*}
    \centering
    \includegraphics[trim={1.0cm 0cm 1.75cm 1.75cm},clip,width=\columnwidth]{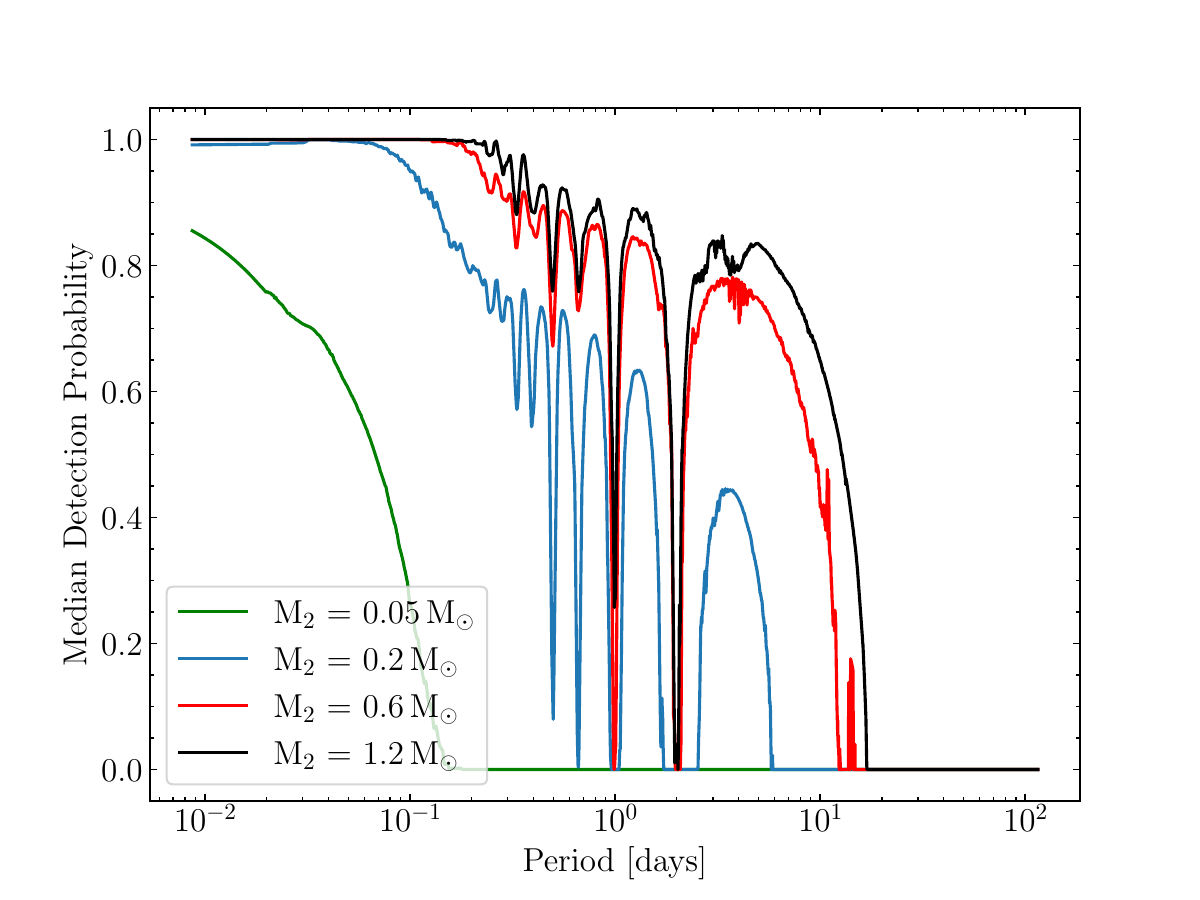}
    \includegraphics[trim={1.0cm 0cm 1.75cm 1.75cm},clip,width=\columnwidth]{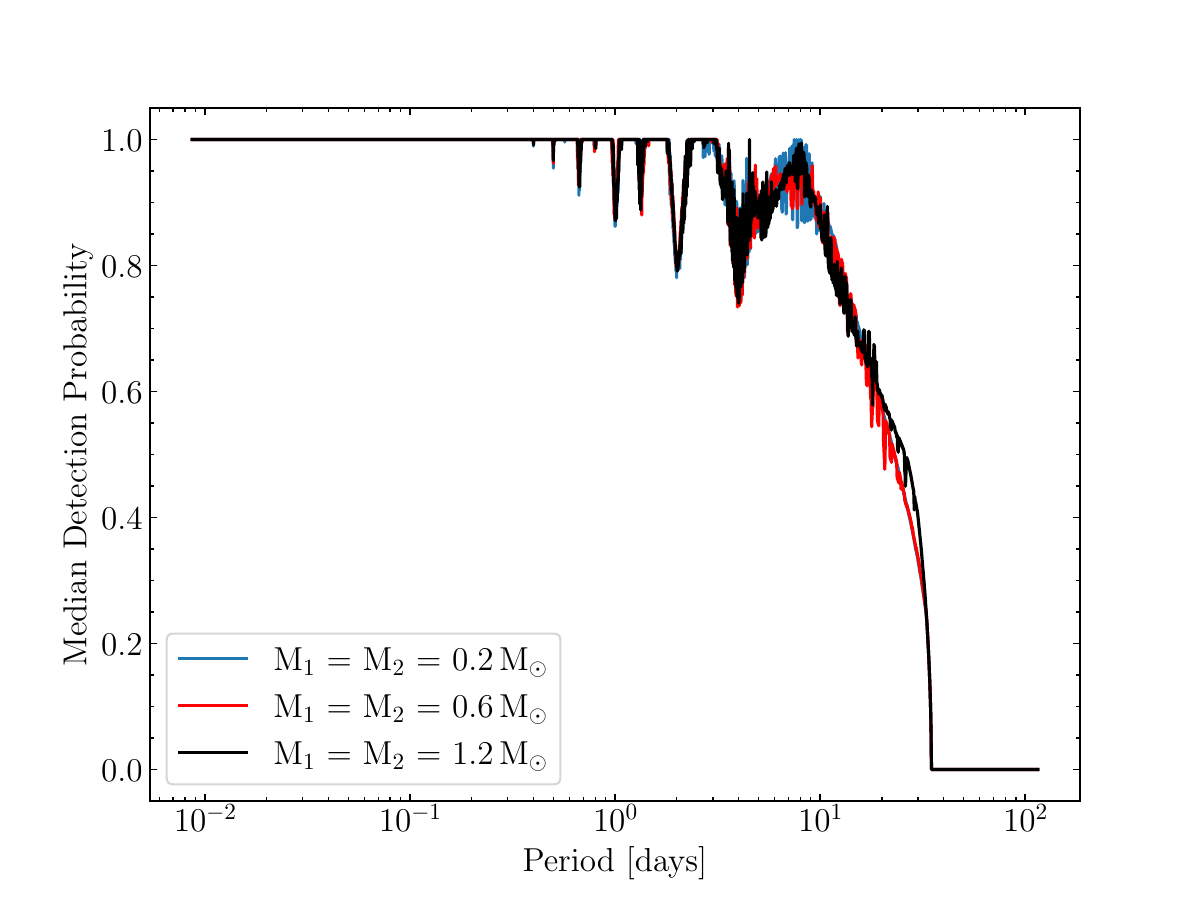}
    \caption{Left: The detection efficiency for a single-lined binary. Plotted is the detection probability for a 4$\sigma$ or larger significance of a non-constant RV over a range of orbital periods with our WHT observing strategy. A brighter WD with mass 0.6\,M$_\odot$ is assumed and the detection efficiency of a companion WD or brown dwarf with mass M$_2$ is plotted for a system inclination of 60$^\circ$. The detection efficiency for a larger companion (e.g. a neutron star) is near identical to the efficiency for a 1.2\,M$_\odot$ companion. The plotted detection probability is the median of all systems with three or more WHT identification epochs using the timestamps of the observations. Right: The same as the left, but for a double-lined DWD where the two stars have a similar flux contribution. Here, the difference between the RV of the two stars to be deemed separated at $\textrm{R}=8\,000$ (0.81\AA) is 60\,km\,s${^{-1}}$, and this requirement only needs satisfying on one epoch to result in a detection. Repeating dips in the detection probability are predominantly aliases of a $\approx$1\,day period, arising due to the fact that ground-based observations are only possible at nighttime. There is likely a small difference in the detection efficiency dependent on the masses of the stars due to H$\alpha$ being broader for larger values of surface gravity, which hence makes the double-lined signature less likely to be detected, but this is difficult to quantify and is unaccounted for since the surface gravity of the stars in the sample are suspected be in a narrow range of $\log\,g\approx7.75$--8.25\,dex (see also Fig.~\ref{fig:appendixSensitivityLogg} for the temperature and surface gravity sensitivity for double-lined systems).}
    \label{fig:DetectionEfficiency}
\end{figure*}
With the final candidate sample outlined in Section~\ref{subsec:SampleSelection}, we obtained identification spectra of 117 candidates; 27 were observed for a single epoch, 9 for two epochs, and 88 candidates for 3 or more. Typically, more than 3 identification epochs were taken if the candidate was immediately identified as a promising double-lined candidate from the first 3 spectra. Our general observing strategy was to ensure that each candidate was observed on at least two separate nights spaced two nights apart; one of these nights with a single spectrum and the other night with two consecutive exposures to be sensitive to any ultra-compact systems with rapid RV variation. Generally speaking, compact DWDs have a median orbital period of approximately 1\,day, meaning that the two day spacing granted sensitivity orbital periods up to a few days (Fig.~\ref{fig:DetectionEfficiency}), comprising over 90\% of compact DWD systems \citep{Nelemans2001closeWDs}.

These identification spectra were taken with the Intermediate-dispersion Spectrograph and Imaging System (ISIS) mounted on the 4.2\,m William Herschel Telescope (WHT) across 20 nights, being 28 August 2018 - 5 September 2018, 13-15 February 2019, 19 February 2019, 15-17 April 2019, 10 \& 11 June 2019 and 6 \& 7 July 2019. ISIS utilises a dichroic filter to be able to observe in a blue and a red arm simultaneously, where we used the R600B grism for the blue arm centred on 4400\,\AA~giving a wavelength coverage of $\approx$3770--5030\,\AA~and the R1200R grism for the red arm centred on 6562\,\AA~giving a wavelength coverage of $\approx$6250--6870\,\AA. We used a 1.0\arcsec slit width on all nights in 2018 and 1.1\arcsec for the others, having a spectral resolution of 0.81\,\AA~and 0.74\,\AA~in the red arm and 2.2\,\AA~and 2\,\AA~in the blue arm, respectively. 
A CuNe+CuAr arc exposure was taken for each new telescope pointing after acquiring a target for wavelength calibration, where the error in wavelength calibration for each arc frame in the blue arm was approximately 3\,km\,s$^{-1}$ and 2\,km\,s$^{-1}$ in the red arm. The science exposure time was always maintained at under 30\,min not to suffer greatly from orbital smearing and a S/N ratio greater than 25 in the wings and 15 in the line core of H$\alpha$ was sought after.


Bias frames and dome flats were taken at the beginning of each night and used in the data reduction. Spectrophotometric standard stars were also taken at the start or the end of each night to flux calibrate the science exposures and correct for the instrumental response function. When multiple arc frames were taken for a target, the arc closest in time to the science frame was used for wavelength calibration. All data were reduced using the \textsc{Molly} package \citep{Marsh2019Molly} and spectra were extracted using the method outlined in \citet{Marsh1989optimalExtraction}.

The detection probability of a candidate being RV variable or double-lined, based off of the identification spectra alone, is displayed in Fig.~\ref{fig:DetectionEfficiency}. We determine a single-lined system to be RV variable if there is significant deviation from the mean velocity of all exposures (Section~\ref{subsubsec:RVvariability}), while a double-lined system with two stars that contribute a similar flux is identifiable if the difference in velocity of the two stars is above 60\,km\,s$^{-1}$ for the resolution at H$\alpha$.

\section{WD-BASS}
\label{sec:WDBASS}
We created a custom fitting code specialised for the analysis of RV variable WDs with time-series spectra and introduce it in this work to fit the full DBL~\Romannum{1} survey sample: the \textit{White Dwarf Binary And Single Star} (WD-BASS) fitting package \citep{Munday2024WDBASS}. The code is specialised for DA, DB and DC WD fitting and is flexible to include any grid of synthetic spectra.

WD-BASS is divided into 4 primary branches: 1) one-star spectral line fitting, 2) two-star spectral line fitting, 3) simultaneous or independent photometric SED fitting, and 4) RV fitting with a fixed atmospheric solution. Each branch will now be explained.

Access to the code is public and a full description of the code's usage in its most recent version can be found on github\footnote{\url{https://github.com/JamesMunday98/WD-BASS}}.

\subsection{One-star fitting}
Single star fitting in WD-BASS is performed by normalising and fitting specific spectral lines of interest. The user specifies the sets of wavelength ranges on each edge of the line to be used for a linear normalisation and the wavelength range that is fitted. The resolution of the observations at the spectral line of choice is also user supplied. Multiple spectra can be fit simultaneously or independently. One has the capability to fix or vary an atmospheric solution, fix or vary the RV of a star for each spectrum or, if the system is in a binary and the user supplies an orbital period (P) and reference epoch (T$_0$), one can solve for a Keplerian orbit by varying the semi-amplitude (K) and/or the systemic velocity offset (V$_\gamma$). In the latter case, the eccentricity is assumed to be zero with a circularised orbit. Multiple spectral lines in a single spectrum can be fit independently or simultaneously with a fixed/varied RV for each line.

A Markov Chain Monte Carlo (MCMC) algorithm is employed in the code using the \textsc{emcee} python package \citep{ForemanMackey2013emcee} to probe a user-supplied parameter space and maximise the likelihood (which is equal to $-0.5\chi^2$) between the set of normalised observations and the normalised model fits sequentially. The interpolated spectrum for a trial solution is convolved to match the resolution of the observations with a Gaussian kernel.

\subsection{Two-star fitting}
\label{subsec:WDBASS_binary}
The binary star fitting module of WD-BASS functions exactly the same as single star fitting but with the combination of two different model stars that can be of any spectral type, each star having a unique RV. Similar to before, any combination of the semi-amplitude of each star, $K_1$, $K_2$ or their velocity offset, $V_{\gamma, 1}$, $V_{\gamma, 2}$ can be input as fixed/fitted constraints (the velocity offset is not the same as the systemic velocity and is unique for the each star because of gravitational redshift).

Two stars in a binary have unique atmospheric parameters and radii, therefore in a normalised spectrum there is a relative flux scaling between the two. WD-BASS includes the options to fit for the relative flux scaling component, for it to be fixed, or for the scaling to be inferred through the radius of the two stars with T$_{\textrm{eff}}$-$\log\,g$-radius relationships. In its current version, the radius of a DA star is inferred using the evolutionary track models of \citet{Istrate2016} when $\mathrm{M}\leq0.393\,\mathrm{M}_\odot$, \citet{Althaus2013} when $0.393<\mathrm{M}<0.45\,\mathrm{M}_\odot$ and the hydrogen-rich evolutionary sequences of \citet{Bedard2020} otherwise, the assumption being that a star $<0.45\,\mathrm{M}_\odot$ is likely a He-core WD and larger masses have a carbon-oxygen core. Hybrid carbon-oxygen/helium core composition WDs may exist in the mass range of $\approx0.3$--$0.6$\,M$_\odot$ \citep{Zenati2019}, or the hydrogen envelope could be large in some cases \citep{Romero2019}, but a spectrum that is poorly fit in this mass range can be overcome by allowing the scaling factor to vary freely. That said, the scaling is always interpolated in this study as these issues never arose in the fitting, which has often been the case for detached systems \citep{Parsons2017}. For helium-atmosphere DB/DC WDs, WD-BASS currently relies on a linear interpolation of the He-rich evolutionary sequences of \citet{Bedard2020} to predict the radius of the WD (M$_\textrm{H}/\textrm{M}_\textrm{WD}=10^{-10}$).

The model Eddington flux is converted to a flux observed at Earth ($F_{\textrm{obs}}$) with the stellar radius ($R$, inferred from $\log\,g$--T$_{\textrm{eff}}$--radius relationships) and the parallax, following
\begin{equation}
    F_{\textrm{obs}}(\lambda) = \frac{4\pi}{D^2} \left(F_1(\lambda) R_1^2 + F_2(\lambda) R_2^2 \right)
    \label{eqn:EddingtonToObserved}
\end{equation}
where $D=1/\text{parallax}$, $\lambda$ is the wavelength, $F(\lambda)$ is the flux and subscripts 1, 2 represent the brighter and dimmer star, respectively. The fluxes themselves are dependent on T$_{\textrm{eff},1}$, $\log g_1$, T$_{\textrm{eff},2}$, $\log g_2$.

\subsection{Photometric SED fitting}
In the advent of all-sky photometric surveys in the last decades, the astronomical community has a plethora of photometric magnitudes available in multiple wavebands that is publicly available and flux-calibrated. Yet, only recently have the parallaxes measured in the \textit{Gaia} mission proved revolutionary in placing said stars distance scale \citep{GaiaEDR3_2021, GaiaDR3_2023}. For sources up to $G=17$\,mag in \textit{Gaia} DR3 belonging to the 359\,000 high-confidence WDs in the catalogue of \citet{NicolaGaia2021}, 99.96\% of the WDs have a \textsc{parallax/parallax\_error} greater than 10, such that there is now the opportunity to exploit simultaneous fitting of spectral lines with well-flux-calibrated photometry to grant more precise atmospheric parameters. This is highly advantageous not only in single-lined systems, but especially so for double-lined binaries where the unique signature of each star may be difficult to disentangle, also allowing a more accurate system solution \citep{Bedard2017}. When handling photometry from satellite sources, air wavelengths in the model spectrum are converted to vacuum wavelengths.

To fit the photometry, synthetic spectra are created with equation~\ref{eqn:EddingtonToObserved} on an Eddington flux scale and can be reddened in WD-BASS with a supplied E(B-V) extinction constant and an assumed extinction curve with $R_V$=3.1, following the extinction curves of \citet{Gordon2023}. If fitting photometry simultaneously with spectroscopy, the same model is used for both datasets. The model is then integrated over the transmission spectrum of the respective filter to obtain the predicted observed flux in the bandpass. In the MCMC algorithm itself, a Gaussian prior on the parallax is applied and weighted by the supplied parallax error. The parallax, together with the radius inferred from $\log(g)$ and T$_\textrm{eff}$, fixes the normalisation.

\subsection{Improved radial velocity fitting}
It is possible to determine RVs as discussed in the previous sub-sections and this is the standard procedure in WD-BASS. In case of a poor fit to the line core and specifically made for medium- to high-resolution data that includes H$\alpha$, where synthetic model spectra may struggle to reproduce the observed signature \citep[][]{Napiwotzki2020spy, Kilic2021HiddenInPlainSight}, WD-BASS also includes the option to take an atmospheric solution and fit an extra Gaussian component to the star(s) model spectrum (spectra). This process is performed by taking the same set of spectra that were used in the one-star (two-star) atmospheric fitting, fixing the temperature(s), $\log\,g$~(s) and RVs found in the previously determined solution and then minimising the $\chi^2$ with an additional Gaussian solution for the star(s), having amplitude $A_G$ and standard deviation $\sigma_G$. An upper limit of $\sigma_G<5$\,\AA\ is enforced to solely improve the line core, rather than any correction to the shape of the wings of the Balmer lines. Although there is no improvement to the atmospheric fit in the optionally employable method, the purpose of this step is to grant an improved RV accuracy and precision when relevant.

\begin{figure*}
    \centering
    \begin{subfigure}[t]{0.5\textwidth}
        \centering
        \includegraphics[width=10cm, keepaspectratio, clip,trim={0cm 0cm 0cm 1.15cm}]{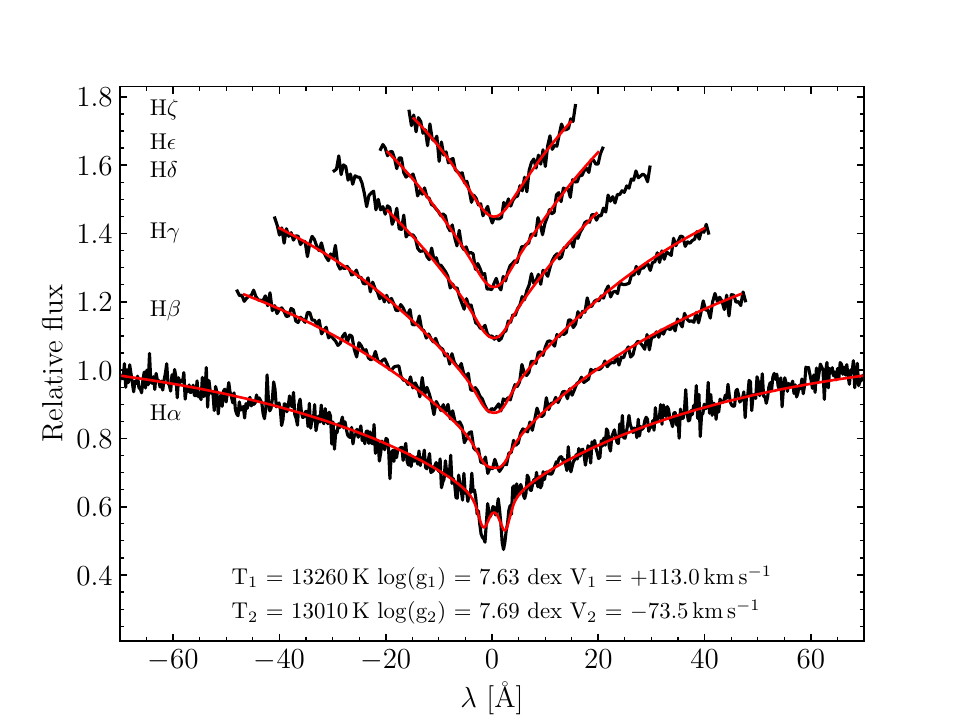}
    \end{subfigure}%
    ~ 
    \begin{subfigure}[t]{0.5\textwidth}
        \centering
        \includegraphics[width=10cm, keepaspectratio, clip,trim={0cm 0cm 0cm 1.15cm}]{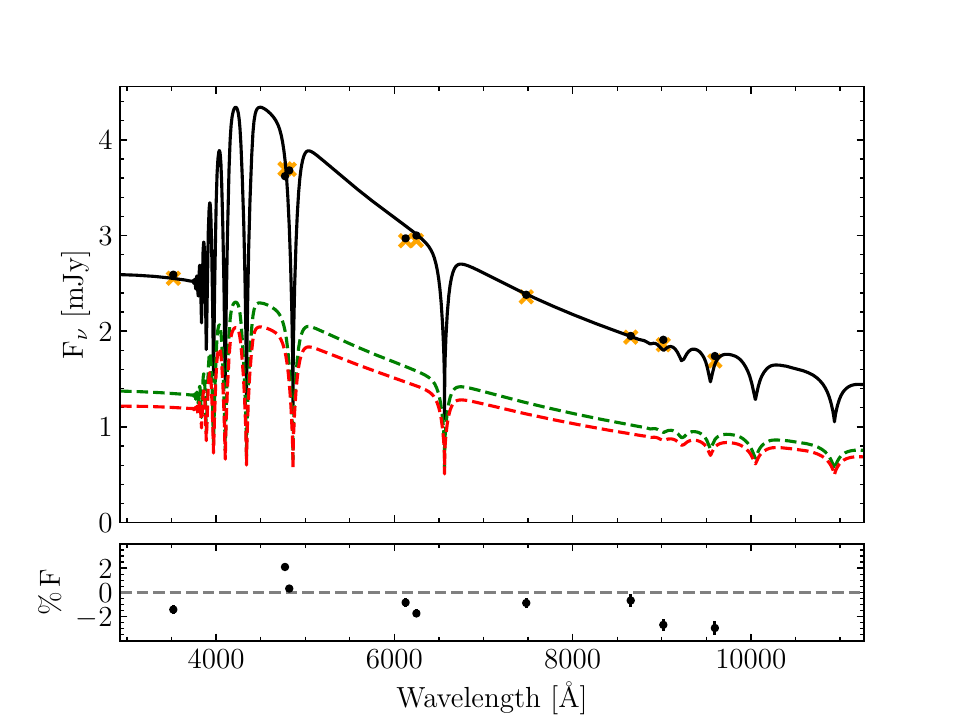}
    \end{subfigure}
    \caption{An example best-fit atmospheric solution for the double-lined system WDJ114446.16+364151.13 utilising simultaneous spectral (left) and photometric (right) fitting. On the left, the observed spectra is plotted in black and the model fit is over-plotted in red. The spectroscopic solution and the RVs of the epoch are written in text below H$\alpha$. On the right (top), the observed data is plotted in black circles, the integrated model flux in each filter in orange crosses, the combined model flux as a solid black line and the flux contributed from the individual stars in dashed red and dashed green. Below is the percentage difference of the observed flux to the model.}
    \label{fig:goldenFit}
\end{figure*}
\section{Methods: Fitting the candidate sample}
\label{sec:Methods}
\subsection{Atmospheric parameters}
For the fitting of all WD atmospheric classifications, we performed a bilinear interpolation in the T$_{\textrm{eff}}$ -- $\log\,g$ space in trial solutions for DA WDs or trilinear in the T$_{\textrm{eff}}$ -- $\log\,g$--H/He space for DB WDs.

The observed spectra were normalised with a linear fit between the two wings of each spectral line and so were the model spectra (the spectral range for normalisation is different depending on the spectral type, detailed in Sections~\ref{subsubsec:MethodsFittingDAWDs}, \ref{subsubsec:MethodsFittingNonDAs}). Our method in choosing the spectral region for normalisation was proportional to the temperature of the WD, with cooler WDs having less-broadened absorption lines. All data within the region of normalisation was included in a $\chi^2$ minimisation with a 4$\sigma$ outlier rejection threshold. We probed trial solutions using the MCMC algorithm with WD-BASS, utilising 100 walkers, a burn-in and a post-burn-in phase. In the case of double-lined systems where multiple exposures of a target were taken, we omit data where both stars are not visible in the spectrum in the atmospheric fitting (originating from an unfavourable orbital phase). In single-lined sources, we model all WHT identification spectra.

Multiple studies have shown a systematic difference between atmospheric constraints predominantly from fitting single WDs photometrically and spectroscopically that varies from survey to survey  \citep{Tremblay2019accuracyDADBgaiaDR2, GenestBeaulieu2019, NicolaGaia2021, ElenaCukanovaite2021_3D_DB, Sahu2023, Izquierdo2023}, stemming from a difference in flux calibration between the surveys and imperfect model atmospheres. However, even with the systematics, all targets within our selection criteria are over-luminous compared to the 0.6\,\,M$_\odot$ WD cooling sequence and, in the vast majority of cases, the spectral features of one star dominate in the spectrum. We consider that the inclusion of photometry to reduce degeneracy in two-star atmospheric solutions outweighs the negatives induced from systematic effects in all cases. Therefore, we incorporate a hybrid fitting technique in WD-BASS of simultaneous spectroscopic and photometric fitting in all systems when available, whereby the combined $\chi^2$ with no weighting for the two datasets is minimised to reach a final solution.

We chose to limit the sources of photometry to Pan-STARRS DR1 (grizy) \citep{Panstarrs} and SDSS DR16 (ugriz) \citep{SDSSdr16} and we chose not to include \textit{Gaia} photometry as the very broad filters of \textit{Gaia} are not optimal for two star fitting,  where any extra flux calibration offset between \textit{Gaia} and the other surveys could hamper the validity of the atmospheric solution. In any case, \textit{Gaia} was still the best starting point for candidate selection due to its increased photometric precision and all-sky coverage. The extinction coefficient, A$_V$, of each source was obtained through the reddening maps of \citet{Lallement2022} at a distance determined from the inverse of the \textit{Gaia} DR3 parallax and was converted to $\textrm{E(B-V)}=\textrm{A}_V/3.1$ before reddening the synthetic spectrum. An example showcasing the quality of fit to H$\alpha$--H8 following all of these steps with simultaneous photometric and spectroscopic hybrid fitting is displayed in Fig.~\ref{fig:goldenFit} for one double-lined DWD.

\subsubsection{DA WDs}
\label{subsubsec:MethodsFittingDAWDs}
\begin{figure}
    \centering
    \includegraphics[keepaspectratio, width=\columnwidth, clip,trim={0.5cm 0cm 1.5cm 1.5cm}]{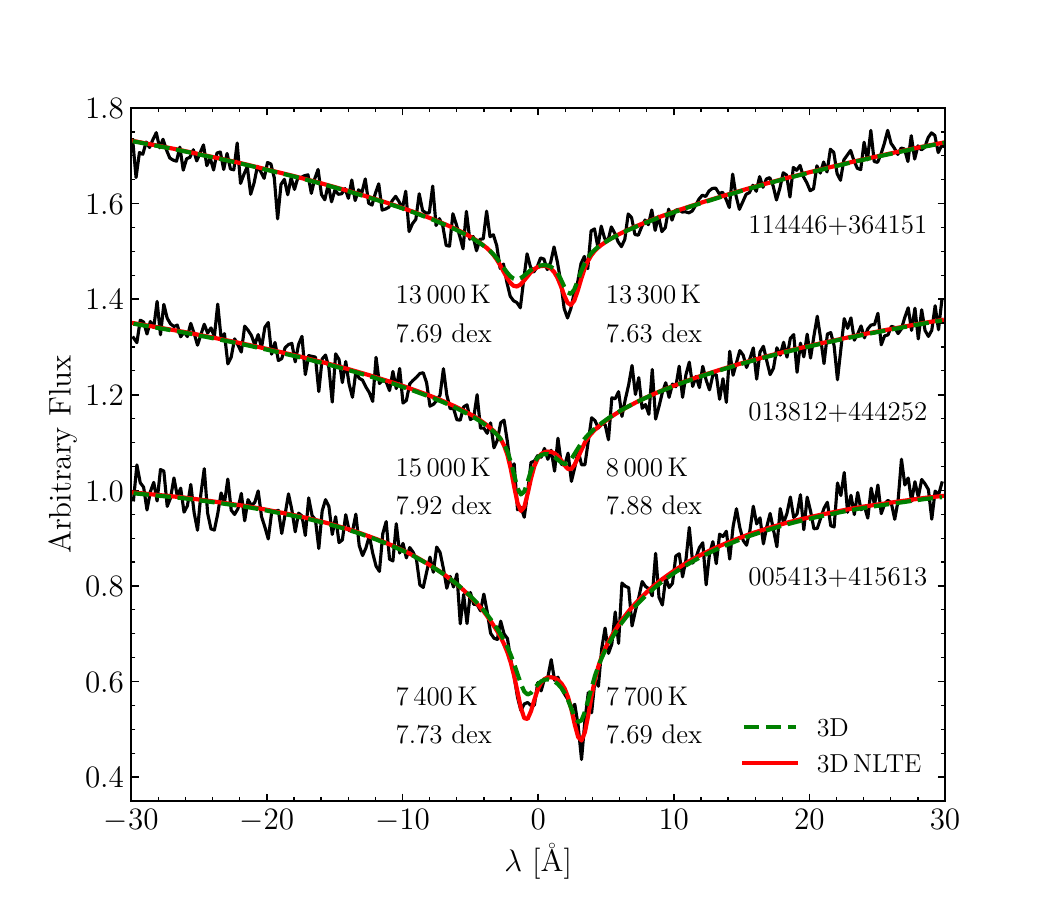}
    \caption{A comparison of the model fit to three double-lined systems with a variety of combinations of effective temperature. In red, the 3D+NLTE grid that was created in this work and used to fit all single- and double-lined systems (see Section~\ref{sec:Methods}). In dashed green, 3D effects only from T15. The lines H$\alpha$-H8 were fitted simultaneously in these systems, but H$\alpha$ is shown alone in the figure since little difference is noticeable at the other Balmer lines because of the lower resolution of the data as well as the lesser impact of NLTE effects to the synthetic spectrum. The relative scaling between stars was included from the appropriate $\log\,g$--T$_\textrm{eff}$--radius relationship (Section~\ref{subsec:WDBASS_binary}).}
    \label{fig:comparison3DNLTEGrids}
\end{figure}
Over the course of this study, we fitted the full sample of DWD binary candidates with two primary grids of model spectra: the 3D local thermodynamic equilibrium (LTE) DA spectra of \citet{Tremblay2013, Tremblay2015} and the 1D non-LTE (NLTE) DA spectra of \citet{Kilic2021HiddenInPlainSight}, which we will refer to as T15 and K21, respectively. Both grids rely on the line profiles of \citet{Tremblay2009}. The NLTE grid of K21 was computed by performing NLTE spectral synthesis on the LTE atmospheric structures of \citet{Tremblay2009} using a modified version of the SYNSPEC code (see \citealt{ Kilic2021HiddenInPlainSight} for details). NLTE effects are especially important in the atmosphere of hot WDs (T$_{\mathrm{eff}} \gtrsim$ 40\,000\,K; \citealt{Napiwotzki1997}) but also remain non-negligible in the upper atmosphere of cooler objects, thus affecting the line cores \citep{Heber1997, Koester1998}. In particular, they are noticeable in high-resolution H$\alpha$ observations of cool WDs in double-lined binaries \citep{Bergeron1989, Kilic2021HiddenInPlainSight}. The same effect of a decreased model flux in the line cores is apparent for the 3D vs 1D and NLTE vs LTE grids; in the former case, because of the cooling effects that come with the inclusion of convective overshoot \citep{Tremblay2013} and in the latter because of an increased opacity (increased NLTE lower level population) in the upper atmosphere of a WD where the lines are formed.

We observe that NLTE effects on H$\alpha$ line cores are noticeable at medium resolution for the full range of effective temperatures and surface gravities in our sample. Both the grids of K21 and T15 performed well for all other Balmer lines observed at lower resolution, where the line core impact diminishes due to the increased broadening from the response of the blue grating. That said and even with the use of NLTE physics in the models, the synthetic line cores were too shallow. While the poorer fit caused in this case may not be detrimental to or hardly noticeable in single star fitting of an isolated WD, the impact is amplified in spectra with two DA WD components as the stars are blended, such that the line cores are the only components where a unique signature of each star is prominent and hence significantly impact the validity of the fit.

We were therefore encouraged to create a new synthetic spectral grid that approximately include both the 3D and NLTE effects on the line cores; a hybrid ``3D-NLTE'' grid that we introduce for the first time in this work. We started by dividing the NLTE and LTE grids of K21, which include otherwise identical input physics, yielding a NLTE correction factor $\mathrm{f}\left(\lambda, T_{\textrm{eff}}, \log\,g\right)$. This factor was then interpolated onto the wavelength grid of T15 and applied to their 3D synthetic spectra. The final step was to broaden the product\footnote{Convective (turbulent) velocity broadening is not included by default in the synthetic $\left< {\rm 3D} \right>$ spectral grid in T15.} with the velocity vector in figure~11 of \citet{Tremblay2013aPureH3DmodelsWDs} at optical depth $\log(\tau_R)=-4$. This generated a NLTE correction to a 3D grid with resulting synthetic spectra (that will herein be referred to as 3D-NLTE) that incorporate 3D and NLTE physics without a full computational effort to include both effects simultaneously.

A comparison of T15 with the 3D-NLTE grid including simultaneous photometric fitting for three double-lined systems in the sample is displayed in Fig.~\ref{fig:comparison3DNLTEGrids}. In addition, we compared the T15, K21 and 3D-NLTE grids by fitting our full observed sample with each. We found that the difference in the reported atmospheric parameters of each double-lined system is small ($<500$\,K, $<0.05$\,dex for both stars) and smaller still for single-star fits to single-lined systems ($<200$\,K, $<0.03$\,dex), yet we found that the 3D-NLTE grid outperformed the others by better fitting the observed spectra, especially at H$\alpha$. For that reason, we have utilised the 3D-NLTE spectral grid to fit all sources presented in this study. We note that for a minority of double-lined systems and 
specifically in the temperature range 9\,000--10\,000\,K and surface gravity range 7.6--8.0\,dex, we witness that the 3D-NLTE synthetic spectra slightly under-predict the flux at H$\alpha$ compared to the observations; the synthetic line cores become too deep instead of too shallow (see e.g. WDJ212935.23+001332.26 in Appendix~\ref{fig:appendixDoubleLiners1}). We accredit the flux under-prediction to either small imperfections in the T15 grid since 3D effects are strongest in this temperature regime, or to an artefact of not creating a grid with the inclusion of 3D and NLTE effects being modelled together instead of our multiplicative factor approach.

In the process of atmospheric fitting and in general for DAs, we chose to linearly normalise Balmer lines using data in the range of $\pm(4\,500-5\,000)$\,km\,s$^{-1}$ around H$\alpha$, H$\beta$, H$\gamma$, H$\delta$ and $\pm(2\,000-2\,500)$\,km\,s$^{-1}$ around H$\epsilon$ and H8 when T$_{\textrm{eff}}\geq8\,000$\,K. In WDs cooler than 8\,000\,K, inferred from the temperature quoted in \citet{NicolaGaia2021} from the fitting of \textit{Gaia} photometry, we reduced the normalisation range to be approximately $\pm(2\,250-2\,500)$\,km\,s$^{-1}$ and $\pm(1\,750-2\,000)$\,km\,s$^{-1}$, for the respective groups of Balmer lines. The narrower prescription for cooler WDs was chosen to restrict the contribution of the continuum to the fit. Fine tuning of a couple of angstroms either side of the absorption features was performed to improve the normalisation and spectroscopic fit if a flux-calibration artefact was within the normalisation range.

Two DA systems in the sample are unique from the rest in the way that their spectra show a sharp emission at H$\alpha$ only originating from background emission in the galactic plane, not from the system itself. These are WDJ183442.33+170028.00 and WDJ193845.80+264751.85, and to perform atmospheric fitting we mask the narrow hydrogen emission in both sources (see WDJ183442.33+170028.00 in Fig.~\ref{fig:appendixDoubleLiners1}). No photometric data is fitted for these sources.

\subsubsection{Non-DA WDs}
\label{subsubsec:MethodsFittingNonDAs}
For the couple of DB WDs in our observed sample, we utilised synthetic spectra from \citet{ElenaCukanovaite2021_3D_DB}, incorporating 3D effects when possible. The observed and synthetic spectra were continuum normalised and all visible helium absorption features fitted. When including a helium-rich atmosphere DC WD component, we use the same synthetic spectra from \citet{ElenaCukanovaite2021_3D_DB} but with a fixed H/He fraction of 10$^{-5}$ \citep[][]{Bergeron2019,McCleery2020_40pcNorth} and normalise spectra with a DC in the same way as for DA WDs. Hydrogen-rich atmosphere DC WDs exist below approximately 5\,000\,K and we continue using the T15 synthetic spectra to model hydrogen-rich DCs.

Some systems within our survey are not DA/DB/DC WDs (see Table~\ref{tab:non-DAs}). For the sample of interest, these systems are viewed as contaminants and as such we do not provide a full analysis of their atmospheric constraints in this paper. Notable targets of interest were 1) WDJ002215.19+423642.15 having a faint and shifted H$\alpha$ towards the blue, perhaps indicative of a cool and magnetic star, 2) WDJ010343.47+555941.53 which could be an evolved CV or a subdwarf with an F/G/K star companion given the CH absorption band (G-band) at 4305\AA\ (in \textit{Gaia} DR3, this source has no parallax or proper motion, whereas it does in \textit{Gaia} DR2). An analysis of time-series data is particularly encouraged for the last two systems.

\subsubsection{Fitting single-lined DWD candidates}
\label{subsubsec:FittingSingleLinedDWDcandidates}
Irrelevant of the spectral type, our full observed sample is composed of DWD candidates. When we see unique signatures of both stars in the spectra and can identify individual spectral types, assigning the relevant atmosphere is simple. On the other hand, single-lined targets may still have an additional flux component that jeopardises the accuracy of a single-star solution.

To combat this dilemma, we perform two-star fits to all single-lined DA targets that show a missing flux component of a second WD, noticeable from overly shallow model fluxes in the line cores of the Balmer lines, a poor fit to the photometry or a fitted parallax outlier that deviates more than 3$\sigma$ from the expected value. The vast majority of targets from the DBL survey show Balmer lines, and it is expected that in a poorly-fitting target for a single-star DA model that the missing component is another DA WD. Hence, we assume that the most common companion would be another star of type DA. The second option that we consider is that the companion is a DC WD with a helium-rich atmosphere\footnote{A hydrogen atmosphere DC companion is considered when fitting DA synthetic spectra when the temperature is less than approximately 5\,000\,K as DAs become DCs} and a H/He fraction (by number) of 10$^{-5}$ \citep{ElenaCukanovaite2021_3D_DB}. It is expected that a DC WD would have $\log(g)\approx8.0$\,dex and we fix the surface gravity of the DC to remove parameter degeneracy in the fitting, before comparing a DA+DA and DA+DC ($\log(g)=8$) best-fit by eye. A visual comparison then allowed us to exclude a combination, if possible. As a last and separate check, we allowed the surface gravity of the DC to be free and comment on whether a DC WD with $\log(g)=$7.5--9.0 would reproduce the observations and note if this is the case.

For any target where a single-star model fits well and we see little to no improvement in the $\chi^2$ of the hybrid fitting when including a second star, we consider the situation entirely degenerate and report an atmospheric solution with just the one WD, but we advise that caution is taken when interpreting the solution. If these systems are binaries, the temperature of the brighter component will be slightly overestimated and the surface gravity slightly underestimated.

\subsection{Radial velocities}
\label{subsec:RadialVelocities}
\subsubsection{Fitting}
To determine RVs, we fitted to H$\alpha$ alone because of the much higher resolution in this spectral range for our observations, whereas the addition of data from the lower-resolution blue arm data would add confusion noise. Occasionally, and especially when the observing conditions worsened, the continuum S/N ratio can be relatively low (S/N<20), leading to a poor spectrum normalisation. We use the best atmospheric fit rather than the data to calculate the normalisation factors for the observed spectrum and then minimise the $\chi^2$ in the wavelength range within $40$\,\AA~of H$\alpha$. Applying this re-normalisation slightly improved the accuracy and precision of the RVs in general and noticeably for the low S/N observations. Data within the $\pm5-40$\,\AA~range was sigma-clipped with a threshold of 4$\sigma$; excluding data near the line cores themselves. RVs and RV errors are reported from the median and standard deviation of 1000 bootstrapping iterations.

In double-lined DWD and single WD fitting, the model flux from each star is used in search for independent RVs. For the RVs of single-lined DWDs, a wide Gaussian component (to cover the broad wings of H$\alpha$) and a narrow Lorentizian component (to model the NLTE core) with a common RV were fit to all spectra to get a best model, and this model was then maintained fixed to search for independent RVs.

\subsubsection{RV variability}
\label{subsubsec:RVvariability}
From the identification spectra of double-lined systems, we can use the atmospheric solution and the measured RV of each component to assign each system a maximum orbital period. If we assume the unlikely case that the epoch of observation for all double-lined detection occurred when the stars were moving at a maximum orbital velocity and that the binary is angled edge-on ($i=90^\circ$), the binary mass function can be rearranged to be
\begin{equation}
    P_{\textrm{max}} = \frac{2\pi G M_1^3}{K_{2,\textrm{min}}^3  (M_1+M_2)^2} = \frac{2\pi G M_2^3}{K_{1,\textrm{min}}^3  (M_1+M_2)^2}
\end{equation}
where we can take a rudimentary assumption that $K_{1,\textrm{min}}$ and $K_{2,\textrm{min}}$ are half of the maximum RV difference of the two stars. With an atmospheric solution bringing photometric masses of M$_1$ and M$_2$, the maximum period can be solved for.

While spotting a double-lined signature is often obvious upon inspection of the data, a statistically significant RV variability for a single-lined system is not. To search for single-lined RV variability the mean of all extracted RVs for a source is taken and a null-hypothesis that the RV is a constant with respect to the mean is tested. We compute the $\chi^2$ of all measurements compared to the mean and use the relevant $\chi^2$-distribution for the number of degrees of freedom to calculate the probability that a source is not RV variable. For a system to be considered a candidate RV variable system, we require that it passes the 1\% $\left(\log_{10}p_{\textrm{bin}}<-2\right)$ threshold and 0.01\% $\left(\log_{10}p_{\textrm{bin}}<-4\right)$ to be considered as a WD binary with an unseen companion. All double-lined systems are naturally RV variable. All RV errors were propagated with an additional wavelength calibration error of 2\,km\,s$^{-1}$ in quadrature  (Section~\ref{sec:Observations}).

\section{Results}
\label{sec:results}

\begin{table*}
    \centering
    \begin{tabular}{c|r|c|c|r|c|c|c|l|c|l|c|c}
         WDJ name & T$_{\textrm{eff}, 1}$ &  $\log$\,g$_1$ & M$_1$ & T$_{\textrm{eff}, 2}$ &  $\log$\,g$_2$ & M$_2$ & M$_T$ & D & Exp & $\Delta$RV$_{\textrm{max}}$ & P$_{\textrm{max}}$ & Ref\\
         & [kK] & [dex] & [\(\textup{M}_\odot\)] & [kK] & [dex] & [\(\textup{M}_\odot\)] & [\(\textup{M}_\odot\)] & [pc] & \# & [km\,s$^{-1}$] & [d]\\
     \hline
J000319.54+022623.28 & 18.2$^{+0.3}_{-0.3}$ & 7.69$^{+0.05}_{-0.04}$ & 0.47$^{+0.02}_{-0.02}$ & 7.5$^{+0.3}_{-0.3}$ & 7.48$^{+0.06}_{-0.06}$ & 0.38$^{+0.02}_{-0.03}$ & 0.85$^{+0.03}_{-0.04}$ & 158.3& 3 & $154\pm13$ & 3.0 &  -\\
J002602.29$-$103751.86 & 10.7$^{+0.2}_{-0.2}$ & 7.74$^{+0.04}_{-0.04}$ & 0.47$^{+0.02}_{-0.02}$ & 5.8$^{+0.1}_{-0.1}$ & 7.60$^{+0.04}_{-0.04}$ & 0.42$^{+0.01}_{-0.02}$ & 0.88$^{+0.02}_{-0.03}$ & 88.3& 3 & $204\pm9$ & 1.2 &  -\\
J005413.14+415613.73 & 7.7$^{+0.2}_{-0.1}$ & 7.69$^{+0.10}_{-0.05}$ & 0.43$^{+0.05}_{-0.03}$ & 7.4$^{+0.2}_{-0.3}$ & 7.73$^{+0.05}_{-0.10}$ & 0.45$^{+0.03}_{-0.05}$ & 0.88$^{+0.06}_{-0.06}$ & 54.0& 2 & $192\pm4$ & 1.2 &  -\\
J013446.42+282616.83 & 13.1$^{+0.2}_{-0.2}$ & 7.60$^{+0.04}_{-0.06}$ & 0.41$^{+0.02}_{-0.02}$ & 9.3$^{+0.2}_{-0.4}$ & 7.72$^{+0.05}_{-0.05}$ & 0.45$^{+0.03}_{-0.03}$ & 0.86$^{+0.03}_{-0.03}$ & 176.3& 3 & $226\pm12$ & 0.7 &  -\\
J013812.93+444252.10 & 15.0$^{+0.2}_{-0.2}$ & 7.92$^{+0.04}_{-0.04}$ & 0.57$^{+0.02}_{-0.02}$ & 8.0$^{+0.1}_{-0.1}$ & 7.88$^{+0.05}_{-0.05}$ & 0.53$^{+0.03}_{-0.03}$ & 1.10$^{+0.04}_{-0.04}$ & 81.5& 3 & $169\pm8$ & 2.4 &  -\\
J014202.72+262354.58 & 12.2$^{+0.2}_{-0.2}$ & 7.86$^{+0.06}_{-0.05}$ & 0.53$^{+0.03}_{-0.03}$ & 8.3$^{+0.2}_{-0.2}$ & 7.72$^{+0.07}_{-0.06}$ & 0.45$^{+0.03}_{-0.03}$ & 0.97$^{+0.04}_{-0.04}$ & 171.4& 4 & $214\pm25$ & 1.3 &  -\\
J020119.40$-$050748.59 & 8.3$^{+0.1}_{-0.1}$ & 7.80$^{+0.05}_{-0.05}$ & 0.49$^{+0.03}_{-0.02}$ & 6.7$^{+0.1}_{-0.1}$ & 7.91$^{+0.06}_{-0.06}$ & 0.54$^{+0.03}_{-0.03}$ & 1.02$^{+0.04}_{-0.04}$ & 85.0& 3 & $175\pm7$ & 1.6 &  -\\
J020847.22+251409.97 & 21.2$^{+0.3}_{-0.3}$ & 7.86$^{+0.04}_{-0.04}$ & 0.55$^{+0.02}_{-0.02}$ & 11.6$^{+0.2}_{-0.6}$ & 8.21$^{+0.04}_{-0.05}$ & 0.74$^{+0.03}_{-0.03}$ & 1.29$^{+0.04}_{-0.04}$ & 39.1& 4 & $175\pm10$ & 1.5 &  -\\
J080856.79+461300.08 & 14.0$^{+0.2}_{-0.2}$ & 7.99$^{+0.05}_{-0.05}$ & 0.60$^{+0.03}_{-0.03}$ & 10.1$^{+0.2}_{-0.2}$ & 7.76$^{+0.05}_{-0.05}$ & 0.47$^{+0.02}_{-0.02}$ & 1.08$^{+0.04}_{-0.04}$ & 118.2& 3 & $154\pm41$ & 4.0 &  -\\
J084457.81+453632.94 & 9.8$^{+0.2}_{-0.2}$ & 7.97$^{+0.06}_{-0.05}$ & 0.58$^{+0.04}_{-0.03}$ & 5.9$^{+0.1}_{-0.2}$ & 7.71$^{+0.06}_{-0.05}$ & 0.43$^{+0.03}_{-0.03}$ & 1.01$^{+0.05}_{-0.04}$ & 60.9& 4 & $138\pm18$ & 5.7 &  -\\
J114446.16+364151.13 & 13.3$^{+0.2}_{-0.2}$ & 7.63$^{+0.05}_{-0.06}$ & 0.42$^{+0.02}_{-0.02}$ & 13.0$^{+0.2}_{-0.2}$ & 7.69$^{+0.12}_{-0.05}$ & 0.45$^{+0.06}_{-0.02}$ & 0.87$^{+0.06}_{-0.03}$ & 89.7& 3 & $183\pm6$ & 1.3 &  -\\
J130014.82+181734.41 & 10.9$^{+0.2}_{-0.2}$ & 8.13$^{+0.05}_{-0.05}$ & 0.68$^{+0.03}_{-0.03}$ & 6.7$^{+0.1}_{-0.1}$ & 7.75$^{+0.06}_{-0.05}$ & 0.46$^{+0.03}_{-0.03}$ & 1.14$^{+0.04}_{-0.04}$ & 84.3& 3 & $177\pm22$ & 3.5 &  -\\
J135342.35+165651.75 & 9.6$^{+0.2}_{-0.3}$ & 7.76$^{+0.05}_{-0.11}$ & 0.47$^{+0.03}_{-0.05}$ & 7.6$^{+0.2}_{-0.2}$ & 7.70$^{+0.13}_{-0.05}$ & 0.43$^{+0.07}_{-0.03}$ & 0.90$^{+0.07}_{-0.06}$ & 101.4& 3 & $151\pm17$ & 2.9 &  -\\
J141625.94+311600.55 & 13.3$^{+0.3}_{-0.3}$ & 7.74$^{+0.06}_{-0.05}$ & 0.47$^{+0.03}_{-0.03}$ & 12.8$^{+0.2}_{-0.2}$ & 7.62$^{+0.05}_{-0.05}$ & 0.42$^{+0.02}_{-0.02}$ & 0.89$^{+0.04}_{-0.04}$ & 115.5& 5 & $189\pm16$ & 1.6 &  -\\
J141632.84+111003.85 & 10.5$^{+0.2}_{-0.2}$ & 7.76$^{+0.06}_{-0.06}$ & 0.47$^{+0.03}_{-0.03}$ & 7.5$^{+0.2}_{-0.2}$ & 7.60$^{+0.05}_{-0.05}$ & 0.42$^{+0.02}_{-0.02}$ & 0.89$^{+0.03}_{-0.03}$ & 129.1& 3 & $158\pm20$ & 2.6 &  -\\
J151109.90+404801.18 & 9.1$^{+0.1}_{-0.1}$ & 8.12$^{+0.05}_{-0.05}$ & 0.67$^{+0.03}_{-0.03}$ & 7.6$^{+0.1}_{-0.1}$ & 7.71$^{+0.05}_{-0.05}$ & 0.44$^{+0.02}_{-0.03}$ & 1.11$^{+0.04}_{-0.04}$ & 54.9& 2 & $167\pm17$ & 4.1 &  -\\
J152038.37+390349.32 & 9.6$^{+0.1}_{-0.1}$ & 8.02$^{+0.04}_{-0.05}$ & 0.61$^{+0.03}_{-0.03}$ & 5.4$^{+0.1}_{-0.1}$ & 7.35$^{+0.05}_{-0.05}$ & 0.32$^{+0.03}_{-0.01}$ & 0.93$^{+0.04}_{-0.03}$ & 94.4& 3 & $212\pm16$ & 2.1 &  -\\
J153615.83+501350.98 & 9.6$^{+0.1}_{-0.2}$ & 7.73$^{+0.05}_{-0.05}$ & 0.46$^{+0.02}_{-0.02}$ & 7.3$^{+0.2}_{-0.2}$ & 7.79$^{+0.06}_{-0.06}$ & 0.48$^{+0.03}_{-0.03}$ & 0.94$^{+0.04}_{-0.04}$ & 68.2& 2 & $137\pm17$ & 3.3 & 1\\
J160822.19+420543.44 & 14.0$^{+0.2}_{-0.2}$ & 7.89$^{+0.05}_{-0.05}$ & 0.55$^{+0.03}_{-0.03}$ & 11.0$^{+0.2}_{-0.2}$ & 7.76$^{+0.04}_{-0.05}$ & 0.47$^{+0.02}_{-0.02}$ & 1.02$^{+0.03}_{-0.03}$ & 43.3& 5 & $165\pm6$ & 2.7 & 2\\
J163441.85+173634.09 & 11.4$^{+0.2}_{-0.2}$ & 7.69$^{+0.04}_{-0.04}$ & 0.44$^{+0.02}_{-0.02}$ & 8.1$^{+0.1}_{-0.1}$ & 7.80$^{+0.05}_{-0.04}$ & 0.48$^{+0.03}_{-0.02}$ & 0.93$^{+0.03}_{-0.03}$ & 25.6& 3 & $161\pm11$ & 1.9 & 3\\
J165935.59+620934.03 & 13.0$^{+0.2}_{-0.2}$ & 7.80$^{+0.04}_{-0.04}$ & 0.50$^{+0.02}_{-0.02}$ & 8.2$^{+0.3}_{-0.3}$ & 8.17$^{+0.07}_{-0.05}$ & 0.70$^{+0.04}_{-0.03}$ & 1.20$^{+0.05}_{-0.04}$ & 110.9& 3 & $143\pm16$ & 2.3 &  -\\
J170120.99$-$191527.57 & 20.5$^{+0.3}_{-0.3}$ & 8.08$^{+0.05}_{-0.05}$ & 0.67$^{+0.03}_{-0.03}$ & 13.5$^{+0.2}_{-0.2}$ & 7.75$^{+0.05}_{-0.05}$ & 0.48$^{+0.02}_{-0.02}$ & 1.15$^{+0.04}_{-0.04}$ & 96.7& 3 & $258\pm14$ & 1.1 &  -\\
J180115.37+721848.76 & 18.0$^{+0.3}_{-0.3}$ & 7.88$^{+0.04}_{-0.04}$ & 0.55$^{+0.02}_{-0.02}$ & 11.2$^{+0.2}_{-0.2}$ & 8.02$^{+0.05}_{-0.05}$ & 0.62$^{+0.03}_{-0.03}$ & 1.17$^{+0.04}_{-0.04}$ & 128.0& 3 & $197\pm20$ & 1.3 &  -\\
J180150.89+103401.08 & 22.0$^{+0.3}_{-0.3}$ & 7.92$^{+0.04}_{-0.04}$ & 0.59$^{+0.02}_{-0.02}$ & 8.3$^{+0.2}_{-0.2}$ & 7.89$^{+0.05}_{-0.05}$ & 0.53$^{+0.03}_{-0.03}$ & 1.12$^{+0.04}_{-0.04}$ & 114.9& 2 & $132\pm17$ & 5.4 &  -\\
J181058.67+311940.94 & 20.0$^{+0.3}_{-0.4}$ & 8.16$^{+0.04}_{-0.04}$ & 0.72$^{+0.03}_{-0.03}$ & 16.7$^{+0.3}_{-0.3}$ & 8.35$^{+0.04}_{-0.04}$ & 0.83$^{+0.03}_{-0.03}$ & 1.55$^{+0.04}_{-0.04}$ & 48.9& 6 & $186\pm5$ & 1.9 & 4*\\
J182606.04+482911.30 & 14.4$^{+0.3}_{-0.3}$ & 7.72$^{+0.12}_{-0.07}$ & 0.47$^{+0.06}_{-0.03}$ & 11.3$^{+0.8}_{-0.3}$ & 7.89$^{+0.09}_{-0.11}$ & 0.54$^{+0.05}_{-0.06}$ & 1.01$^{+0.08}_{-0.07}$ & 135.6& 2 & $154\pm16$ & 2.1 &  -\\
J183442.33$-$170028.00 & 8.2$^{+0.1}_{-0.1}$ & 7.59$^{+0.05}_{-0.05}$ & 0.42$^{+0.02}_{-0.02}$ & 7.0$^{+0.2}_{-0.1}$ & 7.76$^{+0.06}_{-0.06}$ & 0.46$^{+0.03}_{-0.03}$ & 0.88$^{+0.03}_{-0.03}$ & \underline{96.7} & 6 & $294\pm12$ & 0.3 &  -\\
J192002.51$-$184442.99 & 20.4$^{+0.4}_{-0.4}$ & 8.17$^{+0.08}_{-0.06}$ & 0.73$^{+0.05}_{-0.04}$ & 12.0$^{+0.3}_{-0.3}$ & 8.08$^{+0.07}_{-0.08}$ & 0.65$^{+0.04}_{-0.05}$ & 1.38$^{+0.07}_{-0.06}$ & 156.1& 2 & $243\pm24$ & 1.1 &  -\\
J192420.74+070135.14 & 16.7$^{+0.4}_{-0.7}$ & 7.96$^{+0.07}_{-0.07}$ & 0.59$^{+0.04}_{-0.04}$ & 14.1$^{+0.7}_{-0.3}$ & 8.03$^{+0.08}_{-0.07}$ & 0.63$^{+0.05}_{-0.04}$ & 1.22$^{+0.06}_{-0.06}$ & 161.0& 3 & $179\pm17$ & 1.9 &  -\\
J211327.98+720814.03 & 11.1$^{+0.2}_{-0.2}$ & 7.63$^{+0.05}_{-0.05}$ & 0.42$^{+0.02}_{-0.02}$ & 7.0$^{+0.2}_{-0.2}$ & 7.50$^{+0.05}_{-0.05}$ & 0.38$^{+0.02}_{-0.02}$ & 0.79$^{+0.03}_{-0.03}$ & 96.4& 3 & $114\pm20$ & 6.0 &  -\\
J212935.23+001332.26 & 9.2$^{+0.4}_{-0.2}$ & 7.69$^{+0.08}_{-0.07}$ & 0.44$^{+0.04}_{-0.04}$ & 7.9$^{+0.6}_{-0.2}$ & 7.64$^{+0.09}_{-0.05}$ & 0.44$^{+0.02}_{-0.02}$ & 0.87$^{+0.05}_{-0.04}$ & 65.5& 2 & $214\pm25$ & 0.9 &  -\\
J214323.95$-$175413.00 & 14.2$^{+0.2}_{-0.2}$ & 8.05$^{+0.05}_{-0.05}$ & 0.64$^{+0.03}_{-0.03}$ & 14.0$^{+0.2}_{-0.2}$ & 8.04$^{+0.05}_{-0.05}$ & 0.64$^{+0.03}_{-0.03}$ & 1.28$^{+0.04}_{-0.04}$ & 118.2& 3 & $142\pm16$ & 4.3 &  -\\
J221209.01+612906.96 & 8.1$^{+0.1}_{-0.2}$ & 7.90$^{+0.06}_{-0.05}$ & 0.54$^{+0.03}_{-0.03}$ & 7.0$^{+0.2}_{-0.1}$ & 7.93$^{+0.06}_{-0.06}$ & 0.55$^{+0.04}_{-0.03}$ & 1.09$^{+0.05}_{-0.05}$ & 64.5& 3 & $88\pm9$ & 14.9 &  -\\
J231404.30+552814.11 & 14.0$^{+0.2}_{-0.2}$ & 8.08$^{+0.05}_{-0.05}$ & 0.66$^{+0.03}_{-0.03}$ & 8.6$^{+0.1}_{-0.1}$ & 7.50$^{+0.05}_{-0.04}$ & 0.38$^{+0.03}_{-0.01}$ & 1.04$^{+0.04}_{-0.03}$ & 105.6& 3 & $137\pm11$ & 7.9 &  -\\

    \end{tabular}
    \caption{The best-fitting parameters to the definite double-lined DWDs. Subscripts 1 and 2 represent the hotter and cooler star, respectively. Errors on the atmospheric constraints of the stars were obtained by combining in quadrature the statistical error from $\chi^2$ fitting with an external error of 1.4\% for T$_{\text{eff}}$ and 0.042\,dex for $\log g$ \citep[][]{Liebert2005}, however this should be considered a minimum due to extra degeneracy in two-star fitting. We emphasise caution again on interpretation of the masses for stars with a temperature below 5\,500\,K. The `Exp' column shows the total number of ID spectra exposures. The distance column includes the distance derived from fitting of the parallax with the inclusion of SDSS and PAN-STARRS photometry and a Gaussian prior using the \textit{Gaia} parallax. The one exception where no photometric data was used was WDJ183442.33-170028.00 because of its very crowded field in the lining of the Milky Way and a distance (D) from the \textit{Gaia} parallax is given and underlined. A reference is provided in the `Ref' column when applicable if the system has already been discovered as a double-lined DWD or with an asterisk if the system was previously discovered to be a compact DWD but was not shown to have a double-lined signature. References: 1) \citet{Kilic2021HiddenInPlainSight}, 2) \citet{Kilic2020twoDoubleLined}, 3) \citet{Kilic2021HiddenInPlainSight}, 4) \citet{Sahu2023}. The rest of the sample including candidate double-lined DWDs are listed in Tables~\ref{tab:lowSNRdoublelined}--\ref{tab:non-DAs}.}
    \label{tab:doubleLinedParams}
\end{table*}

The best-fit atmospheric solutions are reported in Tables~\ref{tab:doubleLinedParams}--\ref{tab:singleLinedParams} as the 50th percentile of the post-burnin chains and errors as the 16th and 84th percentiles of this distribution. Within, subscripts 1 and 2 represent the hotter and cooler star, respectively.

\begin{figure}
    \centering
    \includegraphics[trim={0.5cm 0cm 0.75cm 1.5cm},clip,width=\columnwidth]{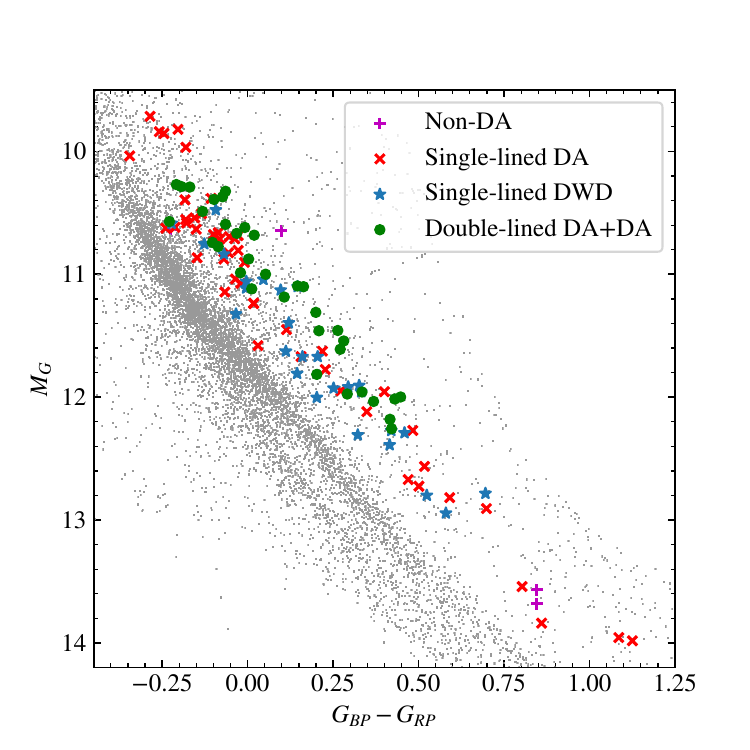}
    \caption{The same as Fig.~\ref{fig:sampleSelection}, but further zoomed-in on the search selection area and with annotated object types of each target. Within are all double-lined DWDs (green circles), single-lined DWD binaries discovered from a poor spectroscopic or photometric solution with a single star model (blue stars), single-lined DA WDs that fit well for a single star model (red crosses) and non-DA WDs (purple pluses).}
    \label{fig:HRsinglesAndDoubles}
\end{figure}

34 systems are double-lined DWDs. Another 11 are double-lined DWD candidates as they show a small hint of a double-lined signature, however we can not confirm these as double-lined DWDs due to the S/N ratio of the data or perhaps an unfavourable orbital phase sampling. 27 single-lined DWDs were found where one star is visible in the spectrum but the source require an additional flux component for spectroscopic and photometric consistency, 38 sources appear as a single-lined DA WD with no companion present and 7 non-DA sources were found (Table \ref{tab:non-DAs}). The location of these systems on the HR diagram separated by category is depicted in Fig.~\ref{fig:HRsinglesAndDoubles}. 

As the spectroscopic surface gravity of each star is largely defined by the relative depth of all Balmer lines compared to one another \citep{Tremblay2009}, we emphasise that care should be taken when interpreting the parameters of hybrid fits for stars with a temperature below approximately 5\,500\,K as the only Balmer lines visible in the data are H$\alpha$ and at times H$\beta$; all other Balmer lines vanish at lower temperatures. Regardless, the temperatures of the stars are expected to be accurate given the inclusion to the fit of flux-calibrated photometry. The error on these stars stays relatively low because the surface gravity is also constrained with the photometric data.

\subsection{Double-lined DA DWDs}
\label{subsec:resultsDoubleLined}
\begin{figure*}
    \centering
    \includegraphics[trim={1.5cm 1.5cm 2.25cm 3.5cm},clip,width=15cm,keepaspectratio]{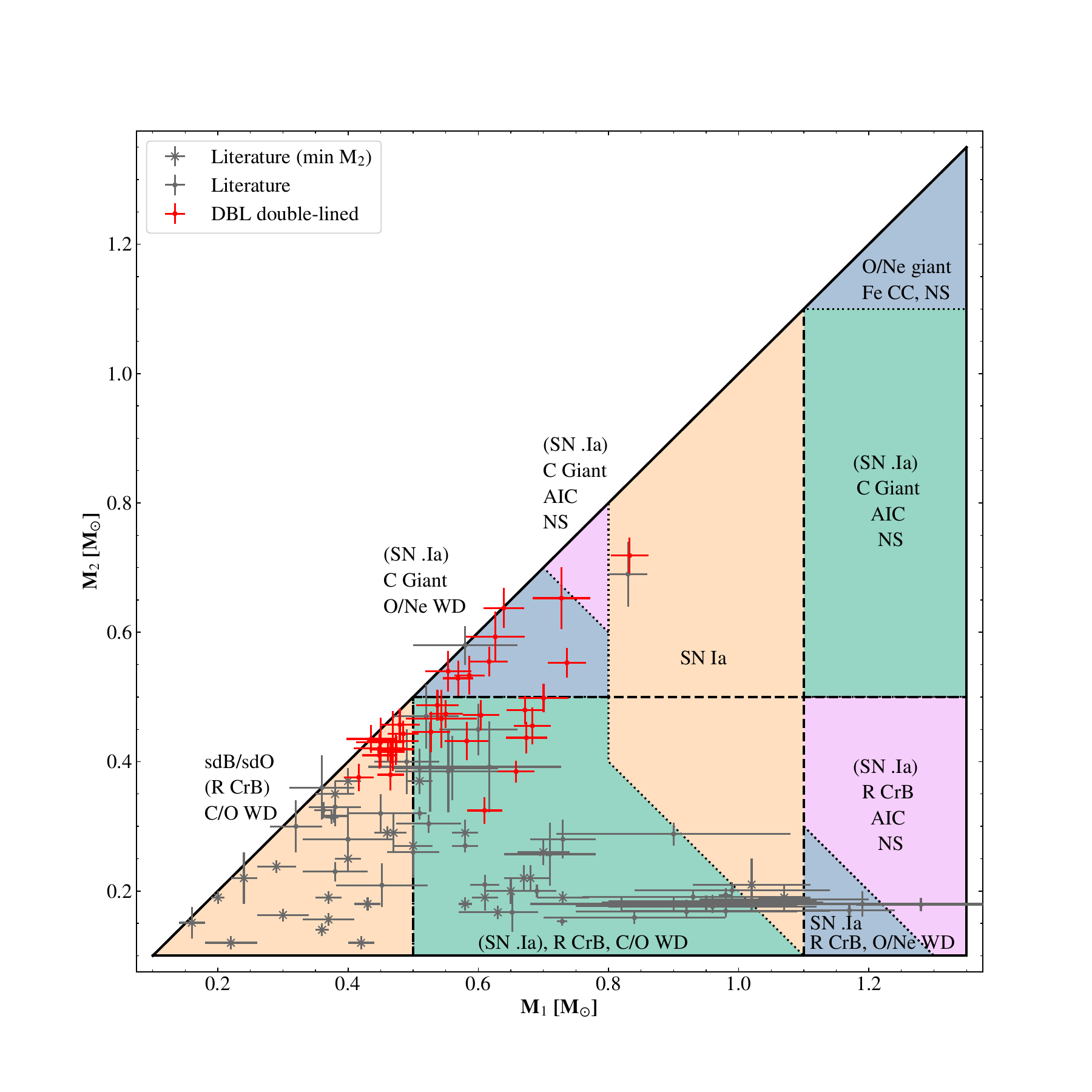}
        \caption{The mass distribution of compact DWD binaries where both star masses are quoted. The diagram is a reproduction of figure~3 of \citet{Shen2015}. For this figure alone, M$_1$ is the larger mass of the two stars for each system and M$_2$ the smaller; all other mentions of star 1 or 2 in this study address indicate the hotter or cooler component, respectively. Systems plotted are DWDs in the literature where the mass of both components have an error better than 20\% of the mass of the star (see the database upkept at github.com/JamesMunday98/CloseDWDbinaries for individual systems and references within). No filtering is applied on the orbital period for neither the literature sample nor the sample from our DBL survey, such that some objects will undergo the categorised events in over a Hubble time. The suggested evolutionary path for each category should be viewed as an approximate guideline only since stable mass transfer and AM~CVn evolution are omitted. Acronyms first mentioned in the figure are as follows. sdB/sdO: hot subdwarf type B/O. R~CrB: R Coronae Borealis. AIC: Accretion-induced collapse. NS: Neutron star. Fe CC: Iron core-collapse. SN: supernova.}
    \label{fig:massDistributionEvolution}
\end{figure*}
34 of our targets listed in Table~\ref{tab:doubleLinedParams} were found to be double-lined, double-DA DWDs, signifying at least a 29\% detection efficiency in the survey. A further 3 targets found in Table~\ref{tab:lowSNRdoublelined} are deemed to be likely double-lined DWDs but the S/N ratio of the data leads to a lack of certainty, and a further 8 targets show a faint hint of a double-lined signature. If all are double-lined DWDs, the detection efficiency of the survey rises to 32\% and 38\%, respectively. These detection efficiencies are minima given the fact that the inclusion of \textit{Gaia}~DR3 parallaxes and colours have removed some systems from the original sample selection criteria. In the case that these other 11 candidates are not double-lined DWDs, they would be single-lined DWD binaries owing to the photometric and spectroscopic fit requiring the flux of a second DA WD for a good Balmer line and/or photometric fit, which then could be compact or unresolved, wide binaries. 

Since our initial identification spectra, a couple of systems have since been discovered by other authors and shown to be double-lined DWDs. WDJ153615.83+501350.98 was earlier noted to be a double-degenerate candidate by \citet{Zuckerman2003}, but since has been independently confirmed and discovered by \citet[][WD 1534+503]{Kilic2021HiddenInPlainSight} to be a double-lined DWD with much higher resolution observations at H$\alpha$. Using the WHT identification spectrum where the two stars are most separated (see Fig.~\ref{fig:appendixDoubleLiners2}), we obtain a 1$\sigma$ consistent atmospheric solution with the solution in \citet{Kilic2021HiddenInPlainSight} where our model reproduces H$\beta$--H8, the photometric SED and the \textit{Gaia} parallax excellently and H$\alpha$ well. \citet{Kilic2021HiddenInPlainSight} do not fit to any data at H$\alpha$ for atmospheric parameters and use H$\beta$--H8, nevertheless, we recommend the usage of the system parameters described in their study owing to a slightly better consistency with the orbital solution for the photometric masses derived. Published in the same study, WDJ163441.85+173634.09 was independently discovered by \citet[][PG 1632+177]{Kilic2021HiddenInPlainSight} as a double-lined DWD and we again find 1$\sigma$-consistent atmospheric solution for both stars, where the fit in their study 
was also obtained using H$\beta$--H8. WDJ160822.19+420543.44 was independently discovered by \citet[][WD 1606+422]{Kilic2020twoDoubleLined} and we find a near-identical atmospheric solution for both stars as well. Besides these 3 double-lined systems, we believe that all of the other 31 double-lined DWDs presented in this study are new discoveries.

While not a clear double-lined DWD discovery, WDJ181058.67+311940.94 was flagged by \citet{Sahu2023} as a strong double-degenerate candidate from a poor spectral fit to its Lyman-$\alpha$ profile. These authors do not extend their study to introduce two-star fits to the outliers in the data, meaning that the surface gravity of their single star solution is under-estimated and vastly different to our solution, but this is entirely expected due to the observed flux of the second star needing to be recovered by artificially inflating the radius in a single star fit. We have now confirmed the system as a DWD binary by detecting a double-lined feature at H$\alpha$.

Our solutions for the confirmed double-lined systems report that the masses of the hotter component range from 0.4--0.75\,$\mathrm{M}_\odot$ with a median mass of $0.53\,\mathrm{M}_\odot$ and that the dimmer companion has a median mass of $0.45\,\mathrm{M}_\odot$. These measurements are slightly less than the 0.6\,$\mathrm{M}_\odot$ canonical mass of a WD, but this is expected given that most if not all of these compact double-lined DWDs have underwent a phase of mass loss in the past from binary interactions.

WDJ183442.33$-$170028.00, which as noted in Section~\ref{subsubsec:MethodsFittingDAWDs} shows a sharp hydrogen emission that does not originate from the system itself but from a background source in the galactic plane, is double-lined at the lower resolution H$\beta$ in two exposures and is highly asymmetric at H$\alpha$ after masking the hydrogen cloud emission. Our observations of this DWD binary indicate a very large RV$_{\textrm{max}}=294\pm12$\,km\,s$^{-1}$, which is the largest velocity difference detected amongst the full observed sample and the system is composed of two low-mass WDs of $0.42\pm0.02$\,$\mathrm{M}_\odot$ and $0.46\pm0.03$\,$\mathrm{M}_\odot$, giving rise to a maximum orbital period of 0.4\,d. It is possible that there are more compact systems than these in the observed sample based on our random phase sampling, and phase-resolved spectroscopy is the means to find out.

The new double-lined systems discovered seemingly show no preference for the cooler star being the more massive of the two. This challenges the scenario that the systems are formed via two common envelope phases that each led to significant shrinkage of the orbit, because in that case the more massive object would always have the largest cooling age.
A feasible evolutionary path is instead that the less massive WD was first formed via stable mass transfer and cooled typically. Then, the other star evolved towards a WD, the system underwent a common envelope phase and exposed the star's core. The new WD has a larger mass than the old one, but is still hotter being fresh out of the envelope, while the older, less-massive WD is cooler.

An interesting feature of the DBL survey is that we see a build-up of double-lined DWDs that runs approximately 0.5\,mag above and parallel to the 0.6\,\,M$_\odot$ WD cooling track, as is prevalent in Fig.~\ref{fig:HRsinglesAndDoubles}. The difference of 0.5\,mag equates to roughly half of the flux, and so the natural conclusion is that double-lined DWD binaries are easiest to identify in our survey for two stars of similar brightness. While this is an unavoidable detection bias, this emphasises a strength of the DBL survey and an important point for the future of double-lined DWD detection outside of the DBL survey: 10s to 100s of double-lined DWDs in this strip of the HR diagram are undetectable from RV variability in lower resolution data where the two stars are inseparable and the two signals average to a net zero velocity change; at higher resolution, these are easy to identify.

\subsection{Single-lined DWDs}
27 sources in Table~\ref{tab:spectroscopicOrPhotometricFitDWDs} were found to be single-lined, but a poor fit to the photometric and/or spectroscopic solution indicates that they are likely DWD binaries. Together with the double-lined systems and the double-lined candidate systems, this means that 72 of the 117 systems surveyed in the initial search selection using the \textit{Gaia} DR2 astrometric solution data show a noticeable spectroscopic and/or photometric contribution from a WD companion.

Three of the single-lined DWDs, WDJ024323.67$-$143957.37 ($\Delta$RV$_{\textrm{max}}=67\pm7$\,km\,s$^{-1}$), WDJ172740.51+102337.94 ($\Delta$RV$_{\textrm{max}}=57\pm10$\,km\,s$^{-1}$) and WDJ193833.62$-$092519.87 ($\Delta$RV$_{\textrm{max}}=42\pm8$\,km\,s$^{-1}$), show high RV variable log-probabilities of $-99.0$, $-9.9$ and $-6.7$, respectively. Further spectroscopic observations of these three DWDs to trace the RV of the brighter star are encouraged to search for an orbital period. WDJ231519.82-052900.27 is a candidate for being RV variable with a log-probability of $-$2.3. All other sources with multiple spectra are consistent with a non-variable RV and most if not all are expected to be binaries with an orbital period greater than 10\,days, where the only prospects of finding an orbital period are from a larger time baseline of astrometric or spectroscopic observations. Some of the non-RV variable candidates could have periods that fall in the aliases that we are not sensitive to in our search, which are predominantly aliases of 1\,day (Fig.~\ref{fig:DetectionEfficiency}).

Overall in the sample of single- or double-lined DWDs, no system shows any signs of magnetism. If present, we would be sensitive to a magnetic field strength greater than approximately 50\,kG with the identification spectra. The only DWD that has a strong magnetic field is NLTT~12758 \citep[][]{Kawka2017}, and the dynamo theory proposed by \citet{Schreiber2021natureBfield} has been shown to support the lack of magnetic DWD binaries in the observed population and can explain the apparently chance timing of detecting the magnetic WD in NLTT~12758 \citep{Schreiber2022}. If true, a lack of magnetic DWDs in our survey is unsurprising, and the fact that we have specifically targeted systems with a 0.6\,$\mathrm{M}_\odot$ star and a dimmer, likely more massive, companion \citep[much like the 0.83+0.69\,$\mathrm{M}_\odot$ pairing of NLTT~12758,][]{Kawka2017} alludes that we may have been more sensitive to a magnetic component with our survey criteria, but none were found.

\subsection{Single star DA WDs}
\label{subsec:resultsSingleLined}
38 targets in Table~\ref{tab:singleLinedParams} appear as single-lined, DA WDs where a single-star model fits the photometric and spectroscopic data well. In many of these systems, the reason is likely that they are isolated, low mass WDs or that one star strongly outshines its companion, which is highlighted by the grouping of single star DAs with $G_{\textrm{mag}}\approx10$\,mag in Fig.~\ref{fig:HRsinglesAndDoubles}. Alternative reasons for identifying so many single star DA WDs in this region are that an additional unseen and red flux component makes the target appear redder and over-luminous, that the sources are indeed single WDs and a hybrid helium/carbon-oxygen core composition makes them appear over-luminous, that they are the product of an extremely-low-mass WD merger, or that some candidates are outliers in the \textit{Gaia} database. Some of the sources classified as single DA WDs have crowded fields or issues with the \textit{Gaia} DR3 parallax (see Table~\ref{tab:singleLinedParams}), meaning that they are spectroscopic fits instead of a hybrid. A few sources in Table~\ref{tab:singleLinedParams} have M$\gtrsim$0.55\,$\mathrm{M}_\odot$, and these are all cases where the location of the star on the HR diagram is on the lower edge of our search selection or when no photometry was fit. The minimum single star WD mass is approximately 0.3\,$\mathrm{M}_\odot$, as was expected from the upper limit of the selection in Table~\ref{tab:SelectionCuts}. The median mass of the single star DA WDs in Table~\ref{tab:singleLinedParams} is 0.43\,$\mathrm{M}_\odot$.

We searched the single DA WDs for RV variations and a few targets displayed clear variability. The largest variation was in WDJ131913.72+450509.81 which exhibited an $75\pm5$\,km\,s$^{-1}$ shift between two exposures, strongly indicating that this source has an unseen component and is a WD binary. WDJ031242.85+221828.36 is also a clear case where an unseen component is present, having a maximum shift of $70\pm10$\,km\,s$^{-1}$. WDJ191927.69+395839.48 and WDJ230831.77+454212.19 exhibit a 30\,km\,s$^{-1}$ shift and both pass the threshold criteria of $\log_{10}(\textrm{p}_\textrm{bin})<-4$, making them RV variable systems. We promote follow-up observations of all to determine the orbital periods.

\subsection{Gaia astrometric solutions}
We cross-matched the \textit{Gaia} non-single star catalogue \citep{Halbwachs2023gaiaNSS} with our full observed sample and we found that 5 in the observed sample have entries. One system that we flag as being a probable double-lined DWD but the S/N ratio leads to a lack of clarity, WDJ225123.02+293944.49, has an astrometric period of $278.0\pm0.3$\,d, indicative that it may be a triple star system. The same can be said for the candidate double-lined DWD WDJ211345.93+262133.27 which has an astrometric period of 219.7$\pm$0.2\,d. Three single-lined DWDs have detected astrometric periods, being WDJ023117.04+285939.88 ($103.9\pm0.07$\,d), WDJ205650.56+062149.68 ($81.4\pm0.3$\,d) and WDJ232519.87+140339.61 ($249.4\pm1.1$\,d). The most likely situation for these three is that they are wide DWDs and it is hence unsurprising that we do not witness a significant RV variability in any. Finally, WDJ232519.87+140339.61, a source that we have categorised as a single-lined WD with no significant flux contribution from a companion, has an astrometric period of $249.4\pm1.1$\,d.

\section{The fate of the observed DWD sample}
The end state of all of DWD binaries is largely dependent on the masses of the two constituents and core compositions. All compact WDs will gradually inspiral and initiate mass transfer from a loss of orbital angular momentum predominantly from the radiation of gravitational waves. The outcomes of a system are then a merger scenario, a detonation event or an orbital expansion following a brief period of mass transfer with a stripped companion. For a merger, exotic merger by-products such as the population of R~CrB stars or hot subdwarfs \citep{Schwab2018,Schwab2019} in addition to the existence of some high-mass magnetic WDs \citep{Kilic2023massive} are predicted to be at least partially explained by DWD mergers \citep{Shen2012longTermEvolutionDWD, Schwab2012evolutionToWDmergerRemnants}. The long suspected type \Romannum{1}a supernova detonation event, whereby mass accretion causes the carbon to ignite in degenerate conditions, either because the density increases due to mass accretion or due to shocks, can occur and in the last decade much interest has been sparked in the double-detonation sub-Chandrasekhar Supernova theory with .\Romannum{1}a supernovae \citep{Bildsten2007}. Both have been observationally supported by the discovery of hyper-velocity runaway stars in the Milky Way \citep{Shen2018d6,ElBadry2023d6}. 
Yet, up to now, very few high-mass WDs in compact binaries have been found \citep{Geier2007,Brown2016,2017MagneticDWDsuperChandra,Pelisoli2021type1aWDsubdwarf,Pallathadka2023}, drawing ambiguity on whether DWD binaries can replicate the observed rates of type \Romannum{1}a supernovae \citep{Maoz2012supernovaReview}. Orbital expansion may happen when a system becomes an AM CVn \citep{Soleheim2010, Kupfer2024}, although survival of a period minimum to produce an out-spiraling population of AM~CVns remains unclear owing to the short lived period of time in this orbital phase and the large observational biases with the population.

The full series of photometric masses of double-lined DWD binaries identified in our work is plotted in Fig.~\ref{fig:massDistributionEvolution}, that shows the assessment of \citet{Shen2015} of the most likely outcomes for mass combination but in many cases may be quite uncertain. Here and only in this figure, subscripts 1 and 2 are the more and less massive components, respectively. Single-lined DWDs are omitted since many could be wide binaries. As mentioned earlier, the vast majority of DWD binaries in the literature that constitute the observed population come in the form of low-mass WD binaries, which is far from reflective of the population as a whole \citep[see e.g. ][]{Toonen2012type1aCommonEnvelope}. An immediate change in the observed population of DWDs is now obvious from our uncovered sample, with the majority of our systems having larger combined total masses. None of our sample falls into the low-mass regime and the majority of systems are dominated by a more massive star of approximately 0.55\,$\mathrm{M}_\odot$.

One double-lined system in the observed sample exceeds the Chandrasekhar mass limit, being WDJ181058.67+311940.94 with a total mass of $1.55\pm0.04$\,$\mathrm{M}_\odot$ and the maximum RV separation between the two stars indicates a maximum orbital period of 2.0 days. This represents the second super-Chandrasekhar mass DWD discovered to date after NLTT~12758 \citep{Kawka2017}. Continued efforts to resolve the orbital period of WDJ181058.67+311940.94 may reveal a more compact nature and will provide a time estimate for the type \Romannum{1}a detonation. Furthermore, from the identification spectra alone, the maximum orbital periods confirm that some systems have orbital periods of less than $\approx$10\,hrs signifying a merger time of less than a Hubble time. Continued work will strive to resolve the orbital dynamics of the sample (Munday et al, in prep).

\begin{table*}
    \centering
    \begin{tabular}{c|r|c|c|r|c|c|c|l|c|l|c|c}
         WDJ name & T$_{\textrm{eff}, 1}$ &  $\log$\,g$_1$ & M$_1$ & T$_{\textrm{eff}, 2}$ &  $\log$\,g$_2$ & M$_2$ & M$_T$ & D & Exp & $\Delta$RV$_{\textrm{max}}$ & P$_{\textrm{max}}$ & Ref\\
         & [kK] & [dex] & [\(\textup{M}_\odot\)] & [kK] & [dex] & [\(\textup{M}_\odot\)] & [\(\textup{M}_\odot\)] & [pc] & \# & [km\,s$^{-1}$] & [d]\\
         \hline

         J062538.73$-$162132.02 & 14.0$^{+0.2}_{-0.2}$ & 7.99$^{+0.07}_{-0.07}$ & 0.60$^{+0.04}_{-0.04}$ & 11.3$^{+0.2}_{-0.2}$ & 7.87$^{+0.06}_{-0.06}$ & 0.53$^{+0.03}_{-0.03}$ & 1.14$^{+0.06}_{-0.05}$ & 149.8& 1 & $127\pm47$ & 6.4 &  -\\
J211927.07$-$030002.38 & 11.1$^{+0.2}_{-0.5}$ & 8.31$^{+0.06}_{-0.08}$ & 0.80$^{+0.04}_{-0.05}$ & 6.5$^{+0.2}_{-0.3}$ & 7.56$^{+0.10}_{-0.09}$ & 0.40$^{+0.04}_{-0.02}$ & 1.19$^{+0.05}_{-0.06}$ & 95.0& 3 & $188\pm7$ & 4.1 &  -\\
J225123.02+293944.49 & 6.3$^{+0.1}_{-0.1}$ & 8.56$^{+0.07}_{-0.06}$ & 0.95$^{+0.04}_{-0.04}$ & 5.1$^{+0.1}_{-0.1}$ & 7.79$^{+0.05}_{-0.06}$ & 0.46$^{+0.03}_{-0.03}$ & 1.41$^{+0.05}_{-0.05}$ & 19.9& 3 & $104\pm20$ & 29.0 &  -\\

    \end{tabular}
    \caption{The same as Table~\ref{tab:doubleLinedParams}, but double-lined DWD candidates where the lower S/N ratio of the spectra of these targets is too difficult to conclude that they are double-lined definitively, although such a signature appears to be apparent. Continued observation and affirmation is encouraged. WDJ211927.07$-$030002.38 and WDJ225123.02+293944.49 had 1 exposure taken on one day followed by 2 on the other, and in both cases they appear double-lined in the first two exposures and slightly the third. The errors on the atmospheric constraints include an external error of 1.4\% for T$_{\text{eff}}$ and 0.042\,dex for $\log g$ \citep[][]{Liebert2005}, however this should be considered a minimum due to extra degeneracy in two-star fitting.}
    \label{tab:lowSNRdoublelined}
\end{table*}

\begin{table*}
    \centering
    \begin{tabular}{c|r|c|c|r|c|c|c|l|c|l|c}
         WDJ name & T$_{\textrm{eff}, 1}$ &  $\log$\,g$_1$ & M$_1$ & T$_{\textrm{eff}, 2}$ &  $\log$\,g$_2$ & M$_2$ & M$_T$ & D & Exp & $\Delta$RV$_{\textrm{max}}$ & Ref\\
         & [kK] & [dex] & [\(\textup{M}_\odot\)] & [kK] & [dex] & [\(\textup{M}_\odot\)] & [\(\textup{M}_\odot\)]  & [pc] & \# & [km\,s$^{-1}$]\\
         \hline
J053316.88$-$075049.72 & 11.1$^{+0.5}_{-0.5}$ & 7.22$^{+0.09}_{-0.09}$ & 0.32$^{+0.05}_{-0.05}$ & 7.7$^{+0.4}_{-0.3}$ & 7.16$^{+0.12}_{-0.13}$ & 0.29$^{+0.03}_{-0.04}$ & 0.61$^{+0.06}_{-0.07}$ & \underline{80.5} & 1 & $101\pm13$ &  -\\
J141354.17+153020.71 & 15.2$^{+0.3}_{-0.3}$ & 7.61$^{+0.04}_{-0.10}$ & 0.42$^{+0.02}_{-0.02}$ & 12.1$^{+0.5}_{-2.5}$ & 8.36$^{+0.29}_{-0.08}$ & 0.83$^{+0.18}_{-0.06}$ & 1.25$^{+0.18}_{-0.06}$ & 145.6& 2 & $37\pm7$ &  -\\
J145011.93$-$191408.67 & 8.1$^{+0.1}_{-0.1}$ & 7.66$^{+0.05}_{-0.04}$ & 0.42$^{+0.02}_{-0.02}$ & 5.3$^{+0.3}_{-0.2}$ & 7.78$^{+0.08}_{-0.08}$ & 0.46$^{+0.05}_{-0.05}$ & 0.87$^{+0.05}_{-0.05}$ & 48.6& 5 & $173\pm38$ & 1\\
J180600.36$-$002720.92 & 25.0$^{+0.4}_{-0.4}$ & 7.60$^{+0.04}_{-0.04}$ & 0.45$^{+0.02}_{-0.02}$ & 11.2$^{+0.5}_{-1.3}$ & 7.94$^{+0.07}_{-0.09}$ & 0.57$^{+0.04}_{-0.05}$ & 1.03$^{+0.04}_{-0.06}$ & 150.4& 3 & $16\pm49$ &  -\\
J193642.54$-$054744.38 & 10.2$^{+0.2}_{-0.2}$ & 7.83$^{+0.05}_{-0.05}$ & 0.51$^{+0.03}_{-0.02}$ & 7.1$^{+0.3}_{-0.3}$ & 7.98$^{+0.06}_{-0.07}$ & 0.58$^{+0.03}_{-0.04}$ & 1.09$^{+0.04}_{-0.04}$ & 102.2& 3 & $177\pm17$ &  -\\
J204922.92+262517.50 & 16.9$^{+0.3}_{-0.4}$ & 7.83$^{+0.05}_{-0.07}$ & 0.53$^{+0.02}_{-0.03}$ & 14.6$^{+0.3}_{-0.5}$ & 8.06$^{+0.06}_{-0.05}$ & 0.65$^{+0.04}_{-0.03}$ & 1.17$^{+0.04}_{-0.05}$ & 141.7& 4 & $34\pm5$ &  -\\
J211345.93+262133.27 & 8.6$^{+0.1}_{-0.1}$ & 7.79$^{+0.04}_{-0.04}$ & 0.48$^{+0.02}_{-0.02}$ & 5.8$^{+0.1}_{-0.1}$ & 7.83$^{+0.08}_{-0.05}$ & 0.49$^{+0.04}_{-0.03}$ & 0.97$^{+0.05}_{-0.03}$ & 31.4& 3 & $150\pm32$ & 2\\
J234929.57+102255.57 & 14.4$^{+0.2}_{-0.2}$ & 7.82$^{+0.04}_{-0.04}$ & 0.51$^{+0.02}_{-0.02}$ & 7.4$^{+0.2}_{-0.2}$ & 7.96$^{+0.06}_{-0.07}$ & 0.57$^{+0.04}_{-0.04}$ & 1.09$^{+0.04}_{-0.05}$ & 116.1& 3 & $106\pm39$ &  -\\

    \end{tabular}
    \caption{The same as Table~\ref{tab:doubleLinedParams}, but candidate double-lined DWDs where the dimmer star is very slightly noticeable in the spectra, but could easily be confused with single-lined DWD binary. These double-lined candidates are much less probable than those in Table~\ref{tab:lowSNRdoublelined}. High resolution and high S/N observations is again promoted to confirm/deny a double-lined state. The errors on the atmospheric constraints include an external error of 1.4\% for T$_{\text{eff}}$ and 0.042\,dex for $\log g$ \citep[][]{Liebert2005}, however this should be considered a minimum due to extra degeneracy in two-star fitting. References: 1) \citet{Kilic2020twoDoubleLined}, who find the source to be single-lined and RV variable with a period of 1.79\,d. 2) \citet{Bedard2017}.}
    \label{tab:candidateTinyDetectionDblLined}
\end{table*}

\begin{table*}
    \centering
    \begin{tabular}{c|r|c|r|c|r|c|r|l|c|c|c|c|c|c|c}
         WDJ name & T$_{1,\textrm{AA}}$ &  $\log$\,g$_{1,\textrm{AA}}$ & T$_{2,\textrm{AA}}$ &  $\log$\,g$_{2,\textrm{AA}}$ & T$_{{1,\textrm{AC}}}$ &  $\log$\,g$_{1,\textrm{AC}}$ & T$_{2,\textrm{AC}}$ & D & Exp & p$_{\textrm{bin}}$ & Ref\\
         & [kK] & [dex] & [kK] & [dex] & [kK] & [dex] & [kK] & [pc] & \# & $\log_{10}$ \\
         \hline
         
J001353.60+204852.65 & 17.7$^{+0.3}_{-0.4}$ & 7.90$^{+0.07}_{-0.06}$ & 8.3$^{+0.6}_{-0.3}$ & 7.83$^{+0.09}_{-0.11}$ & 16.8$^{+0.3}_{-0.3}$ & 7.78$^{+0.05}_{-0.05}$ & 7.2$^{+0.3}_{-0.3}$ & 291.0 & 3 & -0.6 &  - \\
J003045.78+273333.36 & 16.1$^{+0.2}_{-0.2}$ & 7.86$^{+0.04}_{-0.04}$ & 7.6$^{+0.2}_{-0.2}$ & 7.74$^{+0.05}_{-0.06}$ & 14.6$^{+0.2}_{-0.2}$ & 7.70$^{+0.04}_{-0.04}$ & * & 165.1 & 4 & -0.2 &  - \\
J014511.23+313243.56 & 9.6$^{+0.1}_{-0.1}$ & 7.87$^{+0.05}_{-0.05}$ & 6.4$^{+0.1}_{-0.1}$ & 8.00$^{+0.04}_{-0.04}$ & 9.6$^{+0.1}_{-0.2}$ & 7.86$^{+0.05}_{-0.05}$ & 6.5$^{+0.1}_{-0.1}$ & 36.7 & 1 &  - & 1\\
J023117.04+285939.88 & 7.5$^{+0.1}_{-0.1}$ & 7.96$^{+0.04}_{-0.04}$ & 5.8$^{+0.1}_{-0.1}$ & 8.08$^{+0.05}_{-0.04}$ & 7.5$^{+0.1}_{-0.1}$ & 7.82$^{+0.05}_{-0.05}$ & * & 27.9 & 5 & -0.0 &  - \\
J024323.67$-$143957.37 & 9.2$^{+0.3}_{-0.2}$ & 7.72$^{+0.06}_{-0.08}$ & 5.1$^{+0.2}_{-0.1}$ & 7.61$^{+0.13}_{-0.14}$ & 9.1$^{+0.7}_{-0.1}$ & 7.70$^{+0.06}_{-0.04}$ & * & 80.2 & 3 & -99.0 &  - \\
J080739.31+132110.65 & 11.6$^{+0.2}_{-0.2}$ & 7.68$^{+0.04}_{-0.05}$ & 7.3$^{+0.3}_{-0.3}$ & 7.80$^{+0.07}_{-0.07}$ & 11.1$^{+0.2}_{-0.2}$ & 7.61$^{+0.04}_{-0.05}$ & * & 115.0 & 1 &  - &  - \\
J084417.70+750008.79 & 14.5$^{+0.2}_{-0.2}$ & 8.04$^{+0.04}_{-0.04}$ & 7.0$^{+0.2}_{-0.2}$ & 8.01$^{+0.05}_{-0.06}$ & 13.3$^{+0.2}_{-0.2}$ & 7.96$^{+0.05}_{-0.04}$ & 7.1$^{+0.2}_{-0.1}$ & 102.4 & 4 & -0.0 &  - \\
J084634.41+194505.18 & 10.9$^{+0.3}_{-0.3}$ & 7.65$^{+0.09}_{-0.08}$ & 7.7$^{+0.3}_{-0.5}$ & 7.61$^{+0.08}_{-0.08}$ & 10.1$^{+0.2}_{-0.2}$ & 7.44$^{+0.04}_{-0.05}$ & * & 146.5 & 1 &  - &  - \\
J102459.83+044610.50 & 12.4$^{+0.6}_{-0.2}$ & 7.62$^{+0.04}_{-0.05}$ & 5.7$^{+0.3}_{-0.2}$ & 7.64$^{+0.35}_{-0.09}$ & 12.1$^{+0.2}_{-0.2}$ & 7.60$^{+0.04}_{-0.04}$ & * & 43.1 & 1 &  - &  - \\
J104709.19+345346.65 & 9.2$^{+0.1}_{-0.1}$ & 7.70$^{+0.05}_{-0.05}$ & 5.4$^{+0.1}_{-0.2}$ & 7.69$^{+0.05}_{-0.06}$ & 9.0$^{+0.2}_{-0.1}$ & 7.63$^{+0.04}_{-0.04}$ & * & 79.4 & 1 &  - &  - \\
J113100.20+493826.27 & 11.6$^{+0.2}_{-0.2}$ & 7.91$^{+0.04}_{-0.05}$ & 6.9$^{+0.1}_{-0.1}$ & 7.71$^{+0.06}_{-0.05}$ & 10.7$^{+0.3}_{-0.2}$ & 7.72$^{+0.04}_{-0.04}$ & 7.3$^{+0.1}_{-0.1}$ & 99.3 & 1 &  - &  - \\
J113347.81+624313.29 & - & - & - & - & 6.4$^{+0.1}_{-0.1}$ & 7.92$^{+0.04}_{-0.04}$ & 6.0$^{+0.1}_{-0.1}$ & 45.8 & 1 &  - & 2\\
J131257.90+580511.29 & 11.6$^{+0.2}_{-0.2}$ & 8.01$^{+0.04}_{-0.04}$ & 6.7$^{+0.1}_{-0.1}$ & 8.12$^{+0.05}_{-0.05}$ & 11.2$^{+0.2}_{-0.2}$ & 7.93$^{+0.04}_{-0.04}$ & 6.4$^{+0.1}_{-0.9}$ & 30.9 & 2 & -0.0 &  - \\
J135738.69$-$025819.41 & 17.1$^{+0.6}_{-0.4}$ & 7.92$^{+0.11}_{-0.08}$ & 14.8$^{+0.5}_{-1.1}$ & 7.98$^{+0.13}_{-0.12}$ & 16.0$^{+1.3}_{-1.3}$ & 7.57$^{+0.07}_{-0.05}$ & 6.9$^{+1.3}_{-0.8}$ & 137.2 & 2 & -0.2 &  - \\
J154214.21$-$034131.29 & 11.5$^{+0.2}_{-0.2}$ & 8.07$^{+0.05}_{-0.04}$ & 7.9$^{+0.2}_{-0.2}$ & 8.07$^{+0.04}_{-0.05}$ & 10.8$^{+0.2}_{-0.2}$ & 7.89$^{+0.07}_{-0.11}$ & 5.5$^{+1.0}_{-0.2}$ & 51.9 & 1 &  - &  - \\
J172740.51+102337.94 & - & - & - & - & 10.5$^{+0.2}_{-0.2}$ & 8.23$^{+0.04}_{-0.05}$ & 7.8$^{+0.2}_{-0.1}$ & 59.0 & 5 & -9.9 &  - \\
J191329.91+163822.12 & 8.3$^{+0.1}_{-0.1}$ & 7.88$^{+0.04}_{-0.05}$ & 6.7$^{+0.1}_{-0.1}$ & 8.03$^{+0.05}_{-0.05}$ & 8.4$^{+0.1}_{-0.1}$ & 7.78$^{+0.05}_{-0.04}$ & * & 77.0 & 5 & -1.0 &  - \\
J192359.24+214103.62 & 9.4$^{+0.1}_{-0.2}$ & 7.81$^{+0.05}_{-0.04}$ & 6.1$^{+0.2}_{-0.1}$ & 7.81$^{+0.05}_{-0.06}$ & 9.5$^{+0.1}_{-0.1}$ & 7.80$^{+0.04}_{-0.05}$ & 6.4$^{+0.1}_{-0.9}$ & 35.3 & 3 & -0.0 &  - \\
J193833.62$-$092519.87 & 15.9$^{+0.2}_{-0.2}$ & 8.08$^{+0.05}_{-0.04}$ & 8.0$^{+0.2}_{-0.1}$ & 7.66$^{+0.04}_{-0.04}$ & 12.5$^{+0.2}_{-0.2}$ & 7.73$^{+0.04}_{-0.04}$ & 8.7$^{+0.1}_{-0.1}$ & 114.8 & 3 & -6.7 &  - \\
J204517.85+810503.40 & 8.7$^{+0.1}_{-0.1}$ & 8.19$^{+0.06}_{-0.05}$ & 7.8$^{+0.1}_{-0.1}$ & 8.01$^{+0.05}_{-0.05}$ & 8.9$^{+0.1}_{-0.1}$ & 7.93$^{+0.04}_{-0.04}$ & * & 67.0 & 3 & -1.7 & 1\\
J205650.56+062149.68 & 11.0$^{+0.2}_{-0.2}$ & 7.98$^{+0.06}_{-0.05}$ & 6.5$^{+0.2}_{-0.2}$ & 7.67$^{+0.05}_{-0.07}$ & 10.1$^{+0.2}_{-0.1}$ & 7.77$^{+0.04}_{-0.04}$ & 6.7$^{+0.1}_{-0.1}$ & 95.0 & 3 & -0.6 &  - \\
J213616.39$-$131834.50 & 10.3$^{+0.1}_{-0.1}$ & 7.85$^{+0.04}_{-0.04}$ & 5.0$^{+0.1}_{-0.1}$ & 8.03$^{+0.04}_{-0.05}$ & - & - & - & 23.4 & 3 & -0.1 &  - \\
J214632.37+155039.08 & 8.5$^{+2.0}_{-0.3}$ & 7.83$^{+0.25}_{-0.14}$ & 6.9$^{+4.1}_{-0.3}$ & 7.96$^{+0.20}_{-0.52}$ & 8.6$^{+0.1}_{-0.1}$ & 7.79$^{+0.04}_{-0.05}$ & * & 74.5 & 3 & -0.2 &  - \\
J221052.87$-$045540.80 & 7.9$^{+0.1}_{-0.1}$ & 7.84$^{+0.04}_{-0.04}$ & 6.9$^{+0.1}_{-0.1}$ & 7.90$^{+0.05}_{-0.05}$ & - & - & - & 63.5 & 3 & -0.1 &  - \\
J231443.05$-$073417.85 & 14.0$^{+0.2}_{-0.2}$ & 7.99$^{+0.04}_{-0.04}$ & 9.5$^{+0.2}_{-0.2}$ & 8.10$^{+0.04}_{-0.05}$ & - & - & - & 75.1 & 3 & -0.1 &  - \\
J231519.82$-$052900.27 & 6.5$^{+0.1}_{-0.2}$ & 7.65$^{+0.21}_{-0.05}$ & 5.9$^{+0.1}_{-0.1}$ & 7.88$^{+0.07}_{-0.26}$ & 6.3$^{+0.1}_{-0.1}$ & 7.48$^{+0.04}_{-0.04}$ & * & 54.4 & 3 & -2.3 &  - \\
J233041.67+110206.43 & 21.7$^{+0.3}_{-0.3}$ & 7.97$^{+0.04}_{-0.04}$ & 7.0$^{+0.3}_{-0.3}$ & 8.10$^{+0.07}_{-0.07}$ & 21.5$^{+0.3}_{-0.3}$ & 7.94$^{+0.04}_{-0.04}$ & 6.4$^{+0.2}_{-0.6}$ & 99.2 & 1 &  - &  - \\

    \end{tabular}
    \caption{The same as Table~\ref{tab:doubleLinedParams}, but single-lined DWD binaries inferred from the spectroscopic fit, photometric fit or both fits to the data. References are provided if a study has performed a spectroscopic fit and obtained temperature estimates to both stars. To be identified from the photometric fit, the slope of the SED had to be significantly inconsistent with a single WD not to be confused with an inaccurate reddening coefficient, or the parallax had to be more than a +3$\sigma$ outlier while still showing a poor solution (meaning that a closer distance is sought after because of the missing flux of a required second star). Two scenarios are considered; parameters subscripted by ``AA'' are for a DA+DA combination (or a $<5000$\,K hydrogen-rich atmosphere DC) while parameters subscripted by ``AC'' are for a DA+DC where the DC has a helium-rich atmosphere and $\log(g_2)=8.0$\,dex is assumed and fixed. If there are hyphens in the `AC' columns, this represents that the fit to the photometry and/or spectroscopy with a DC model is poor and that the DA+DA solution should be adopted. If there is an asterisk in the column, this represents that a helium-rich atmosphere DC WD would fit the data only if the surface gravity is varied between $\log(g)=7.5$--$9.0$\,dex (see Section~\ref{subsubsec:FittingSingleLinedDWDcandidates}), but we consider this to be improbable and that the DA+DA solution to likely be correct. The errors on the atmospheric constraints include an external error of 1.4\% for T$_{\text{eff}}$ and 0.042\,dex for $\log g$ \citep[][]{Liebert2005}, however this should be considered a minimum due to extra degeneracy in two-star fitting. $\textrm{p}_\textrm{bin}$ was inferred from RVs that were obtained using the model flux of both stars with a common RV for the DA+DA stellar type combination. References: 1) \citet{Bedard2017}. 2) \citet{Limoges2015}.}
    \label{tab:spectroscopicOrPhotometricFitDWDs}
\end{table*}

\begin{table*}
    \centering
    \begin{tabular}{c|r|c|c|l|c|c|c|c|c}
         WDJ name & T$_{\textrm{eff}}$ &  $\log$\,g  & M & D & Spec & p$_{\textrm{bin}}$ & Ref & Other Name \\
         & [kK] & [cm\,s$^{-2}$] & [\(\textup{M}_\odot\)] & [pc] & \# & $\log_{10}$\\
         \hline
J001321.07+282019.83 & 26.2$^{+0.4}_{-0.4}$ & 7.84$^{+0.04}_{-0.04}$ & 0.56$^{+0.02}_{-0.02}$ & 135.3 & 13 & -0.5 & 1 & PG 0010+281\\
J003508.27+135045.32 & 22.9$^{+0.4}_{-0.7}$ & 7.47$^{+0.09}_{-0.06}$ & 0.44$^{+0.02}_{-0.02}$ & 246.8 & 3 & -0.2 & 2 & WD 0032+135\\
J015437.45+374145.37 & 12.9$^{+0.2}_{-0.2}$ & 7.53$^{+0.04}_{-0.04}$ & 0.42$^{+0.01}_{-0.01}$ & 136.6 & 3 & -0.1 & - & -\\
J022631.25+203106.04 & 21.5$^{+0.3}_{-0.3}$ & 7.89$^{+0.04}_{-0.04}$ & 0.57$^{+0.02}_{-0.02}$ & 164.9 & 3 & -0.1 & - & -\\
J031242.85+221828.36 & 6.3$^{+0.1}_{-0.1}$ & 7.36$^{+0.04}_{-0.04}$ & 0.33$^{+0.03}_{-0.02}$ & 63.2 & 3 & -99.0 & 3, 4 & LP 355-39\\
J042127.88+701419.37 & 12.4$^{+0.2}_{-0.6}$ & 7.60$^{+0.04}_{-0.04}$ & 0.41$^{+0.02}_{-0.02}$ & 54.5 & 1 &  - & 1 & GD 429\\
J074837.63+004011.72 & 15.5$^{+1.2}_{-0.6}$ & 7.57$^{+0.06}_{-0.06}$ & 0.40$^{+0.03}_{-0.03}$ & 184.9 & 3 & -0.8 & - & -\\
J081706.44+054954.62 & 11.3$^{+0.2}_{-0.2}$ & 7.61$^{+0.04}_{-0.04}$ & 0.41$^{+0.02}_{-0.02}$ & 133.3 & 1 &  - & - & -\\
J091914.80+772350.79 & 9.1$^{+0.1}_{-0.1}$ & 7.56$^{+0.04}_{-0.04}$ & 0.41$^{+0.01}_{-0.01}$ & 46.5 & 6 & -0.0 & 3 & -\\
J093709.98+650746.91 & 18.1$^{+0.4}_{-0.3}$ & 7.92$^{+0.04}_{-0.04}$ & 0.58$^{+0.02}_{-0.02}$ & 105.5 & 4 & -0.3 & - & SDSS J093709.96+650746.6\\
J101606.87$-$011917.14 & 8.1$^{+0.1}_{-0.1}$ & 7.63$^{+0.04}_{-0.04}$ & 0.43$^{+0.02}_{-0.02}$ & \underline{46.3} & 1 &  - & 3, 5 & EGGR 2532\\
J114604.37+051401.54 & 6.6$^{+0.1}_{-0.1}$ & 7.43$^{+0.04}_{-0.04}$ & 0.35$^{+0.01}_{-0.01}$ & 62.3 & 1 &  - & 3, 4, 5 & WD 1143+055\\
J130313.03$-$032323.92 & 7.2$^{+0.1}_{-0.1}$ & 7.56$^{+0.05}_{-0.04}$ & 0.40$^{+0.02}_{-0.02}$ & 64.6 & 1 &  - & 3, 5 & WD 1300-031\\
J131913.72+450509.81 & 13.1$^{+0.2}_{-0.2}$ & 7.43$^{+0.04}_{-0.04}$ & 0.40$^{+0.01}_{-0.02}$ & 49.0 & 2 & -99.0 & 5 & EGGR 188\\
J134503.00$-$110434.10 & 21.0$^{+2.8}_{-0.8}$ & 7.57$^{+0.08}_{-0.12}$ & 0.42$^{+0.04}_{-0.04}$ & 246.6 & 2 & -1.0 & - & -\\
J142047.04+465601.58 & 10.2$^{+0.5}_{-0.1}$ & 7.58$^{+0.05}_{-0.04}$ & 0.42$^{+0.02}_{-0.02}$ & 98.6 & 2 & -0.1 & 3, 5 & WD 1418+471\\
J150402.99+345440.76 & 18.8$^{+0.3}_{-0.3}$ & 8.07$^{+0.04}_{-0.04}$ & 0.66$^{+0.03}_{-0.03}$ & - & 6 & -0.2 & 1 & PG 1502+351\\
J152849.51$-$012853.74 & 12.8$^{+0.2}_{-0.2}$ & 7.93$^{+0.04}_{-0.04}$ & 0.57$^{+0.02}_{-0.02}$ & 52.4 & 3 & -1.5 & - & -\\
J155840.22+162556.04 & 12.3$^{+0.2}_{-0.2}$ & 7.47$^{+0.05}_{-0.04}$ & 0.40$^{+0.01}_{-0.02}$ & 106.3 & 1 &  - & 1 &2MASS J15584020+1625556\\
J160647.76$-$152740.19 & 31.9$^{+0.8}_{-0.5}$ & 7.98$^{+0.06}_{-0.14}$ & 0.64$^{+0.03}_{-0.07}$ & \underline{194.4} & 4 & -3.4 & - & -\\
J175644.61$-$020847.66 & 16.1$^{+1.4}_{-0.3}$ & 7.52$^{+0.09}_{-0.04}$ & 0.43$^{+0.01}_{-0.01}$ & 157.8 & 4 & -0.0 & - & -\\
J181304.62+222449.79 & 21.4$^{+0.3}_{-0.3}$ & 7.91$^{+0.04}_{-0.04}$ & 0.58$^{+0.02}_{-0.02}$ & 209.3 & 2 & -0.8 & - & -\\
J181339.31+255058.16 & 19.1$^{+1.1}_{-0.3}$ & 7.53$^{+0.05}_{-0.10}$ & 0.40$^{+0.02}_{-0.02}$ & 183.5 & 3 & -0.5 & - & -\\
J182444.29+600159.40 & 12.7$^{+0.2}_{-0.2}$ & 7.70$^{+0.04}_{-0.04}$ & 0.45$^{+0.02}_{-0.02}$ & 96.2 & 3 & -0.1 & \\
J183752.49$-$125257.45 & 37.8$^{+0.6}_{-0.5}$ & 8.21$^{+0.04}_{-0.04}$ & 0.78$^{+0.03}_{-0.03}$ & 179.5 & 3 & -0.2 & - & -\\
J185640.86+120844.61 & 13.6$^{+1.5}_{-0.4}$ & 7.53$^{+0.08}_{-0.05}$ & 0.42$^{+0.01}_{-0.01}$ & 156.3 & 3 & -0.0 & - & -\\
J190401.01$-$102305.20 & 16.2$^{+0.2}_{-0.3}$ & 7.92$^{+0.04}_{-0.04}$ & 0.57$^{+0.02}_{-0.02}$ & 103.0 & 3 & -0.1 & - & -\\
J191927.69+395839.48 & 18.9$^{+0.3}_{-0.3}$ & 7.73$^{+0.04}_{-0.04}$ & 0.49$^{+0.02}_{-0.02}$ & 105.8 & 3 & -15.1 & 6 & KIC 4829241\\
J192817.81+354442.60 & 20.5$^{+0.3}_{-0.3}$ & 7.92$^{+0.04}_{-0.04}$ & 0.58$^{+0.02}_{-0.02}$ & 138.2 & 1 &  - & - & -\\
J193845.80+264751.85 & 22.2$^{+0.4}_{-0.3}$ & 7.93$^{+0.04}_{-0.05}$ & 0.59$^{+0.02}_{-0.02}$ & \underline{97.0} & 5 & -0.2 & - & -\\
J195314.95$-$105417.89 & 22.6$^{+0.7}_{-0.6}$ & 7.74$^{+0.07}_{-0.08}$ & 0.50$^{+0.04}_{-0.03}$ & \underline{145.0} & 1 &  - & - & -\\
J195622.94+641359.19 & 18.3$^{+0.3}_{-0.3}$ & 7.77$^{+0.04}_{-0.04}$ & 0.50$^{+0.02}_{-0.02}$ & 85.5 & 3 & -0.2 & - & -\\
J205020.65+263040.76 & 5.2$^{+0.1}_{-0.1}$ & 8.54$^{+0.04}_{-0.04}$ & 0.94$^{+0.03}_{-0.03}$ & 19.1 & 1 &  - & 3, 4, 7, 8 & GJ 4166\\
J230831.77+454212.19 & 13.6$^{+0.2}_{-0.2}$ & 7.48$^{+0.04}_{-0.04}$ & 0.41$^{+0.01}_{-0.02}$ & 103.1 & 3 & -6.8 & - & -\\
J231406.68+233343.06 & 7.1$^{+0.1}_{-0.1}$ & 7.45$^{+0.04}_{-0.04}$ & 0.36$^{+0.02}_{-0.02}$ & 68.5 & 3 & -0.2 & 3 & -\\
J232519.87+140339.61 & 4.9$^{+0.1}_{-0.1}$ & 7.29$^{+0.04}_{-0.04}$ & 0.30$^{+0.02}_{-0.02}$ & 23.5 & 1 &  - & 3, 4, 7, 8 & WD 2322+137\\
J232557.82+255222.39 & 5.6$^{+0.1}_{-0.1}$ & 7.49$^{+0.04}_{-0.04}$ & 0.38$^{+0.03}_{-0.03}$ & 46.0 & 3 & -0.0 & 3, 4 & EGGR 380\\
J235313.18+205117.58 & 7.4$^{+0.1}_{-0.1}$ & 7.64$^{+0.04}_{-0.05}$ & 0.43$^{+0.02}_{-0.02}$ & 57.0 & 3 & -0.2 & 3, 4 & WD 2350+205\\

    \end{tabular}
    \caption{The best fits to systems that appear as a single DA WD, where no/insufficient contribution of a second star is distinguishable in the spectroscopy and photometry. No photometric fitting was applied to underlined targets because they lie in a crowded field (hence only a spectroscopic fit was performed), and the distances underlined are instead obtained from the \textit{Gaia} parallaxes. In the case of WDJ150402.99+345440.76, the \textit{Gaia} solution reports a common proper motion pair at inconsistent distances to one another and neither are consistent with a photometrically fit parallax, and as such we ignore the \textit{Gaia} parallax. The errors on the atmospheric constraints include an external error of 1.4\% for T$_{\text{eff},1}$ and 0.042\,dex for $\log g_1$ \citep[][]{Liebert2005}. References are included when spectroscopic/hybrid fitting has been performed with the source within the last 5 years or the most recent study otherwise, under the assumption that there are common datasets between studies (e.g. SDSS). References: 1) \citet{Gianninas2011} 2) \citet{Kleinman2013} 3) \citet{Caron2023} 4) \citet{Blouin2019} 5) \citet{Kilic2020sdss100pc} 6) \citet{Guo2015} 7) \citet{OBrien2024} 8) \citet{McCleery2020_40pcNorth}.}
    \label{tab:singleLinedParams}
\end{table*}

\begin{table*}
    \centering
    \begin{tabular}{c|l|c|c}
        WDJ name & D &  SpT & Note \\
         \hline
         J002215.19+423642.15 & \underline{34.5} & DC & Perhaps H$\alpha$ absorption at 6554\AA.\\
         J004502.15$-$040710.05 & \underline{43.2} &  DQ & Swan bands\\
         J010343.47+555941.53  & \underline{26.1} &   Not a WD & Evolved CV or sd+F/G/K star. Parallax in \textit{Gaia} DR2 not DR3\\
         J021243.27$-$080216.23 & \underline{119.7} & MS & No WD signatures. Common proper motion pair MS \& blue object\\
         J182138.99+144158.19 & 89.7 & DB & T$_{\textrm{eff}}=21.3\pm0.3$\,kK, $\log\,g=8.02\pm0.04$ dex\\
         J195513.90+222458.79 & \underline{14.8} & MS & No signature of WD. Parallax in DR2 not DR3\\
         J201437.22+231607.23 & \underline{29.8} & MS & No signature of WD. Parallax in \textit{Gaia} DR2 not DR3 \\
    \end{tabular}
    \caption{Sources in the sample with no DA WD present. Spectral types of main sequence (MS) stars, cataclysmic variable (CV) systems and systems with a subdwarf (sd) are included. Underlined distances are obtained from a \textit{Gaia} parallax (DR3 if available, otherwise DR2). The spectrum of WDJ010343.47+555941.53 is also plotted in Fig.~\ref{fig:AppendixEvolvedCV}.}
    \label{tab:non-DAs}
\end{table*}

\section{Conclusions}
\label{sec:Conclusions}
We have presented a pilot study including the first results from the DBL survey based on 20 nights of observations with the WHT. We surveyed a large sample of 117 DWD binary candidates that reside above the 0.6\,\,M$_\odot$ WD cooling track through the exploitation of the \textit{Gaia}~DR2 HR diagram, randomly sampling candidates from a magnitude limited selection. Then, we fitted and obtained atmospheric solutions for the entire sample when all visible stars in the spectra were of spectral type DA, DB or DC with the custom-made fitting code WD-BASS that is designed for time-series spectroscopy of WDs and is publicly available for use.

The primary goal of the survey was to find double-lined DWDs which allow the obtention of precise masses for both stars in the binary through atmospheric fitting. Our methods demonstrate at least a 29\% double-lined detection rate after a couple of observing epochs and that 73 of the 117 candidates show a separable spectroscopic/photometric flux contribution that must arise from a DWD configuration. A further 6 single-lined WDs are flagged for having an unseen companion based off RV variability. All double-lined DWDs are compact binaries, and the large RV difference between the two stars provides some insight to the orbital period distribution of the systems.

For the first time, we have been able to observationally identify a class of DWDs with system masses $M_T\geq1.0\,$M$_\odot$ that may undergo a type .\Romannum{1}a supernova or merge to become a massive WD. One system (WDJ181058.67+311940.94) located just 49\,pc away hosts two relatively massive WDs of mass $0.72\pm0.03$\,M$_\odot$ and $0.83\pm0.03$\,M$_\odot$, making it the second DWD system confirmed as super-Chandrasekhar mass.

In continued work we will report orbital solutions of many of the double-lined DWDs in this study. Many more double-lined DWDs are waiting to be discovered within the search selection, where we will strive to obtain further completion of the 625 $G_\textrm{mag}$<17\,mag double degenerate candidates.

\section*{Acknowledgements}
We thank Ralf Napiwotzki for their help in inspiring WD-BASS by making their \textsc{fitsb2} double-lined binary fitting code \citep{Napiwotzki2004fitsb2} available to us. JM was supported by funding from a Science and Technology Facilities Council (STFC) studentship. IP acknowledges support from a Royal Society University Research Fellowship (URF\textbackslash R1\textbackslash 231496). This research received funding from the European Research Council under the European Union’s Horizon 2020 research and innovation programme number 101002408 (MOS100PC). TC was supported by NASA through the NASA Hubble Fellowship grant HST-HF2-51527.001-A awarded by the Space Telescope Science Institute, which is operated by the Association of Universities for Research in Astronomy, Inc., for NASA, under contract NAS5-26555. For the purpose of open access, the authors have applied a creative commons attribution (CC BY) licence to any author accepted manuscript version arising. 

Based on observations made with the William Herschel Telescope operated on the island of La Palma by the Isaac Newton Group of Telescopes in the Spanish Observatorio del Roque de los Muchachos of the Instituto de Astrofísica de Canarias.

\section*{Data Availability}
The reduced spectra and extracted RVs are available upon request to the authors. Raw data can be obtained through the Isaac Newton Group data archive with the dates of observations mentioned in text. Our compilation of all discovered DWD binaries is available through Github at \url{https://github.com/JamesMunday98/CloseDWDbinaries}.

\appendix



\bibliographystyle{mnras}
\bibliography{mnras_template} 

\begin{thebibliography}{}
\makeatletter
\relax
\def\mn@urlcharsother{\let\do\@makeother \do\$\do\&\do\#\do\^\do\_\do\%\do\~}
\def\mn@doi{\begingroup\mn@urlcharsother \@ifnextchar [ {\mn@doi@} {\mn@doi@[]}}
\def\mn@doi@[#1]#2{\def\@tempa{#1}\ifx\@tempa\@empty \href {http://dx.doi.org/#2} {doi:#2}\else \href {http://dx.doi.org/#2} {#1}\fi \endgroup}
\def\mn@eprint#1#2{\mn@eprint@#1:#2::\@nil}
\def\mn@eprint@arXiv#1{\href {http://arxiv.org/abs/#1} {{\tt arXiv:#1}}}
\def\mn@eprint@dblp#1{\href {http://dblp.uni-trier.de/rec/bibtex/#1.xml} {dblp:#1}}
\def\mn@eprint@#1:#2:#3:#4\@nil{\def\@tempa {#1}\def\@tempb {#2}\def\@tempc {#3}\ifx \@tempc \@empty \let \@tempc \@tempb \let \@tempb \@tempa \fi \ifx \@tempb \@empty \def\@tempb {arXiv}\fi \@ifundefined {mn@eprint@\@tempb}{\@tempb:\@tempc}{\expandafter \expandafter \csname mn@eprint@\@tempb\endcsname \expandafter{\@tempc}}}

\bibitem[\protect\citeauthoryear{{Abril}, {Schmidtobreick}, {Ederoclite}  \& {L{\'o}pez-Sanjuan}}{{Abril} et~al.}{2020}]{Abril020CV_HRdiagram}
{Abril} J.,  {Schmidtobreick} L.,  {Ederoclite} A.,   {L{\'o}pez-Sanjuan} C.,  2020, \mn@doi [\mnras] {10.1093/mnrasl/slz181}, \href {https://ui.adsabs.harvard.edu/abs/2020MNRAS.492L..40A} {492, L40}

\bibitem[\protect\citeauthoryear{{Adamane Pallathadka} et~al.,}{{Adamane Pallathadka} et~al.}{2024}]{Pallathadka2023}
{Adamane Pallathadka} G.,  et~al., 2024, \mn@doi [\apj] {10.3847/1538-4357/ad3e86}, \href {https://ui.adsabs.harvard.edu/abs/2024ApJ...968...42A} {968, 42}

\bibitem[\protect\citeauthoryear{{Ahumada} et~al.,}{{Ahumada} et~al.}{2020}]{SDSSdr16}
{Ahumada} R.,  et~al., 2020, \mn@doi [\apjs] {10.3847/1538-4365/ab929e}, \href {https://ui.adsabs.harvard.edu/abs/2020ApJS..249....3A} {249, 3}

\bibitem[\protect\citeauthoryear{{Althaus}, {Miller Bertolami}  \& {C{\'o}rsico}}{{Althaus} et~al.}{2013}]{Althaus2013}
{Althaus} L.~G.,  {Miller Bertolami} M.~M.,   {C{\'o}rsico} A.~H.,  2013, \mn@doi [\aap] {10.1051/0004-6361/201321868}, \href {https://ui.adsabs.harvard.edu/abs/2013A&A...557A..19A} {557, A19}

\bibitem[\protect\citeauthoryear{{Amaro-Seoane} et~al.,}{{Amaro-Seoane} et~al.}{2023a}]{LISAwhitepaper2023}
{Amaro-Seoane} P.,  et~al., 2023a, \mn@doi [Living Reviews in Relativity] {10.1007/s41114-022-00041-y}, \href {https://ui.adsabs.harvard.edu/abs/2023LRR....26....2A} {26, 2}

\bibitem[\protect\citeauthoryear{{Amaro-Seoane} et~al.,}{{Amaro-Seoane} et~al.}{2023b}]{AmaroSeoane2023lisaWhitePaper}
{Amaro-Seoane} P.,  et~al., 2023b, \mn@doi [Living Reviews in Relativity] {10.1007/s41114-022-00041-y}, \href {https://ui.adsabs.harvard.edu/abs/2023LRR....26....2A} {26, 2}

\bibitem[\protect\citeauthoryear{{Badenes} \& {Maoz}}{{Badenes} \& {Maoz}}{2012}]{BadenesMaoz2012}
{Badenes} C.,  {Maoz} D.,  2012, \mn@doi [\apjl] {10.1088/2041-8205/749/1/L11}, \href {https://ui.adsabs.harvard.edu/abs/2012ApJ...749L..11B} {749, L11}

\bibitem[\protect\citeauthoryear{{B{\'e}dard}, {Bergeron}  \& {Fontaine}}{{B{\'e}dard} et~al.}{2017}]{Bedard2017}
{B{\'e}dard} A.,  {Bergeron} P.,   {Fontaine} G.,  2017, \mn@doi [\apj] {10.3847/1538-4357/aa8bb6}, \href {https://ui.adsabs.harvard.edu/abs/2017ApJ...848...11B} {848, 11}

\bibitem[\protect\citeauthoryear{{B{\'e}dard}, {Bergeron}, {Brassard}  \& {Fontaine}}{{B{\'e}dard} et~al.}{2020}]{Bedard2020}
{B{\'e}dard} A.,  {Bergeron} P.,  {Brassard} P.,   {Fontaine} G.,  2020, \mn@doi [\apj] {10.3847/1538-4357/abafbe}, \href {https://ui.adsabs.harvard.edu/abs/2020ApJ...901...93B} {901, 93}

\bibitem[\protect\citeauthoryear{{Belokurov} et~al.,}{{Belokurov} et~al.}{2020}]{Belokurov2020ruwe}
{Belokurov} V.,  et~al., 2020, \mn@doi [\mnras] {10.1093/mnras/staa1522}, \href {https://ui.adsabs.harvard.edu/abs/2020MNRAS.496.1922B} {496, 1922}

\bibitem[\protect\citeauthoryear{{Bergeron}, {Wesemael}, {Liebert}  \& {Fontaine}}{{Bergeron} et~al.}{1989}]{Bergeron1989}
{Bergeron} P.,  {Wesemael} F.,  {Liebert} J.,   {Fontaine} G.,  1989, \mn@doi [\apjl] {10.1086/185560}, \href {https://ui.adsabs.harvard.edu/abs/1989ApJ...345L..91B} {345, L91}

\bibitem[\protect\citeauthoryear{{Bergeron}, {Dufour}, {Fontaine}, {Coutu}, {Blouin}, {Genest-Beaulieu}, {B{\'e}dard}  \& {Rolland}}{{Bergeron} et~al.}{2019}]{Bergeron2019}
{Bergeron} P.,  {Dufour} P.,  {Fontaine} G.,  {Coutu} S.,  {Blouin} S.,  {Genest-Beaulieu} C.,  {B{\'e}dard} A.,   {Rolland} B.,  2019, \mn@doi [\apj] {10.3847/1538-4357/ab153a}, \href {https://ui.adsabs.harvard.edu/abs/2019ApJ...876...67B} {876, 67}

\bibitem[\protect\citeauthoryear{{Bildsten}, {Shen}, {Weinberg}  \& {Nelemans}}{{Bildsten} et~al.}{2007}]{Bildsten2007}
{Bildsten} L.,  {Shen} K.~J.,  {Weinberg} N.~N.,   {Nelemans} G.,  2007, \mn@doi [\apjl] {10.1086/519489}, \href {https://ui.adsabs.harvard.edu/abs/2007ApJ...662L..95B} {662, L95}

\bibitem[\protect\citeauthoryear{{Blouin}, {Dufour}, {Thibeault}  \& {Allard}}{{Blouin} et~al.}{2019}]{Blouin2019}
{Blouin} S.,  {Dufour} P.,  {Thibeault} C.,   {Allard} N.~F.,  2019, \mn@doi [\apj] {10.3847/1538-4357/ab1f82}, \href {https://ui.adsabs.harvard.edu/abs/2019ApJ...878...63B} {878, 63}

\bibitem[\protect\citeauthoryear{{Bragaglia}, {Greggio}, {Renzini}  \& {D'Odorico}}{{Bragaglia} et~al.}{1990}]{1990ApJ...365L..13B}
{Bragaglia} A.,  {Greggio} L.,  {Renzini} A.,   {D'Odorico} S.,  1990, \mn@doi [\apjl] {10.1086/185877}, \href {https://ui.adsabs.harvard.edu/abs/1990ApJ...365L..13B} {365, L13}

\bibitem[\protect\citeauthoryear{{Brown}, {Kilic}, {Kenyon}  \& {Gianninas}}{{Brown} et~al.}{2016}]{Brown2016}
{Brown} W.~R.,  {Kilic} M.,  {Kenyon} S.~J.,   {Gianninas} A.,  2016, \mn@doi [\apj] {10.3847/0004-637X/824/1/46}, \href {https://ui.adsabs.harvard.edu/abs/2016ApJ...824...46B} {824, 46}

\bibitem[\protect\citeauthoryear{{Brown} et~al.,}{{Brown} et~al.}{2020}]{Brown2020elmNorthFinal}
{Brown} W.~R.,  et~al., 2020, \mn@doi [\apj] {10.3847/1538-4357/ab63cd}, \href {https://ui.adsabs.harvard.edu/abs/2020ApJ...889...49B} {889, 49}

\bibitem[\protect\citeauthoryear{{Burdge} et~al.,}{{Burdge} et~al.}{2020}]{Burdge2020systematic}
{Burdge} K.~B.,  et~al., 2020, \mn@doi [\apj] {10.3847/1538-4357/abc261}, \href {https://ui.adsabs.harvard.edu/abs/2020ApJ...905...32B} {905, 32}

\bibitem[\protect\citeauthoryear{{Caron}, {Bergeron}, {Blouin}  \& {Leggett}}{{Caron} et~al.}{2023}]{Caron2023}
{Caron} A.,  {Bergeron} P.,  {Blouin} S.,   {Leggett} S.~K.,  2023, \mn@doi [\mnras] {10.1093/mnras/stac3733}, \href {https://ui.adsabs.harvard.edu/abs/2023MNRAS.519.4529C} {519, 4529}

\bibitem[\protect\citeauthoryear{{Chambers} \& {Pan-STARRS Team}}{{Chambers} \& {Pan-STARRS Team}}{2018}]{Panstarrs}
{Chambers} K.,  {Pan-STARRS Team} 2018, in American Astronomical Society Meeting Abstracts \#231. p. 102.01

\bibitem[\protect\citeauthoryear{{Cukanovaite}, {Tremblay}, {Bergeron}, {Freytag}, {Ludwig}  \& {Steffen}}{{Cukanovaite} et~al.}{2021}]{ElenaCukanovaite2021_3D_DB}
{Cukanovaite} E.,  {Tremblay} P.-E.,  {Bergeron} P.,  {Freytag} B.,  {Ludwig} H.-G.,   {Steffen} M.,  2021, \mn@doi [\mnras] {10.1093/mnras/staa3684}, \href {https://ui.adsabs.harvard.edu/abs/2021MNRAS.501.5274C} {501, 5274}

\bibitem[\protect\citeauthoryear{{Cunningham}, {Tremblay}  \& {W. O'Brien}}{{Cunningham} et~al.}{2024}]{Cunningham2024}
{Cunningham} T.,  {Tremblay} P.-E.,   {W. O'Brien} M.,  2024, \mn@doi [\mnras] {10.1093/mnras/stad3275}, \href {https://ui.adsabs.harvard.edu/abs/2024MNRAS.527.3602C} {527, 3602}

\bibitem[\protect\citeauthoryear{{El-Badry} \& {Rix}}{{El-Badry} \& {Rix}}{2018}]{ElBadry2018commonProperMotion}
{El-Badry} K.,  {Rix} H.-W.,  2018, \mn@doi [\mnras] {10.1093/mnras/sty2186}, \href {https://ui.adsabs.harvard.edu/abs/2018MNRAS.480.4884E} {480, 4884}

\bibitem[\protect\citeauthoryear{{El-Badry}, {Rix}  \& {Heintz}}{{El-Badry} et~al.}{2021}]{ElBadry2021edr3properMotion}
{El-Badry} K.,  {Rix} H.-W.,   {Heintz} T.~M.,  2021, \mn@doi [\mnras] {10.1093/mnras/stab323}, \href {https://ui.adsabs.harvard.edu/abs/2021MNRAS.506.2269E} {506, 2269}

\bibitem[\protect\citeauthoryear{{El-Badry} et~al.,}{{El-Badry} et~al.}{2023}]{ElBadry2023d6}
{El-Badry} K.,  et~al., 2023, \mn@doi [The Open Journal of Astrophysics] {10.21105/astro.2306.03914}, \href {https://ui.adsabs.harvard.edu/abs/2023OJAp....6E..28E} {6, 28}

\bibitem[\protect\citeauthoryear{{Foreman-Mackey}, {Hogg}, {Lang}  \& {Goodman}}{{Foreman-Mackey} et~al.}{2013}]{ForemanMackey2013emcee}
{Foreman-Mackey} D.,  {Hogg} D.~W.,  {Lang} D.,   {Goodman} J.,  2013, \mn@doi [\pasp] {10.1086/670067}, \href {https://ui.adsabs.harvard.edu/abs/2013PASP..125..306F} {125, 306}

\bibitem[\protect\citeauthoryear{{Foss}, {Wade}  \& {Green}}{{Foss} et~al.}{1991}]{1991ApJ...374..281F}
{Foss} D.,  {Wade} R.~A.,   {Green} R.~F.,  1991, \mn@doi [\apj] {10.1086/170116}, \href {https://ui.adsabs.harvard.edu/abs/1991ApJ...374..281F} {374, 281}

\bibitem[\protect\citeauthoryear{{Gaia Collaboration} et~al.,}{{Gaia Collaboration} et~al.}{2016}]{GaiaDR1}
{Gaia Collaboration} et~al., 2016, \mn@doi [\aap] {10.1051/0004-6361/201629512}, \href {https://ui.adsabs.harvard.edu/abs/2016A&A...595A...2G} {595, A2}

\bibitem[\protect\citeauthoryear{{Gaia Collaboration} et~al.,}{{Gaia Collaboration} et~al.}{2018}]{GaiaDR2}
{Gaia Collaboration} et~al., 2018, \mn@doi [\aap] {10.1051/0004-6361/201833051}, \href {https://ui.adsabs.harvard.edu/abs/2018A&A...616A...1G} {616, A1}

\bibitem[\protect\citeauthoryear{{Gaia Collaboration} et~al.,}{{Gaia Collaboration} et~al.}{2021}]{GaiaEDR3_2021}
{Gaia Collaboration} et~al., 2021, \mn@doi [\aap] {10.1051/0004-6361/202039657}, \href {https://ui.adsabs.harvard.edu/abs/2021A&A...649A...1G} {649, A1}

\bibitem[\protect\citeauthoryear{{Gaia Collaboration} et~al.,}{{Gaia Collaboration} et~al.}{2023}]{GaiaDR3_2023}
{Gaia Collaboration} et~al., 2023, \mn@doi [\aap] {10.1051/0004-6361/202243940}, \href {https://ui.adsabs.harvard.edu/abs/2023A&A...674A...1G} {674, A1}

\bibitem[\protect\citeauthoryear{{Geier}, {Nesslinger}, {Heber}, {Przybilla}, {Napiwotzki}  \& {Kudritzki}}{{Geier} et~al.}{2007}]{Geier2007}
{Geier} S.,  {Nesslinger} S.,  {Heber} U.,  {Przybilla} N.,  {Napiwotzki} R.,   {Kudritzki} R.~P.,  2007, \mn@doi [\aap] {10.1051/0004-6361:20066098}, \href {https://ui.adsabs.harvard.edu/abs/2007A&A...464..299G} {464, 299}

\bibitem[\protect\citeauthoryear{{Geier}, {Heber}, {Kupfer}  \& {Napiwotzki}}{{Geier} et~al.}{2010}]{2010A&A...515A..37G}
{Geier} S.,  {Heber} U.,  {Kupfer} T.,   {Napiwotzki} R.,  2010, \mn@doi [\aap] {10.1051/0004-6361/200912545}, \href {https://ui.adsabs.harvard.edu/abs/2010A&A...515A..37G} {515, A37}

\bibitem[\protect\citeauthoryear{{Geier}, {Napiwotzki}, {Heber}  \& {Nelemans}}{{Geier} et~al.}{2011}]{2011A&A...528L..16G}
{Geier} S.,  {Napiwotzki} R.,  {Heber} U.,   {Nelemans} G.,  2011, \mn@doi [\aap] {10.1051/0004-6361/201116641}, \href {https://ui.adsabs.harvard.edu/abs/2011A&A...528L..16G} {528, L16}

\bibitem[\protect\citeauthoryear{{Genest-Beaulieu} \& {Bergeron}}{{Genest-Beaulieu} \& {Bergeron}}{2019}]{GenestBeaulieu2019}
{Genest-Beaulieu} C.,  {Bergeron} P.,  2019, \mn@doi [\apj] {10.3847/1538-4357/aafac6}, \href {https://ui.adsabs.harvard.edu/abs/2019ApJ...871..169G} {871, 169}

\bibitem[\protect\citeauthoryear{{Gentile Fusillo} et~al.,}{{Gentile Fusillo} et~al.}{2019}]{NicolaDR2}
{Gentile Fusillo} N.~P.,  et~al., 2019, \mn@doi [\mnras] {10.1093/mnras/sty3016}, \href {https://ui.adsabs.harvard.edu/abs/2019MNRAS.482.4570G} {482, 4570}

\bibitem[\protect\citeauthoryear{{Gentile Fusillo} et~al.,}{{Gentile Fusillo} et~al.}{2021}]{NicolaGaia2021}
{Gentile Fusillo} N.~P.,  et~al., 2021, \mn@doi [\mnras] {10.1093/mnras/stab2672}, \href {https://ui.adsabs.harvard.edu/abs/2021MNRAS.508.3877G} {508, 3877}

\bibitem[\protect\citeauthoryear{{Gianninas}, {Bergeron}  \& {Ruiz}}{{Gianninas} et~al.}{2011}]{Gianninas2011}
{Gianninas} A.,  {Bergeron} P.,   {Ruiz} M.~T.,  2011, \mn@doi [\apj] {10.1088/0004-637X/743/2/138}, \href {https://ui.adsabs.harvard.edu/abs/2011ApJ...743..138G} {743, 138}

\bibitem[\protect\citeauthoryear{{Gordon}, {Clayton}, {Decleir}, {Fitzpatrick}, {Massa}, {Misselt}  \& {Tollerud}}{{Gordon} et~al.}{2023}]{Gordon2023}
{Gordon} K.~D.,  {Clayton} G.~C.,  {Decleir} M.,  {Fitzpatrick} E.~L.,  {Massa} D.,  {Misselt} K.~A.,   {Tollerud} E.~J.,  2023, \mn@doi [\apj] {10.3847/1538-4357/accb59}, \href {https://ui.adsabs.harvard.edu/abs/2023ApJ...950...86G} {950, 86}

\bibitem[\protect\citeauthoryear{{Guo}, {Zhao}, {Tziamtzis}, {Liu}, {Li}, {Zhang}, {Hou}  \& {Wang}}{{Guo} et~al.}{2015}]{Guo2015}
{Guo} J.,  {Zhao} J.,  {Tziamtzis} A.,  {Liu} J.,  {Li} L.,  {Zhang} Y.,  {Hou} Y.,   {Wang} Y.,  2015, \mn@doi [\mnras] {10.1093/mnras/stv2104}, \href {https://ui.adsabs.harvard.edu/abs/2015MNRAS.454.2787G} {454, 2787}

\bibitem[\protect\citeauthoryear{{Halbwachs} et~al.,}{{Halbwachs} et~al.}{2023}]{Halbwachs2023gaiaNSS}
{Halbwachs} J.-L.,  et~al., 2023, \mn@doi [\aap] {10.1051/0004-6361/202243969}, \href {https://ui.adsabs.harvard.edu/abs/2023A&A...674A...9H} {674, A9}

\bibitem[\protect\citeauthoryear{{Heber}, {Napiwotzki}  \& {Reid}}{{Heber} et~al.}{1997}]{Heber1997}
{Heber} U.,  {Napiwotzki} R.,   {Reid} I.~N.,  1997, \aap, \href {https://ui.adsabs.harvard.edu/abs/1997A&A...323..819H} {323, 819}

\bibitem[\protect\citeauthoryear{{Heintz}, {Hermes}, {El-Badry}, {Walsh}, {van Saders}, {Fields}  \& {Koester}}{{Heintz} et~al.}{2022}]{Heintz2022WDageDWD}
{Heintz} T.~M.,  {Hermes} J.~J.,  {El-Badry} K.,  {Walsh} C.,  {van Saders} J.~L.,  {Fields} C.~E.,   {Koester} D.,  2022, \mn@doi [\apj] {10.3847/1538-4357/ac78d9}, \href {https://ui.adsabs.harvard.edu/abs/2022ApJ...934..148H} {934, 148}

\bibitem[\protect\citeauthoryear{{Heintz}, {Hermes}, {Tremblay}, {Baya Ould Rouis}, {Redding}, {Kaiser}  \& {van Saders}}{{Heintz} et~al.}{2024}]{Heintz2024}
{Heintz} T.~M.,  {Hermes} J.~J.,  {Tremblay} P.~E.,  {Baya Ould Rouis} L.,  {Redding} J.~S.,  {Kaiser} B.~C.,   {van Saders} J.~L.,  2024, arXiv e-prints, \href {https://ui.adsabs.harvard.edu/abs/2024arXiv240502423H} {p. arXiv:2405.02423}

\bibitem[\protect\citeauthoryear{{Iben} \& {Tutukov}}{{Iben} \& {Tutukov}}{1984}]{IbenTutukov1984}
{Iben} I. J.,  {Tutukov} A.~V.,  1984, \mn@doi [\apj] {10.1086/162455}, \href {https://ui.adsabs.harvard.edu/abs/1984ApJ...284..719I} {284, 719}

\bibitem[\protect\citeauthoryear{{Istrate}, {Marchant}, {Tauris}, {Langer}, {Stancliffe}  \& {Grassitelli}}{{Istrate} et~al.}{2016}]{Istrate2016}
{Istrate} A.~G.,  {Marchant} P.,  {Tauris} T.~M.,  {Langer} N.,  {Stancliffe} R.~J.,   {Grassitelli} L.,  2016, \mn@doi [\aap] {10.1051/0004-6361/201628874}, \href {https://ui.adsabs.harvard.edu/abs/2016A&A...595A..35I} {595, A35}

\bibitem[\protect\citeauthoryear{{Izquierdo}, {G{\"a}nsicke}, {Rodr{\'\i}guez-Gil}, {Koester}, {Toloza}, {Gentile Fusillo}, {Pala}  \& {Tremblay}}{{Izquierdo} et~al.}{2023}]{Izquierdo2023}
{Izquierdo} P.,  {G{\"a}nsicke} B.~T.,  {Rodr{\'\i}guez-Gil} P.,  {Koester} D.,  {Toloza} O.,  {Gentile Fusillo} N.~P.,  {Pala} A.~F.,   {Tremblay} P.-E.,  2023, \mn@doi [\mnras] {10.1093/mnras/stad282}, \href {https://ui.adsabs.harvard.edu/abs/2023MNRAS.520.2843I} {520, 2843}

\bibitem[\protect\citeauthoryear{{Jim{\'e}nez-Esteban}, {Torres}, {Rebassa-Mansergas}, {Cruz}, {Murillo-Ojeda}, {Solano}, {Rodrigo}  \& {Camisassa}}{{Jim{\'e}nez-Esteban} et~al.}{2023}]{Esteban2023}
{Jim{\'e}nez-Esteban} F.~M.,  {Torres} S.,  {Rebassa-Mansergas} A.,  {Cruz} P.,  {Murillo-Ojeda} R.,  {Solano} E.,  {Rodrigo} C.,   {Camisassa} M.~E.,  2023, \mn@doi [\mnras] {10.1093/mnras/stac3382}, \href {https://ui.adsabs.harvard.edu/abs/2023MNRAS.518.5106J} {518, 5106}

\bibitem[\protect\citeauthoryear{{Karl}, {Napiwotzki}, {Nelemans}, {Christlieb}, {Koester}, {Heber}  \& {Reimers}}{{Karl} et~al.}{2003}]{2003A&A...410..663K}
{Karl} C.~A.,  {Napiwotzki} R.,  {Nelemans} G.,  {Christlieb} N.,  {Koester} D.,  {Heber} U.,   {Reimers} D.,  2003, \mn@doi [\aap] {10.1051/0004-6361:20031278}, \href {https://ui.adsabs.harvard.edu/abs/2003A&A...410..663K} {410, 663}

\bibitem[\protect\citeauthoryear{{Kawka}, {Briggs}, {Vennes}, {Ferrario}, {Paunzen}  \& {Wickramasinghe}}{{Kawka} et~al.}{2017a}]{Kawka2017}
{Kawka} A.,  {Briggs} G.~P.,  {Vennes} S.,  {Ferrario} L.,  {Paunzen} E.,   {Wickramasinghe} D.~T.,  2017a, \mn@doi [\mnras] {10.1093/mnras/stw3149}, \href {https://ui.adsabs.harvard.edu/abs/2017MNRAS.466.1127K} {466, 1127}

\bibitem[\protect\citeauthoryear{{Kawka}, {Briggs}, {Vennes}, {Ferrario}, {Paunzen}  \& {Wickramasinghe}}{{Kawka} et~al.}{2017b}]{2017MagneticDWDsuperChandra}
{Kawka} A.,  {Briggs} G.~P.,  {Vennes} S.,  {Ferrario} L.,  {Paunzen} E.,   {Wickramasinghe} D.~T.,  2017b, \mn@doi [\mnras] {10.1093/mnras/stw3149}, \href {https://ui.adsabs.harvard.edu/abs/2017MNRAS.466.1127K} {466, 1127}

\bibitem[\protect\citeauthoryear{{Keller}, {Breedt}, {Hodgkin}, {Belokurov}, {Wild}, {Garc{\'\i}a-Soriano}  \& {Wise}}{{Keller} et~al.}{2022}]{Keller2022ztfGaia}
{Keller} P.~M.,  {Breedt} E.,  {Hodgkin} S.,  {Belokurov} V.,  {Wild} J.,  {Garc{\'\i}a-Soriano} I.,   {Wise} J.~L.,  2022, \mn@doi [\mnras] {10.1093/mnras/stab3293}, \href {https://ui.adsabs.harvard.edu/abs/2022MNRAS.509.4171K} {509, 4171}

\bibitem[\protect\citeauthoryear{{Kilic}, {Brown}, {Allende Prieto}, {Ag{\"u}eros}, {Heinke}  \& {Kenyon}}{{Kilic} et~al.}{2011}]{Kilic2011a}
{Kilic} M.,  {Brown} W.~R.,  {Allende Prieto} C.,  {Ag{\"u}eros} M.~A.,  {Heinke} C.,   {Kenyon} S.~J.,  2011, \mn@doi [\apj] {10.1088/0004-637X/727/1/3}, \href {https://ui.adsabs.harvard.edu/abs/2011ApJ...727....3K} {727, 3}

\bibitem[\protect\citeauthoryear{{Kilic}, {B{\'e}dard}, {Bergeron}  \& {Kosakowski}}{{Kilic} et~al.}{2020a}]{Kilic2020twoDoubleLined}
{Kilic} M.,  {B{\'e}dard} A.,  {Bergeron} P.,   {Kosakowski} A.,  2020a, \mn@doi [\mnras] {10.1093/mnras/staa466}, \href {https://ui.adsabs.harvard.edu/abs/2020MNRAS.493.2805K} {493, 2805}

\bibitem[\protect\citeauthoryear{{Kilic}, {Bergeron}, {Kosakowski}, {Brown}, {Ag{\"u}eros}  \& {Blouin}}{{Kilic} et~al.}{2020b}]{Kilic2020sdss100pc}
{Kilic} M.,  {Bergeron} P.,  {Kosakowski} A.,  {Brown} W.~R.,  {Ag{\"u}eros} M.~A.,   {Blouin} S.,  2020b, \mn@doi [\apj] {10.3847/1538-4357/ab9b8d}, \href {https://ui.adsabs.harvard.edu/abs/2020ApJ...898...84K} {898, 84}

\bibitem[\protect\citeauthoryear{{Kilic}, {B{\'e}dard}  \& {Bergeron}}{{Kilic} et~al.}{2021}]{Kilic2021HiddenInPlainSight}
{Kilic} M.,  {B{\'e}dard} A.,   {Bergeron} P.,  2021, \mn@doi [\mnras] {10.1093/mnras/stab439}, \href {https://ui.adsabs.harvard.edu/abs/2021MNRAS.502.4972K} {502, 4972}

\bibitem[\protect\citeauthoryear{{Kilic} et~al.,}{{Kilic} et~al.}{2023}]{Kilic2023massive}
{Kilic} M.,  et~al., 2023, \mn@doi [\mnras] {10.1093/mnras/stac3182}, \href {https://ui.adsabs.harvard.edu/abs/2023MNRAS.518.2341K} {518, 2341}

\bibitem[\protect\citeauthoryear{{Kleinman} et~al.,}{{Kleinman} et~al.}{2013}]{Kleinman2013}
{Kleinman} S.~J.,  et~al., 2013, \mn@doi [\apjs] {10.1088/0067-0049/204/1/5}, \href {https://ui.adsabs.harvard.edu/abs/2013ApJS..204....5K} {204, 5}

\bibitem[\protect\citeauthoryear{{Koester}, {Dreizler}, {Weidemann}  \& {Allard}}{{Koester} et~al.}{1998}]{Koester1998}
{Koester} D.,  {Dreizler} S.,  {Weidemann} V.,   {Allard} N.~F.,  1998, \aap, \href {https://ui.adsabs.harvard.edu/abs/1998A&A...338..612K} {338, 612}

\bibitem[\protect\citeauthoryear{{Korol}, {Rossi}, {Groot}, {Nelemans}, {Toonen}  \& {Brown}}{{Korol} et~al.}{2017}]{Korol2017prospects}
{Korol} V.,  {Rossi} E.~M.,  {Groot} P.~J.,  {Nelemans} G.,  {Toonen} S.,   {Brown} A. G.~A.,  2017, \mn@doi [\mnras] {10.1093/mnras/stx1285}, \href {https://ui.adsabs.harvard.edu/abs/2017MNRAS.470.1894K} {470, 1894}

\bibitem[\protect\citeauthoryear{{Korol}, {Koop}  \& {Rossi}}{{Korol} et~al.}{2018}]{Korol2018detectabilityDWDs}
{Korol} V.,  {Koop} O.,   {Rossi} E.~M.,  2018, \mn@doi [\apjl] {10.3847/2041-8213/aae587}, \href {https://ui.adsabs.harvard.edu/abs/2018ApJ...866L..20K} {866, L20}

\bibitem[\protect\citeauthoryear{{Korol}, {Hallakoun}, {Toonen}  \& {Karnesis}}{{Korol} et~al.}{2022a}]{Korol2022}
{Korol} V.,  {Hallakoun} N.,  {Toonen} S.,   {Karnesis} N.,  2022a, \mn@doi [\mnras] {10.1093/mnras/stac415}, \href {https://ui.adsabs.harvard.edu/abs/2022MNRAS.511.5936K} {511, 5936}

\bibitem[\protect\citeauthoryear{{Korol}, {Belokurov}  \& {Toonen}}{{Korol} et~al.}{2022b}]{Korol2022gapDWDseparationGaia}
{Korol} V.,  {Belokurov} V.,   {Toonen} S.,  2022b, \mn@doi [\mnras] {10.1093/mnras/stac1686}, \href {https://ui.adsabs.harvard.edu/abs/2022MNRAS.515.1228K} {515, 1228}

\bibitem[\protect\citeauthoryear{{Kosakowski}, {Kilic}, {Brown}  \& {Gianninas}}{{Kosakowski} et~al.}{2020}]{KosakowskiELMsouth1}
{Kosakowski} A.,  {Kilic} M.,  {Brown} W.~R.,   {Gianninas} A.,  2020, \mn@doi [\apj] {10.3847/1538-4357/ab8300}, \href {https://ui.adsabs.harvard.edu/abs/2020ApJ...894...53K} {894, 53}

\bibitem[\protect\citeauthoryear{{Kosakowski}, {Brown}, {Kilic}, {Kupfer}, {B{\'e}dard}, {Gianninas}, {Ag{\"u}eros}  \& {Barrientos}}{{Kosakowski} et~al.}{2023}]{Kosakowski2023elmSouth}
{Kosakowski} A.,  {Brown} W.~R.,  {Kilic} M.,  {Kupfer} T.,  {B{\'e}dard} A.,  {Gianninas} A.,  {Ag{\"u}eros} M.~A.,   {Barrientos} M.,  2023, \mn@doi [\apj] {10.3847/1538-4357/acd187}, \href {https://ui.adsabs.harvard.edu/abs/2023ApJ...950..141K} {950, 141}

\bibitem[\protect\citeauthoryear{{Kupfer} et~al.,}{{Kupfer} et~al.}{2024}]{Kupfer2024}
{Kupfer} T.,  et~al., 2024, \mn@doi [\apj] {10.3847/1538-4357/ad2068}, \href {https://ui.adsabs.harvard.edu/abs/2024ApJ...963..100K} {963, 100}

\bibitem[\protect\citeauthoryear{{Lallement}, {Vergely}, {Babusiaux}  \& {Cox}}{{Lallement} et~al.}{2022}]{Lallement2022}
{Lallement} R.,  {Vergely} J.~L.,  {Babusiaux} C.,   {Cox} N.~L.~J.,  2022, \mn@doi [\aap] {10.1051/0004-6361/202142846}, \href {https://ui.adsabs.harvard.edu/abs/2022A&A...661A.147L} {661, A147}

\bibitem[\protect\citeauthoryear{{Lamberts}, {Blunt}, {Littenberg}, {Garrison-Kimmel}, {Kupfer}  \& {Sanderson}}{{Lamberts} et~al.}{2019}]{Lamberts2019}
{Lamberts} A.,  {Blunt} S.,  {Littenberg} T.~B.,  {Garrison-Kimmel} S.,  {Kupfer} T.,   {Sanderson} R.~E.,  2019, \mn@doi [\mnras] {10.1093/mnras/stz2834}, \href {https://ui.adsabs.harvard.edu/abs/2019MNRAS.490.5888L} {490, 5888}

\bibitem[\protect\citeauthoryear{{Li}, {Chen}, {Chen}  \& {Han}}{{Li} et~al.}{2019}]{Li2019formationOfELMs}
{Li} Z.,  {Chen} X.,  {Chen} H.-L.,   {Han} Z.,  2019, \mn@doi [\apj] {10.3847/1538-4357/aaf9a1}, \href {https://ui.adsabs.harvard.edu/abs/2019ApJ...871..148L} {871, 148}

\bibitem[\protect\citeauthoryear{{Liebert}, {Bergeron}  \& {Holberg}}{{Liebert} et~al.}{2005}]{Liebert2005}
{Liebert} J.,  {Bergeron} P.,   {Holberg} J.~B.,  2005, \mn@doi [\apjs] {10.1086/425738}, \href {https://ui.adsabs.harvard.edu/abs/2005ApJS..156...47L} {156, 47}

\bibitem[\protect\citeauthoryear{{Limoges}, {Bergeron}  \& {L{\'e}pine}}{{Limoges} et~al.}{2015}]{Limoges2015}
{Limoges} M.~M.,  {Bergeron} P.,   {L{\'e}pine} S.,  2015, \mn@doi [\apjs] {10.1088/0067-0049/219/2/19}, \href {https://ui.adsabs.harvard.edu/abs/2015ApJS..219...19L} {219, 19}

\bibitem[\protect\citeauthoryear{{Maoz} \& {Hallakoun}}{{Maoz} \& {Hallakoun}}{2017}]{Maoz2017spy}
{Maoz} D.,  {Hallakoun} N.,  2017, \mn@doi [\mnras] {10.1093/mnras/stx102}, \href {https://ui.adsabs.harvard.edu/abs/2017MNRAS.467.1414M} {467, 1414}

\bibitem[\protect\citeauthoryear{{Maoz} \& {Mannucci}}{{Maoz} \& {Mannucci}}{2012}]{Maoz2012supernovaReview}
{Maoz} D.,  {Mannucci} F.,  2012, \mn@doi [\pasa] {10.1071/AS11052}, \href {https://ui.adsabs.harvard.edu/abs/2012PASA...29..447M} {29, 447}

\bibitem[\protect\citeauthoryear{{Maoz}, {Mannucci}  \& {Nelemans}}{{Maoz} et~al.}{2014}]{Maoz2014Type1aProgenitors}
{Maoz} D.,  {Mannucci} F.,   {Nelemans} G.,  2014, \mn@doi [\araa] {10.1146/annurev-astro-082812-141031}, \href {https://ui.adsabs.harvard.edu/abs/2014ARA&A..52..107M} {52, 107}

\bibitem[\protect\citeauthoryear{{Maoz}, {Hallakoun}  \& {Badenes}}{{Maoz} et~al.}{2018}]{MaozHallakounBadenes2018}
{Maoz} D.,  {Hallakoun} N.,   {Badenes} C.,  2018, \mn@doi [\mnras] {10.1093/mnras/sty339}, \href {https://ui.adsabs.harvard.edu/abs/2018MNRAS.476.2584M} {476, 2584}

\bibitem[\protect\citeauthoryear{{Marsh}}{{Marsh}}{1989}]{Marsh1989optimalExtraction}
{Marsh} T.~R.,  1989, \mn@doi [\pasp] {10.1086/132570}, \href {https://ui.adsabs.harvard.edu/abs/1989PASP..101.1032M} {101, 1032}

\bibitem[\protect\citeauthoryear{{Marsh}}{{Marsh}}{1995}]{1995MNRAS.275L...1M}
{Marsh} T.~R.,  1995, \mn@doi [\mnras] {10.1093/mnras/275.1.L1}, \href {https://ui.adsabs.harvard.edu/abs/1995MNRAS.275L...1M} {275, L1}

\bibitem[\protect\citeauthoryear{{Marsh}}{{Marsh}}{2019}]{Marsh2019Molly}
{Marsh} T.,  2019, {molly: 1D astronomical spectra analyzer}, Astrophysics Source Code Library, record ascl:1907.012 (\mn@eprint {ascl} {1907.012})

\bibitem[\protect\citeauthoryear{{Marsh}, {Dhillon}  \& {Duck}}{{Marsh} et~al.}{1995}]{1995MNRAS.275..828M}
{Marsh} T.~R.,  {Dhillon} V.~S.,   {Duck} S.~R.,  1995, \mn@doi [\mnras] {10.1093/mnras/275.3.828}, \href {https://ui.adsabs.harvard.edu/abs/1995MNRAS.275..828M} {275, 828}

\bibitem[\protect\citeauthoryear{{Maxted} \& {Marsh}}{{Maxted} \& {Marsh}}{1999}]{MaxtedMarsh1999}
{Maxted} P.~F.~L.,  {Marsh} T.~R.,  1999, \mn@doi [\mnras] {10.1046/j.1365-8711.1999.02635.x}, \href {https://ui.adsabs.harvard.edu/abs/1999MNRAS.307..122M} {307, 122}

\bibitem[\protect\citeauthoryear{{McCleery} et~al.,}{{McCleery} et~al.}{2020}]{McCleery2020_40pcNorth}
{McCleery} J.,  et~al., 2020, \mn@doi [\mnras] {10.1093/mnras/staa2030}, \href {https://ui.adsabs.harvard.edu/abs/2020MNRAS.499.1890M} {499, 1890}

\bibitem[\protect\citeauthoryear{{Moran}, {Marsh}  \& {Bragaglia}}{{Moran} et~al.}{1997}]{1997MNRAS.288..538M}
{Moran} C.,  {Marsh} T.~R.,   {Bragaglia} A.,  1997, \mn@doi [\mnras] {10.1093/mnras/288.2.538}, \href {https://ui.adsabs.harvard.edu/abs/1997MNRAS.288..538M} {288, 538}

\bibitem[\protect\citeauthoryear{Munday}{Munday}{2024}]{Munday2024WDBASS}
Munday J.,  2024, JamesMunday98/WD-BASS: v1.0.0, \mn@doi{10.5281/zenodo.11188044}, \url {https://doi.org/10.5281/zenodo.11188044}

\bibitem[\protect\citeauthoryear{{Nandez}, {Ivanova}  \& {Lombardi}}{{Nandez} et~al.}{2015}]{Nandez2015RecombinationEnergyDWDformation}
{Nandez} J.~L.~A.,  {Ivanova} N.,   {Lombardi} J.~C.~J.,  2015, \mn@doi [\mnras] {10.1093/mnrasl/slv043}, \href {https://ui.adsabs.harvard.edu/abs/2015MNRAS.450L..39N} {450, L39}

\bibitem[\protect\citeauthoryear{{Napiwotzki}}{{Napiwotzki}}{1997}]{Napiwotzki1997}
{Napiwotzki} R.,  1997, \aap, \href {https://ui.adsabs.harvard.edu/abs/1997A&A...322..256N} {322, 256}

\bibitem[\protect\citeauthoryear{{Napiwotzki}, {Edelmann}, {Heber}, {Karl}, {Drechsel}, {Pauli}  \& {Christlieb}}{{Napiwotzki} et~al.}{2001}]{2001A&A...378L..17N}
{Napiwotzki} R.,  {Edelmann} H.,  {Heber} U.,  {Karl} C.,  {Drechsel} H.,  {Pauli} E.~M.,   {Christlieb} N.,  2001, \mn@doi [\aap] {10.1051/0004-6361:20011223}, \href {https://ui.adsabs.harvard.edu/abs/2001A&A...378L..17N} {378, L17}

\bibitem[\protect\citeauthoryear{{Napiwotzki} et~al.,}{{Napiwotzki} et~al.}{2002}]{2002A&A...386..957N}
{Napiwotzki} R.,  et~al., 2002, \mn@doi [\aap] {10.1051/0004-6361:20020361}, \href {https://ui.adsabs.harvard.edu/abs/2002A&A...386..957N} {386, 957}

\bibitem[\protect\citeauthoryear{{Napiwotzki} et~al.,}{{Napiwotzki} et~al.}{2004}]{Napiwotzki2004fitsb2}
{Napiwotzki} R.,  et~al., 2004, in {Hilditch} R.~W.,  {Hensberge} H.,   {Pavlovski} K.,  eds,  Astronomical Society of the Pacific Conference Series Vol. 318, Spectroscopically and Spatially Resolving the Components of the Close Binary Stars. pp 402--410 (\mn@eprint {arXiv} {astro-ph/0403595}), \mn@doi{10.48550/arXiv.astro-ph/0403595}

\bibitem[\protect\citeauthoryear{{Napiwotzki} et~al.,}{{Napiwotzki} et~al.}{2020}]{Napiwotzki2020spy}
{Napiwotzki} R.,  et~al., 2020, \mn@doi [\aap] {10.1051/0004-6361/201629648}, \href {https://ui.adsabs.harvard.edu/abs/2020A&A...638A.131N} {638, A131}

\bibitem[\protect\citeauthoryear{{Nelemans}, {Yungelson}, {Portegies Zwart}  \& {Verbunt}}{{Nelemans} et~al.}{2001}]{Nelemans2001closeWDs}
{Nelemans} G.,  {Yungelson} L.~R.,  {Portegies Zwart} S.~F.,   {Verbunt} F.,  2001, \mn@doi [\aap] {10.1051/0004-6361:20000147}, \href {https://ui.adsabs.harvard.edu/abs/2001A&A...365..491N} {365, 491}

\bibitem[\protect\citeauthoryear{{Nelemans} et~al.,}{{Nelemans} et~al.}{2005}]{2005A&A...440.1087N}
{Nelemans} G.,  et~al., 2005, \mn@doi [\aap] {10.1051/0004-6361:20053174}, \href {https://ui.adsabs.harvard.edu/abs/2005A&A...440.1087N} {440, 1087}

\bibitem[\protect\citeauthoryear{{O'Brien} et~al.,}{{O'Brien} et~al.}{2023}]{OBrien2023_40pcSouth}
{O'Brien} M.~W.,  et~al., 2023, \mn@doi [\mnras] {10.1093/mnras/stac3303}, \href {https://ui.adsabs.harvard.edu/abs/2023MNRAS.518.3055O} {518, 3055}

\bibitem[\protect\citeauthoryear{{O'Brien} et~al.,}{{O'Brien} et~al.}{2024}]{OBrien2024}
{O'Brien} M.~W.,  et~al., 2024, \mn@doi [\mnras] {10.1093/mnras/stad3773}, \href {https://ui.adsabs.harvard.edu/abs/2024MNRAS.527.8687O} {527, 8687}

\bibitem[\protect\citeauthoryear{{Parsons} et~al.,}{{Parsons} et~al.}{2017}]{Parsons2017}
{Parsons} S.~G.,  et~al., 2017, \mn@doi [\mnras] {10.1093/mnras/stx1522}, \href {https://ui.adsabs.harvard.edu/abs/2017MNRAS.470.4473P} {470, 4473}

\bibitem[\protect\citeauthoryear{{Pelisoli} et~al.,}{{Pelisoli} et~al.}{2021}]{Pelisoli2021type1aWDsubdwarf}
{Pelisoli} I.,  et~al., 2021, \mn@doi [Nature Astronomy] {10.1038/s41550-021-01413-0}, \href {https://ui.adsabs.harvard.edu/abs/2021NatAs...5.1052P} {5, 1052}

\bibitem[\protect\citeauthoryear{{Ren}, {Li}, {Ma}, {Cheng}, {Huang}, {Tang}  \& {Hu}}{{Ren} et~al.}{2023}]{Ren2023}
{Ren} L.,  {Li} C.,  {Ma} B.,  {Cheng} S.,  {Huang} S.-J.,  {Tang} B.,   {Hu} Y.-m.,  2023, \mn@doi [\apjs] {10.3847/1538-4365/aca09e}, \href {https://ui.adsabs.harvard.edu/abs/2023ApJS..264...39R} {264, 39}

\bibitem[\protect\citeauthoryear{{Robinson} \& {Shafter}}{{Robinson} \& {Shafter}}{1987}]{1987ApJ...322..296R}
{Robinson} E.~L.,  {Shafter} A.~W.,  1987, \mn@doi [\apj] {10.1086/165725}, \href {https://ui.adsabs.harvard.edu/abs/1987ApJ...322..296R} {322, 296}

\bibitem[\protect\citeauthoryear{{Romero}, {Kepler}, {Joyce}, {Lauffer}  \& {C{\'o}rsico}}{{Romero} et~al.}{2019}]{Romero2019}
{Romero} A.~D.,  {Kepler} S.~O.,  {Joyce} S.~R.~G.,  {Lauffer} G.~R.,   {C{\'o}rsico} A.~H.,  2019, \mn@doi [\mnras] {10.1093/mnras/stz160}, \href {https://ui.adsabs.harvard.edu/abs/2019MNRAS.484.2711R} {484, 2711}

\bibitem[\protect\citeauthoryear{{Saffer}, {Liebert}  \& {Olszewski}}{{Saffer} et~al.}{1988}]{Saffer1988}
{Saffer} R.~A.,  {Liebert} J.,   {Olszewski} E.~W.,  1988, \mn@doi [\apj] {10.1086/166888}, \href {https://ui.adsabs.harvard.edu/abs/1988ApJ...334..947S} {334, 947}

\bibitem[\protect\citeauthoryear{{Sahu} et~al.,}{{Sahu} et~al.}{2023}]{Sahu2023}
{Sahu} S.,  et~al., 2023, \mn@doi [\mnras] {10.1093/mnras/stad2663}, \href {https://ui.adsabs.harvard.edu/abs/2023MNRAS.526.5800S} {526, 5800}

\bibitem[\protect\citeauthoryear{{Schreiber}, {Belloni}, {G{\"a}nsicke}, {Parsons}  \& {Zorotovic}}{{Schreiber} et~al.}{2021}]{Schreiber2021natureBfield}
{Schreiber} M.~R.,  {Belloni} D.,  {G{\"a}nsicke} B.~T.,  {Parsons} S.~G.,   {Zorotovic} M.,  2021, \mn@doi [Nature Astronomy] {10.1038/s41550-021-01346-8}, \href {https://ui.adsabs.harvard.edu/abs/2021NatAs...5..648S} {5, 648}

\bibitem[\protect\citeauthoryear{{Schreiber}, {Belloni}, {Zorotovic}, {Zapata}, {G{\"a}nsicke}  \& {Parsons}}{{Schreiber} et~al.}{2022}]{Schreiber2022}
{Schreiber} M.~R.,  {Belloni} D.,  {Zorotovic} M.,  {Zapata} S.,  {G{\"a}nsicke} B.~T.,   {Parsons} S.~G.,  2022, \mn@doi [\mnras] {10.1093/mnras/stac1076}, \href {https://ui.adsabs.harvard.edu/abs/2022MNRAS.513.3090S} {513, 3090}

\bibitem[\protect\citeauthoryear{{Schwab}}{{Schwab}}{2018}]{Schwab2018}
{Schwab} J.,  2018, \mn@doi [\mnras] {10.1093/mnras/sty586}, \href {https://ui.adsabs.harvard.edu/abs/2018MNRAS.476.5303S} {476, 5303}

\bibitem[\protect\citeauthoryear{{Schwab}}{{Schwab}}{2019}]{Schwab2019}
{Schwab} J.,  2019, \mn@doi [\apj] {10.3847/1538-4357/ab425d}, \href {https://ui.adsabs.harvard.edu/abs/2019ApJ...885...27S} {885, 27}

\bibitem[\protect\citeauthoryear{{Schwab}, {Shen}, {Quataert}, {Dan}  \& {Rosswog}}{{Schwab} et~al.}{2012}]{Schwab2012evolutionToWDmergerRemnants}
{Schwab} J.,  {Shen} K.~J.,  {Quataert} E.,  {Dan} M.,   {Rosswog} S.,  2012, \mn@doi [\mnras] {10.1111/j.1365-2966.2012.21993.x}, \href {https://ui.adsabs.harvard.edu/abs/2012MNRAS.427..190S} {427, 190}

\bibitem[\protect\citeauthoryear{{Shen}}{{Shen}}{2015}]{Shen2015}
{Shen} K.~J.,  2015, \mn@doi [\apjl] {10.1088/2041-8205/805/1/L6}, \href {https://ui.adsabs.harvard.edu/abs/2015ApJ...805L...6S} {805, L6}

\bibitem[\protect\citeauthoryear{{Shen}, {Bildsten}, {Kasen}  \& {Quataert}}{{Shen} et~al.}{2012}]{Shen2012longTermEvolutionDWD}
{Shen} K.~J.,  {Bildsten} L.,  {Kasen} D.,   {Quataert} E.,  2012, \mn@doi [\apj] {10.1088/0004-637X/748/1/35}, \href {https://ui.adsabs.harvard.edu/abs/2012ApJ...748...35S} {748, 35}

\bibitem[\protect\citeauthoryear{{Shen} et~al.,}{{Shen} et~al.}{2018}]{Shen2018d6}
{Shen} K.~J.,  et~al., 2018, \mn@doi [\apj] {10.3847/1538-4357/aad55b}, \href {https://ui.adsabs.harvard.edu/abs/2018ApJ...865...15S} {865, 15}

\bibitem[\protect\citeauthoryear{{Solheim}}{{Solheim}}{2010}]{Soleheim2010}
{Solheim} J.~E.,  2010, \mn@doi [\pasp] {10.1086/656680}, \href {https://ui.adsabs.harvard.edu/abs/2010PASP..122.1133S} {122, 1133}

\bibitem[\protect\citeauthoryear{{Steen}, {Hermes}, {Guidry}, {Paiva}, {Farihi}, {Heintz}, {Ewing}  \& {Berry}}{{Steen} et~al.}{2024}]{Steen2024}
{Steen} M.,  {Hermes} J.~J.,  {Guidry} J.~A.,  {Paiva} A.,  {Farihi} J.,  {Heintz} T.~M.,  {Ewing} B.~B.,   {Berry} N.,  2024, \mn@doi [\apj] {10.3847/1538-4357/ad3e60}, \href {https://ui.adsabs.harvard.edu/abs/2024ApJ...967..166S} {967, 166}

\bibitem[\protect\citeauthoryear{{Temmink}, {Toonen}, {Zapartas}, {Justham}  \& {G{\"a}nsicke}}{{Temmink} et~al.}{2020}]{Temmink2020}
{Temmink} K.~D.,  {Toonen} S.,  {Zapartas} E.,  {Justham} S.,   {G{\"a}nsicke} B.~T.,  2020, \mn@doi [\aap] {10.1051/0004-6361/201936889}, \href {https://ui.adsabs.harvard.edu/abs/2020A&A...636A..31T} {636, A31}

\bibitem[\protect\citeauthoryear{{Toonen}, {Nelemans}  \& {Portegies Zwart}}{{Toonen} et~al.}{2012}]{Toonen2012type1aCommonEnvelope}
{Toonen} S.,  {Nelemans} G.,   {Portegies Zwart} S.,  2012, \mn@doi [\aap] {10.1051/0004-6361/201218966}, \href {https://ui.adsabs.harvard.edu/abs/2012A&A...546A..70T} {546, A70}

\bibitem[\protect\citeauthoryear{{Toonen}, {Hollands}, {G{\"a}nsicke}  \& {Boekholt}}{{Toonen} et~al.}{2017}]{Toonen2017}
{Toonen} S.,  {Hollands} M.,  {G{\"a}nsicke} B.~T.,   {Boekholt} T.,  2017, \mn@doi [\aap] {10.1051/0004-6361/201629978}, \href {https://ui.adsabs.harvard.edu/abs/2017A&A...602A..16T} {602, A16}

\bibitem[\protect\citeauthoryear{{Tremblay} \& {Bergeron}}{{Tremblay} \& {Bergeron}}{2009}]{Tremblay2009}
{Tremblay} P.~E.,  {Bergeron} P.,  2009, \mn@doi [\apj] {10.1088/0004-637X/696/2/1755}, \href {https://ui.adsabs.harvard.edu/abs/2009ApJ...696.1755T} {696, 1755}

\bibitem[\protect\citeauthoryear{{Tremblay}, {Ludwig}, {Steffen}  \& {Freytag}}{{Tremblay} et~al.}{2013a}]{Tremblay2013aPureH3DmodelsWDs}
{Tremblay} P.~E.,  {Ludwig} H.~G.,  {Steffen} M.,   {Freytag} B.,  2013a, \mn@doi [\aap] {10.1051/0004-6361/201220813}, \href {https://ui.adsabs.harvard.edu/abs/2013A&A...552A..13T} {552, A13}

\bibitem[\protect\citeauthoryear{{Tremblay}, {Ludwig}, {Steffen}  \& {Freytag}}{{Tremblay} et~al.}{2013b}]{Tremblay2013}
{Tremblay} P.~E.,  {Ludwig} H.~G.,  {Steffen} M.,   {Freytag} B.,  2013b, \mn@doi [\aap] {10.1051/0004-6361/201322318}, \href {https://ui.adsabs.harvard.edu/abs/2013A&A...559A.104T} {559, A104}

\bibitem[\protect\citeauthoryear{{Tremblay}, {Gianninas}, {Kilic}, {Ludwig}, {Steffen}, {Freytag}  \& {Hermes}}{{Tremblay} et~al.}{2015}]{Tremblay2015}
{Tremblay} P.~E.,  {Gianninas} A.,  {Kilic} M.,  {Ludwig} H.~G.,  {Steffen} M.,  {Freytag} B.,   {Hermes} J.~J.,  2015, \mn@doi [\apj] {10.1088/0004-637X/809/2/148}, \href {https://ui.adsabs.harvard.edu/abs/2015ApJ...809..148T} {809, 148}

\bibitem[\protect\citeauthoryear{{Tremblay}, {Cummings}, {Kalirai}, {G{\"a}nsicke}, {Gentile-Fusillo}  \& {Raddi}}{{Tremblay} et~al.}{2016}]{Tremblay2016}
{Tremblay} P.~E.,  {Cummings} J.,  {Kalirai} J.~S.,  {G{\"a}nsicke} B.~T.,  {Gentile-Fusillo} N.,   {Raddi} R.,  2016, \mn@doi [\mnras] {10.1093/mnras/stw1447}, \href {https://ui.adsabs.harvard.edu/abs/2016MNRAS.461.2100T} {461, 2100}

\bibitem[\protect\citeauthoryear{{Tremblay}, {Cukanovaite}, {Gentile Fusillo}, {Cunningham}  \& {Hollands}}{{Tremblay} et~al.}{2019}]{Tremblay2019accuracyDADBgaiaDR2}
{Tremblay} P.~E.,  {Cukanovaite} E.,  {Gentile Fusillo} N.~P.,  {Cunningham} T.,   {Hollands} M.~A.,  2019, \mn@doi [\mnras] {10.1093/mnras/sty3067}, \href {https://ui.adsabs.harvard.edu/abs/2019MNRAS.482.5222T} {482, 5222}

\bibitem[\protect\citeauthoryear{{Webbink}}{{Webbink}}{1984}]{Webbink1984DWDprogenitorsRCrB}
{Webbink} R.~F.,  1984, \mn@doi [\apj] {10.1086/161701}, \href {https://ui.adsabs.harvard.edu/abs/1984ApJ...277..355W} {277, 355}

\bibitem[\protect\citeauthoryear{{Zenati}, {Toonen}  \& {Perets}}{{Zenati} et~al.}{2019}]{Zenati2019}
{Zenati} Y.,  {Toonen} S.,   {Perets} H.~B.,  2019, \mn@doi [\mnras] {10.1093/mnras/sty2723}, \href {https://ui.adsabs.harvard.edu/abs/2019MNRAS.482.1135Z} {482, 1135}

\bibitem[\protect\citeauthoryear{{Zhang} \& {Jeffery}}{{Zhang} \& {Jeffery}}{2012}]{Zhang2012FormationOfHeRichHotSubdwarfs}
{Zhang} X.,  {Jeffery} C.~S.,  2012, \mn@doi [\mnras] {10.1111/j.1365-2966.2011.19711.x}, \href {https://ui.adsabs.harvard.edu/abs/2012MNRAS.419..452Z} {419, 452}

\bibitem[\protect\citeauthoryear{{Zuckerman}, {Koester}, {Reid}  \& {H{\"u}nsch}}{{Zuckerman} et~al.}{2003}]{Zuckerman2003}
{Zuckerman} B.,  {Koester} D.,  {Reid} I.~N.,   {H{\"u}nsch} M.,  2003, \mn@doi [\apj] {10.1086/377492}, \href {https://ui.adsabs.harvard.edu/abs/2003ApJ...596..477Z} {596, 477}

\bibitem[\protect\citeauthoryear{{van Roestel} et~al.,}{{van Roestel} et~al.}{2022}]{Roestel2022}
{van Roestel} J.,  et~al., 2022, \mn@doi [\mnras] {10.1093/mnras/stab2421}, \href {https://ui.adsabs.harvard.edu/abs/2022MNRAS.512.5440V} {512, 5440}

\makeatother
\end{thebibliography}



\section{All double-lined systems}
All DWDs that exhibit a double-lined feature are displayed in Fig.~\ref{fig:appendixDoubleLiners1} and all of the likely double-lined DWDs that should be confirmed with further observations are found in Figs.~\ref{fig:appendixDoubleLiners2} and \ref{fig:appendixDoubleLiners4}. While hybrid fitting was used to derive atmospheric parameters, which includes the simultaneous fitting of survey photometry and our spectra, the data and model spectrum at H$\alpha$ alone are shown in these figures to emphasise the double-lined nature of each system.

\begin{figure*}
    \centering
    \includegraphics[clip,trim={2cm 3cm 2cm 4.5cm},width=\textwidth,keepaspectratio]{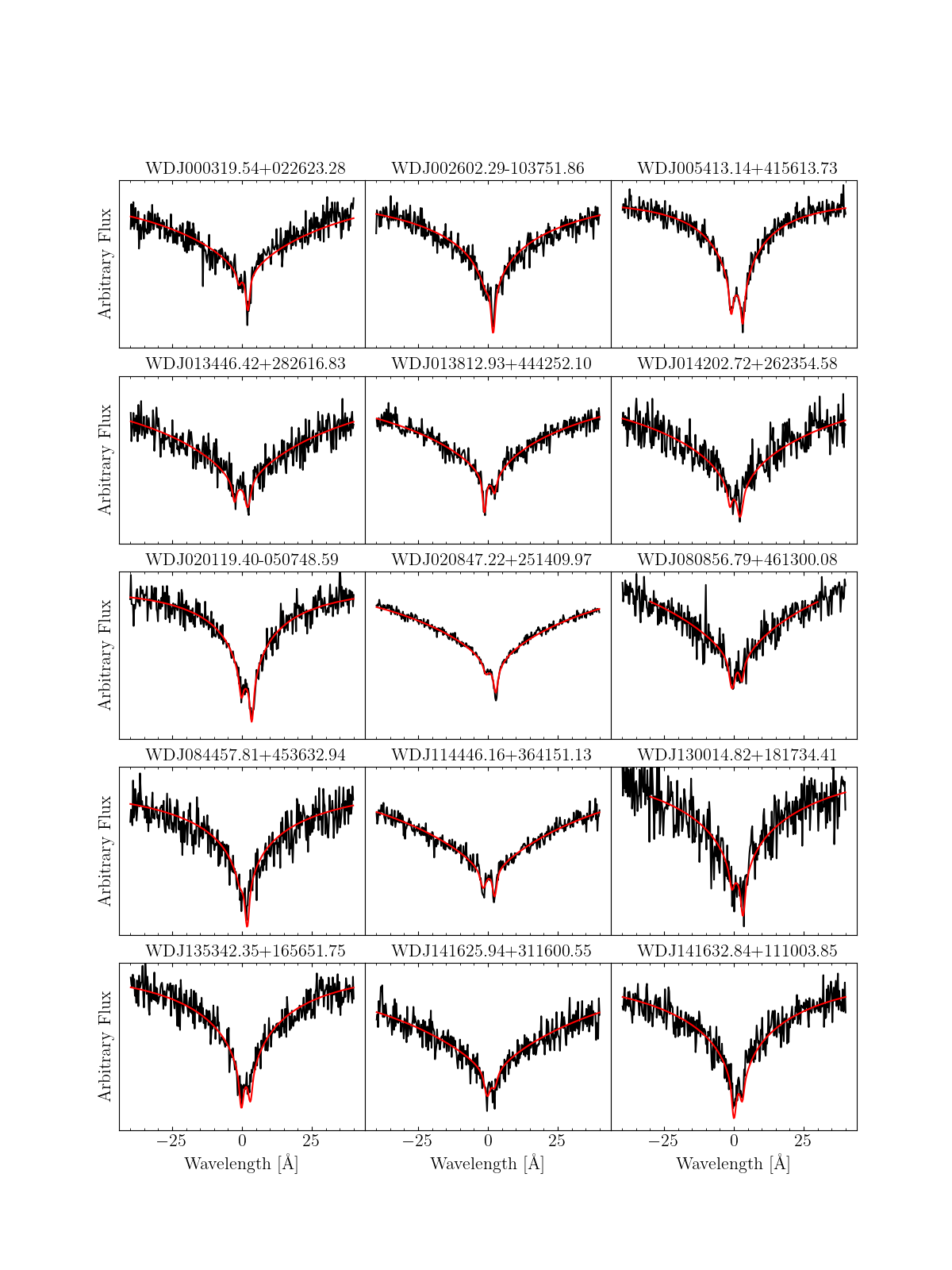}
    \caption{All model fits to the double-lined systems. Other Balmer lines and the photometry were fit simultaneously, but each spectrum is zoomed in at H$\alpha$.}
    \label{fig:appendixDoubleLiners1}
\end{figure*}

\begin{figure*}
    \centering
    \includegraphics[clip,trim={2cm 3cm 2cm 4.5cm},width=\textwidth,keepaspectratio]{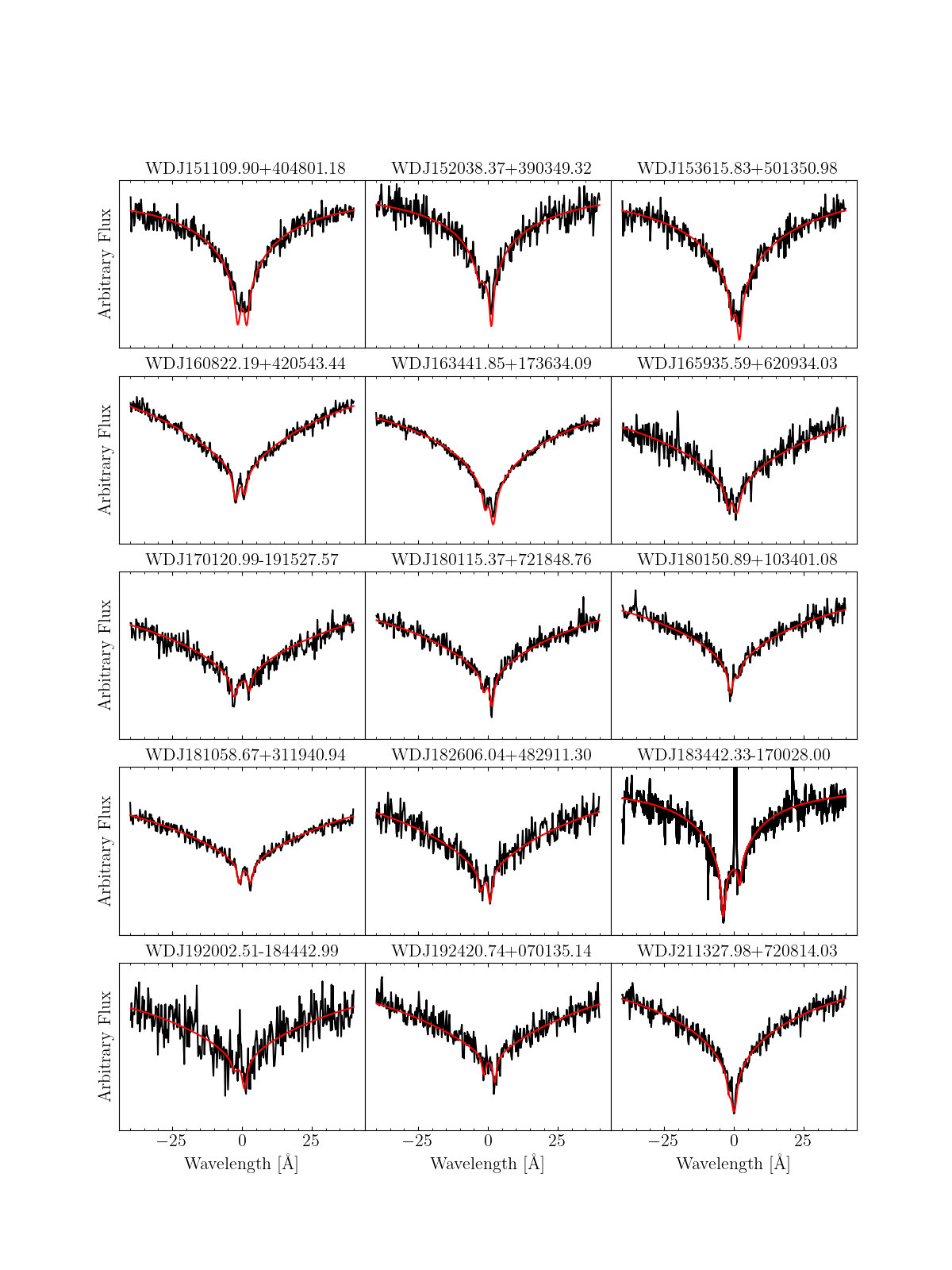}
    \caption*{Continued. The spectrum of WDJ192002.51-184442.99 has a low S/N in the red arm, but it is also double-lined at H$\beta$ with higher S/N data in the blue.}
\end{figure*}

\begin{figure*}
    \centering
    \includegraphics[clip,trim={0cm 0cm 0cm 0cm},width=\textwidth,keepaspectratio]{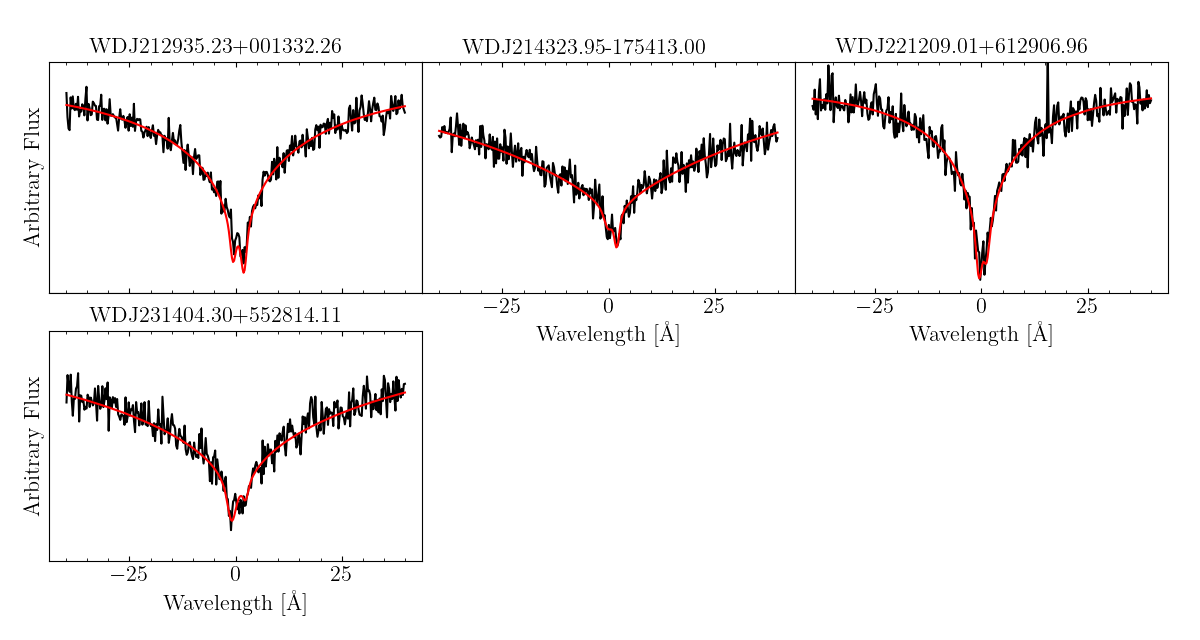}
    \caption*{Continued...}
    \includegraphics[clip,trim={0cm 0cm 0cm 0cm},width=\textwidth,keepaspectratio]{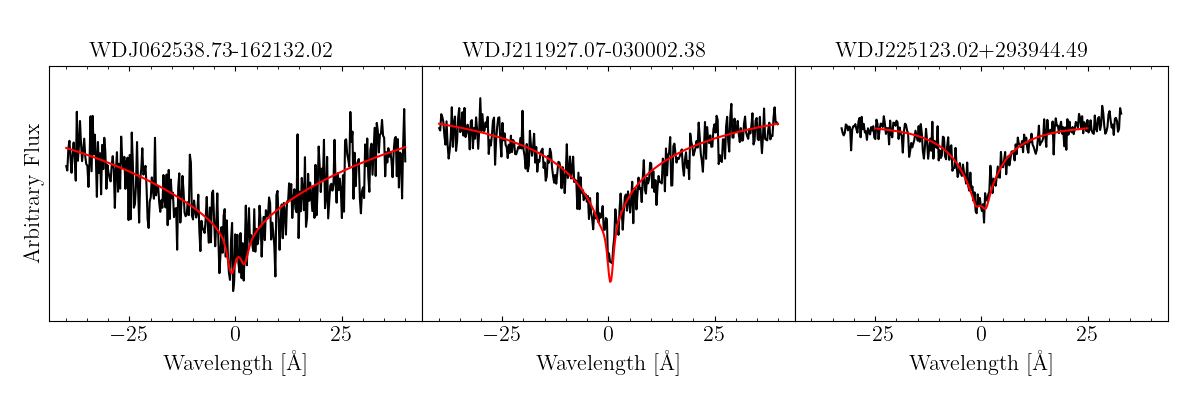}
    \caption{All model fits to likely double-lined systems that should be confirmed with higher S/N ratio data. We note that even though the depth of H$\alpha$ is shallow owing to the cool temperature of the two stars, in multiple exposures do the data and the best model fit hint at double-lined signature.}
    \label{fig:appendixDoubleLiners2}
\end{figure*}

\begin{figure*}
    \centering
    \includegraphics[clip,trim={2cm 1cm 2cm 2.5cm}, width=\textwidth,keepaspectratio]{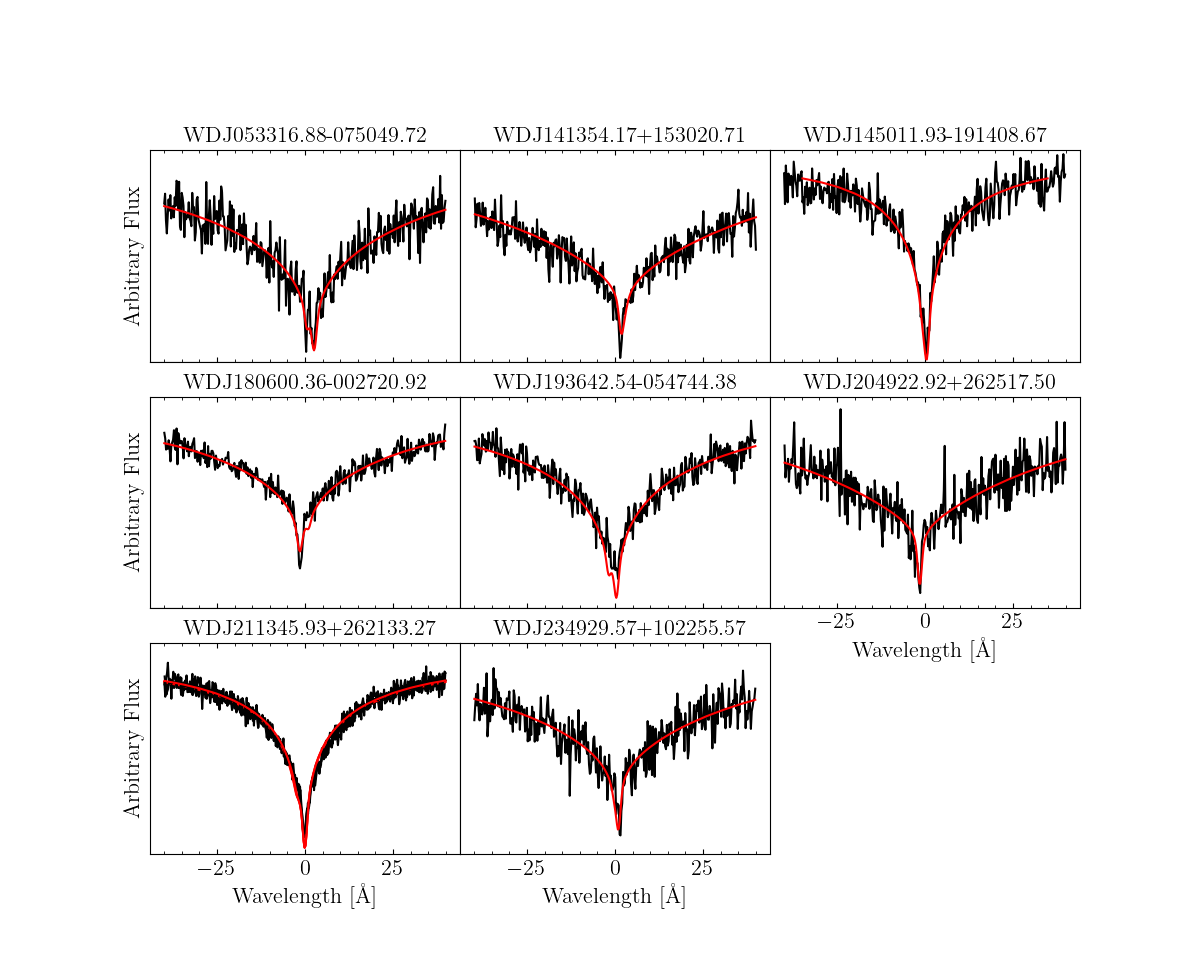}
    \caption{All model fits to the candidate double-lined systems that do not show a clear separation of the two stars but have asymmetric line profiles. Further observation is again encouraged to confirm if the sources are single- or double-lined. In the case that any are single-lined, the flux from an additional component is still required in the spectroscopic/photometric fit, and so these sources would be single-lined DWD binaries.}
    \label{fig:appendixDoubleLiners4}
\end{figure*}

\begin{figure*}
    \centering
    \includegraphics[clip,trim={0cm 0cm 0cm 0cm},width=\textwidth,keepaspectratio]{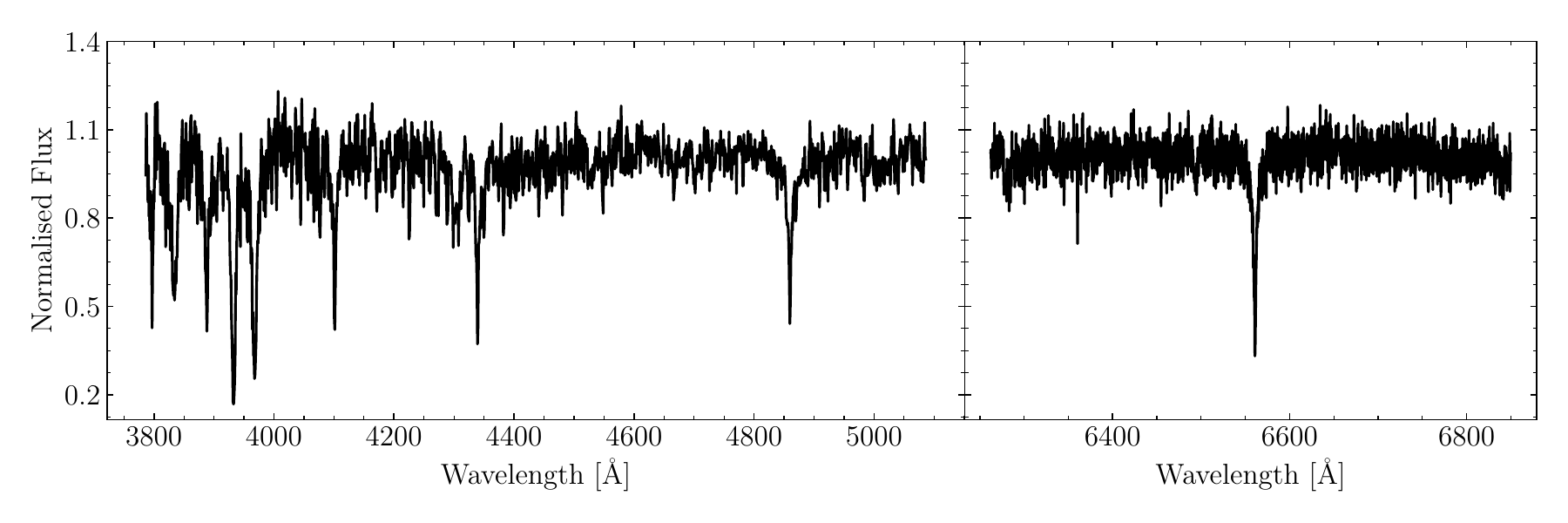}
    \caption{WDJ010343.47+555941.53, an evolved CV or a subdwarf with an F/G/K star companion.}
    \label{fig:AppendixEvolvedCV}
\end{figure*}

\section{Sensitivity to double-lined systems}
\begin{figure*}
    \centering
    \includegraphics[clip,trim={0cm 0cm 1.5cm 0cm},width=8.5cm,keepaspectratio]{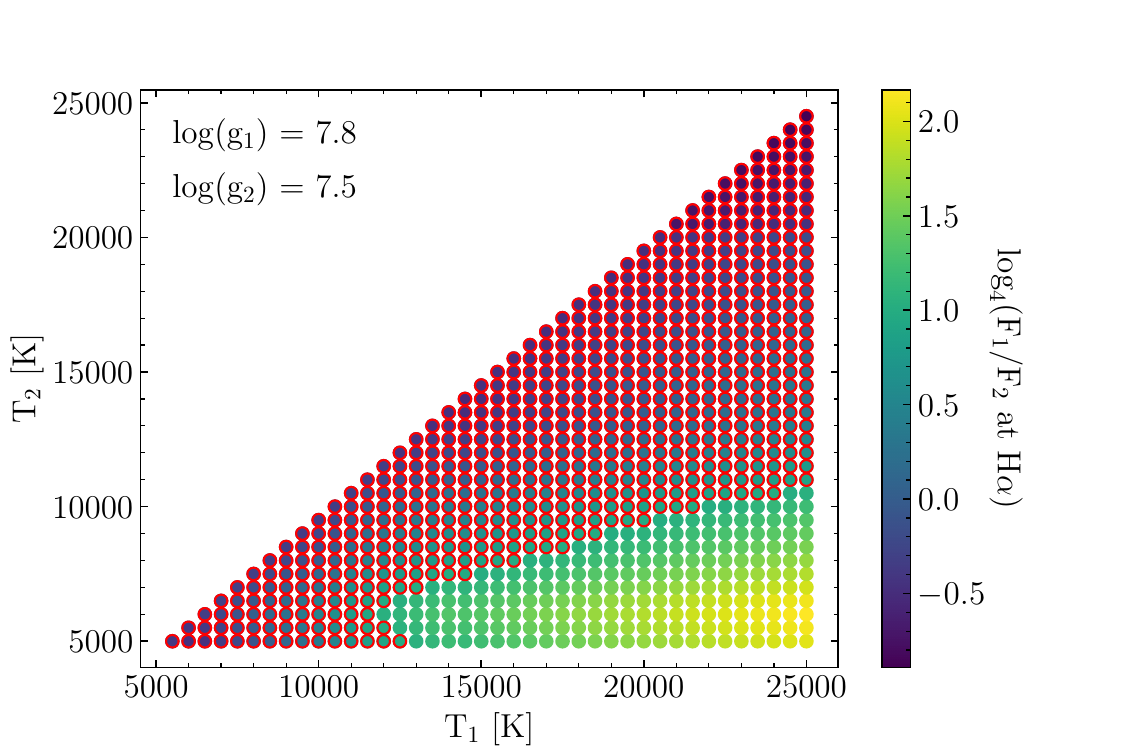}
    \includegraphics[clip,trim={0cm 0cm 1.5cm 0cm},width=8.5cm,keepaspectratio]{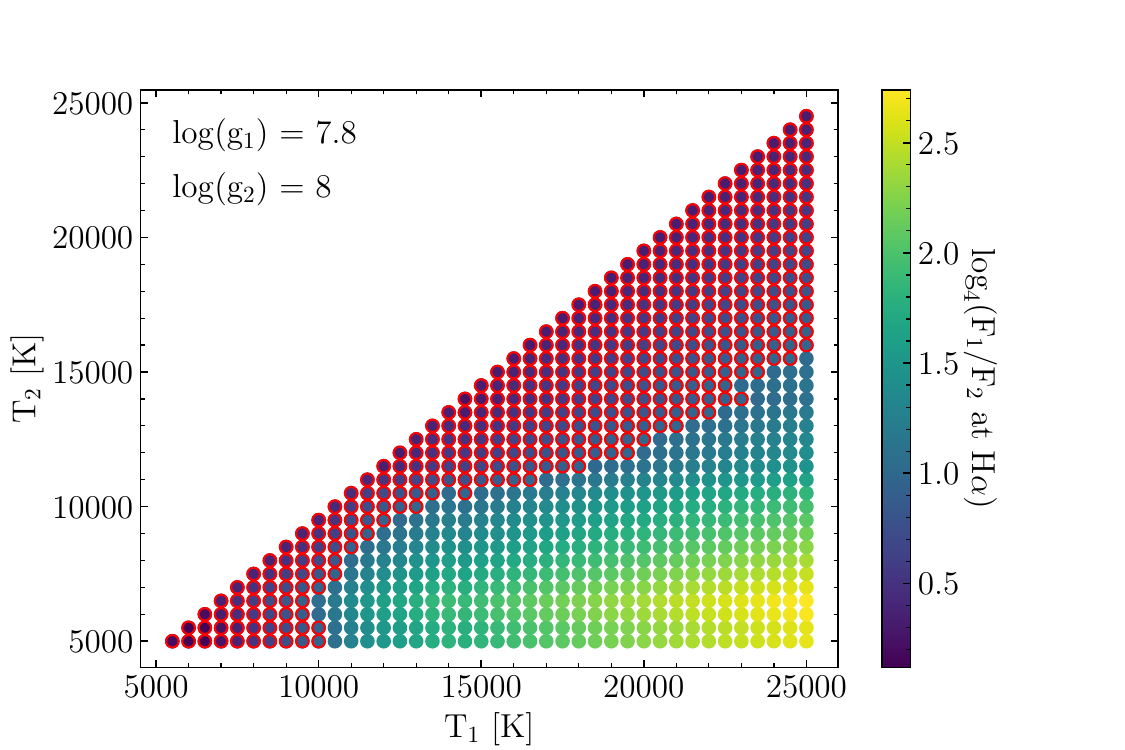}
    \includegraphics[clip,trim={0cm 0cm 1.5cm 0cm},width=8.5cm,keepaspectratio]{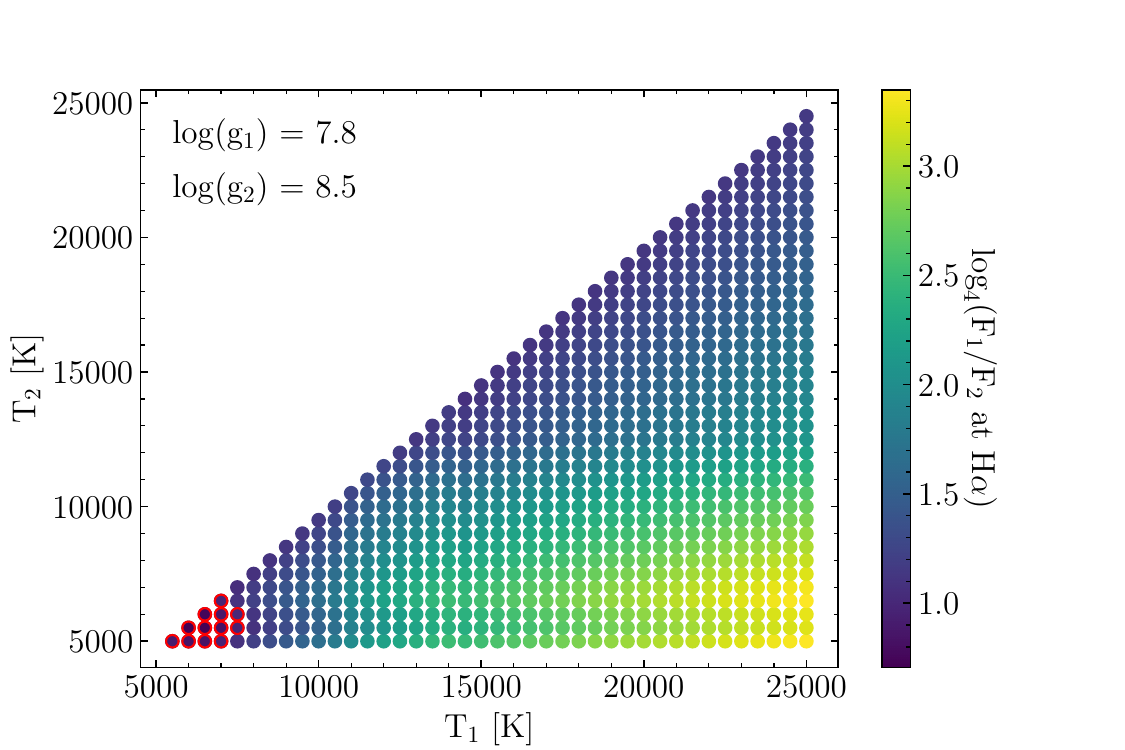}
    \caption{The sensitivity to double-lined systems for a given $\log(g_2)$ purely based on the relative flux contribution of each star from its temperature and radius. A $\log(g_1)=7.8$ is maintained, which is approximately the median surface gravity of the brighter star in the double-lined sample (Table~\ref{tab:doubleLinedParams}). Red circles around individual points indicate when the dimmer component contributes at least 25\% of the flux ($\log_4$ F$_1$/F$_2 =1$), as was desired in the selection criteria of the sample (Section~\ref{subsec:SampleSelection}). The temperatures of both stars are sampled in 500\,K intervals and the scaling from a different radius of each star is included through interpolation of the $\log$(g)-temperature-radius relationships outlined in Section~\ref{subsec:WDBASS_binary}. The relative flux contribution was sampled at the centre of H$\alpha$ and synthetic spectra were convolved to a spectral resolution of $R=8\,000$, matching that of our setup on the WHT. When considering observational biases in the DWD population, the information depicted in these plots should be combined with the detection efficiency in Fig.~\ref{fig:DetectionEfficiency}.}
    \label{fig:appendixSensitivityLogg}
\end{figure*}

The ability to detect a double-lined DWD is dependent on the relative flux contribution of the two stars, and hence there is a dependence on the stars' radii. We depict the impact of this effect in Fig.~\ref{fig:appendixSensitivityLogg} to determine for which combinations of temperature and surface gravity would the companion contribute at least 25\% of the flux at the centre of H$\alpha$ in our data. A minimum flux contribution of 25\% was used to construct the selection criteria in Section~\ref{subsec:SampleSelection}, and below this threshold a second star is unlikely to be detectable at the signal-to-noise ratio of our data.

\bsp	
\label{lastpage}
\end{document}